\documentclass[journal=jpcafh,manuscript=article]{achemso}

\usepackage{chemformula} 
\usepackage[T1]{fontenc} 
\usepackage{placeins}
\usepackage{multirow}
\usepackage{xr}
\usepackage{tensor}
\usepackage{amssymb}
\newcommand{\Mg}{MgB$_2$}
\newcommand{\YBCO}{YBa$_2$Cu$_3$O$_{7-\delta}$}
\newcommand{\TiSe}{$1T$-TiSe$_2$}
\newcommand{\Mgp}{Mg$^{2+}$}
\newcommand{\Bm}{B$^-$}

\newcommand{\eA}{$e\cdot$\AA$^{-3}$}
\newcommand{\eAA}{$e\cdot$\AA$^{-5}$}
\newcommand{\iA}{\AA$^{-1}$}

\newcommand{\Tc}{$T_\mathrm{c}$}

\newcommand{\chiV}{$\chi(T)$}

\newcommand{\rhoc}{$\rho(\mathbf{r}_c)$}

\newcommand{\rhoEHC}{$\rho_\mathrm{EHC}(\mathbf{r})$}
\newcommand{\rhoat}{$\rho_\mathrm{at}(\mathbf{r})$}
\newcommand{\rhocalc}{$\rho_\mathrm{c}(\mathbf{r})$}
\newcommand{\rhoobs}{$\rho_\mathrm{o}(\mathbf{r})$}
\newcommand{\rhoobsT}{$\rho_\mathrm{o}(\mathbf{r}, T)$}
\newcommand{\rhomem}{$\rho_\mathrm{MEM}(\mathbf{r})$}

\newcommand{\Drhototal}{$\Delta \rho_\mathrm{total}(\mathbf{r})$}
\newcommand{\Drholatt}{$\Delta \rho_\mathrm{latt}(\mathbf{r})$}
\newcommand{\DrhoADP}{$\Delta \rho_\mathrm{ADP}(\mathbf{r})$}
\newcommand{\DrhoMP}{$\Delta \rho_\mathrm{MP}(\mathbf{r})$}
\newcommand{\Drhores}{$\Delta \rho_\mathrm{res}(\mathbf{r})$}
\newcommand{\DrhoDFT}{$\Delta \rho_\mathrm{DFT}(\mathbf{r})$}

\newcommand{\Lc}{$\nabla^2 \rho(\mathbf{r}_c)$}

\newcommand{\sthl}{$\sin\theta/\lambda$}
\newcommand{\sthlmax}{$(\sin\theta/\lambda)_\mathrm{max}$}

\author{Jan Langmann}

\author{Hasan Kepenci}

\author{Georg Eickerling}
\email{georg.eickerling@uni-a.de}
\phone{+49 (0)821 598 3362}

\author{Kilian Batke}
\affiliation[Universit\"at Augsburg]
{CPM, Institut f\"ur Physik, Universit\"at Augsburg, 86159
  Augsburg, Germany}

\author{Anton Jesche}
\affiliation[Universit\"at Augsburg]
{Experimentalphysik VI,
  Zentrum f\"ur Elektronische Korrelation und Magnetismus,
  Institut f\"ur Physik, Universit\"at Augsburg,
  86159 Augsburg, Germany}

\author{Mingyu Xu}  
\affiliation[Ames Laboratory]
{The Ames Laboratory, Iowa State University, Ames, Iowa 50011, USA}
\alsoaffiliation[ISU]
{Department of Physics and Astronomy, Iowa State University, Ames, IA 50011, USA}

\author{Paul Canfield}
\affiliation[Ames Laboratory]
{The Ames Laboratory, Iowa State University, Ames, Iowa 50011, USA}
\alsoaffiliation[ISU]
{Department of Physics and Astronomy, Iowa State University, Ames, IA 50011, USA}

\author{Wolfgang Scherer}
\email{wolfgang.scherer@uni-a.de}
\phone{+49 (0)821 598 3350}
\affiliation[Universit\"at Augsburg]
{CPM, Institut f\"ur Physik, Universit\"at Augsburg, 86159
  Augsburg, Germany}

\title{X-ray charge-density studies -- a suitable probe for
  superconductivity?}

\abbreviations{}
\keywords{}

\begin{document}

\begin{tocentry}
  \includegraphics[width=8.25cm]{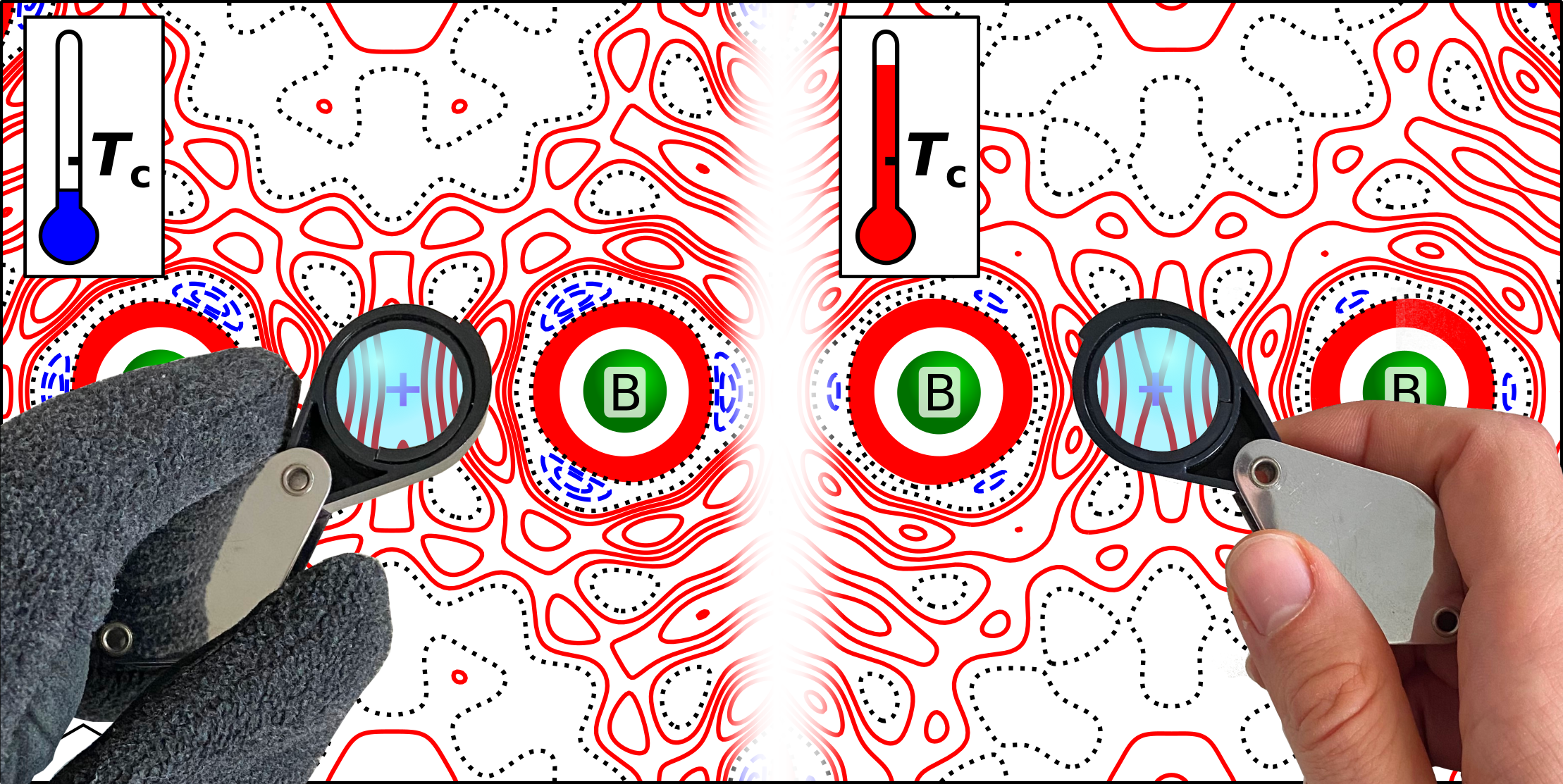}
\end{tocentry}

\begin{abstract}
  Case studies of \TiSe\ and \YBCO\ have demonstrated that x-ray
  diffraction (XRD) studies can be used to trace even subtle
  structural phase transitions which are inherently connected with the
  onset of superconductivity in these benchmark systems. Yet, the
  utility of XRD in the investigation of superconductors like \Mg\
  lacking an additional symmetry-breaking structural phase transition
  is not immediately evident.  Even though, high-resolution powder XRD
  experiments on \Mg\ in combination with maximum entropy method
(MEM)  analyses hinted at differences between the electron density
  distributions at room temperature and 15~K, \textit{i.e.}  below 
  the \Tc\ of approx. 39~K.  The high-resolution single-crystal
  XRD experiments in combination with multipolar refinements presented here can
  reproduce these results, but show that the observed temperature-dependent
  density changes are almost entirely due to a decrease of atomic
  displacement parameters as a natural consequence of reduced thermal
  vibration amplitude with decreasing temperature. Our investigations also shed
  new light on the presence or absence of magnesium vacancies in \Mg\
  samples -- a defect type claimed to control the superconducting
  properties of the compound. We propose that previous reports on the
  tendency of \Mg\ to form non-stoichiometric Mg$_{1-x}$B$_2$ phases
  ($1 - x \sim$~0.95) during high-temperature (HT) synthesis might result from
  the interpretation of XRD data of
  insufficient resolution and/or usage of inflexible refinement
  models. Indeed, advanced refinements based on an Extended Hansen-Coppens (EHC) multipolar model and high-resolution x-ray data, which consider explicitly the
  contraction of core and valence shells of the
  magnesium cations, do not provide any significant evidence for the formation of non-stoichiometric Mg$_{1-x}$B$_2$ phases during HT synthesis.
\end{abstract}

\section{Introduction}

When a crystal enters the superconducting state, its electronic
structure is subjected to fundamental changes. In case of BCS
superconductors, electrons are coupled into Cooper pairs by means of
lattice vibrations.\cite{Alexandrov} This has profound effects on
physical quantities probed by spectroscopic methods. Examples are the
superconducting gap as determined for example via tunneling and
vibrational
spectroscopy\cite{Poole-tunneling-spectroscopy,Poole-vibrational-spectroscopy}
or the Knight shift determined via nuclear magnetic resonance
spectroscopy.\cite{Poole-NMR}

\begin{figure}[htb]
  \centering
  \includegraphics[width=0.4\textwidth]{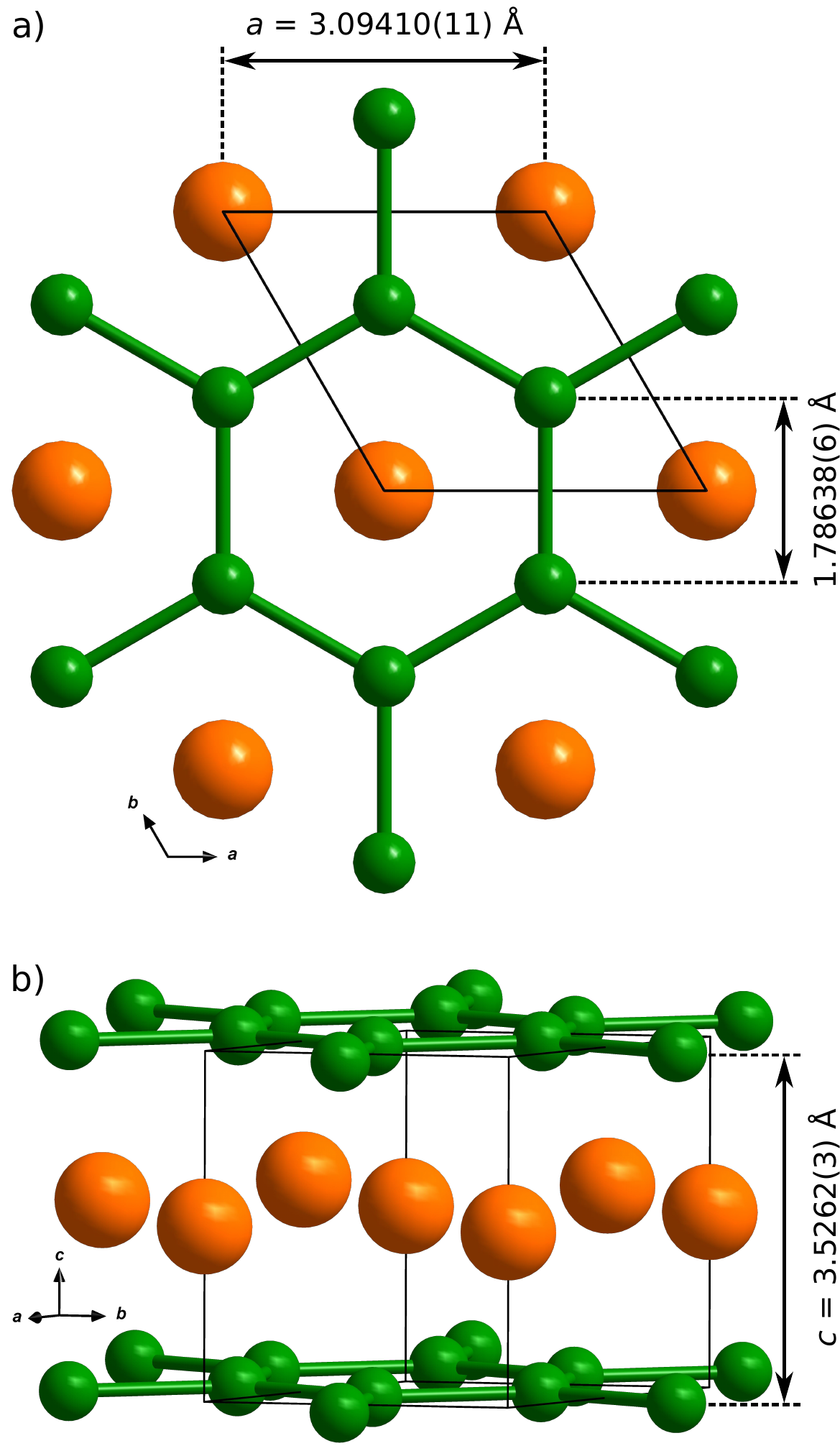}
  \caption{Ball-and-stick model of \Mg\ characterized by
    layers of magnesium (orange) and boron (green) atoms in the $a$-$b$ plane (a) that
    alternate along the stacking direction $c$ (b). Salient
    interatomic distances and lattice parameters at room temperature
    are specified.}
  \label{fig:mgb2-structure}
\end{figure}

It is not directly evident, however, whether the transition from the
normal- to the superconducting state has significant effects on the
(one-)electron density distribution as probed by x-ray diffraction
(XRD) experiments. \textit{A priori} the electron density distribution
in a real crystal is affected by various factors such as chemical
bonding as well as static and dynamic atomic displacements due to
disorder and thermal motion. All of these factors are possibly subject
to changes at the superconducting transition. But up to now,
experimental evidence for differences between normal- and
superconducting electron density distributions from XRD is rather
sparse and inconclusive. This even holds for \Mg\ (see
structural model in Fig.~\ref{fig:mgb2-structure}) as one of the
superconducting benchmark systems characterized by graphite-like boron layers with intercalated Mg cations.  Xue \textit{et
  al.}  reported an anomalous increase of powder XRD reflection
intensities at
$T_\mathrm{c}^\mathrm{onset} \approx$~39~K.\cite{Xue05,Nagamatsu01}
Yet, the authors only consider changes in the phonon spectrum as a
potential cause for their observation and neglect other
factors. Nishibori \textit{et al.}, by contrast, propose a change in
chemical bonding to occur between room temperature and
15~K.\cite{Nishibori01} Their statement is based on maximum entropy
method (MEM) analyses of powder XRD data that show an increasing
electron density at the B--B bond-critical point (BCP) from 0.9~\eA\ to
1.0~\eA\ at decreasing temperatures.\cite{Nishibori01} These BCPs represent saddle-points in the electron density distribution along
the boron-boron bonds (indicated by blue spheres in
Fig.~\ref{fig:mgb2-structure}).  But as the MEM technique is model free and
unable to differentiate between different causes of changes in the
electron density distribution, a modification of thermal smearing
might account for the observations of Nishibori \textit{et al.} as
well.

In this paper, we therefore use multipolar refinements of
temperature-dependent XRD data of single-crystalline \Mg\ samples to
 detect potential electron density redistributions
above and below \Tc,\cite{Xue05,Nishibori01} and to clarify their
origin. In contrast to earlier efforts, the application of flexible
multipolar models allows a decomposition of the total
electron density distribution into different contributions,
\textit{i.e.}  the sample composition, chemical bonding effects and
displacements of atoms from their equilibrium positions due to static
disorder or thermal
motion.\cite{Hansen78,Coppens-MP-model,Gatti-guided-tour}

\section{Methods} \label{sec:methods}

\subsection{Single crystal growth}
\label{sec:methods-single-crystal}

Single-crystalline samples of \Mg\ (crystal~1 and crystal~2; photographic images in
Fig.~\ref{fig:photo-crystal-1} and Fig.~\ref{fig:photo-crystal-2} of
the Supporting Information) were grown by a high-pressure and high-temperature synthesis similar to that described in Ref.~\citenum{Louden2022} (see the Supporting Information for details).

\subsection{Magnetization measurements}
\label{sec:methods-magn}

The temperature-dependent DC magnetization was determined for a \Mg\
single crystal (crystal~2) from the same batch as the one
used in the XRD experiments (see Fig.~\ref{fig:zfc-fc-crystal-2} of
the Supporting Information).  Measurements were performed under
zero-field cooling (ZFC) and field-cooling (FC) conditions in an applied field of
$\mu_0H =$~10~mT employing a QUANTUM-DESIGN MPMS3 SQUID magnetometer. 
Numerical volume integration of a crystal model created with the software
APEX2\cite{APEX} was used to obtain the sample volume for the
calculation of the volume susceptibility \chiV. 
We have found $V = 2.3(1) \cdot 10^{-3}$~mm$^3$.
The superconducting transition temperature was determined to $T_\mathrm{c} =$~37.5~K (onset).

The ZFC value of \chiV~=~-1.03 at $T =$~2~K corresponds to a superconducting volume fraction of close to 100~\% that is overestimated due to neglecting demagnetization effects and the volume error.  
The FC \chiV~amounts to roughly 45~\% of the ZFC value, which is larger than in previous studies (e.g. Ref.~\citenum{Nagamatsu01}) and indicates a lower concentration of pinning centers.

\subsection{XRD data collection and reduction}
\label{sec:methods-xrd-exp}

Two different setups were employed to collect the x-ray diffraction
(XRD) data in this paper: An experiment at 100(2)~K was performed on a
BRUKER Smart-Apex diffractometer featuring a D8 goniometer, an
INCOATEC AgK$_\alpha$ microfocus sealed-tube as x-ray source
($\lambda =$~0.56087~\AA) and a standard OXFORD open-flow N$_2$ cooler
(diffractometer~1).\cite{Cosier86} Experiments at temperatures from
room temperature down to sub-nitrogen cryogenic temperatures relied on
a HUBER four-circle Eulerian cradle goniometer equipped with a DECTRIS
Pilatus CdTe 300K pixel detector, an INCOATEC AgK$_\alpha$ microfocus
sealed-tube x-ray source ($\lambda =$~0.56087~\AA) and an ARS
closed-cycle helium cryocooler with an outer and inner beryllium
vacuum and radiation shield (diffractometer~2).  Data collection at
different temperatures was performed in direct sequence without sample
re-orientation. Recorded Bragg intensities were evaluated using the
\texttt{APEX2}\cite{APEX} (diffractometer~1) or the \texttt{EVAL14}
integration programs\cite{Duisenberg92,Duisenberg03}
(diffractometer~2) and subjected to scaling and absorption corrections
using the program \texttt{SADABS}\cite{Krause15}. More information on
collection and processing of the XRD data can be found in the
Supporting Information.

\subsection{IAM and multipolar refinements}
\label{sec:methods-iam-multipole}

Independent atom model (IAM) and multipolar refinements were performed using the programs
\texttt{JANA2006}\cite{Petricek14} or
\texttt{JANA2020}\cite{Petricek14} (see the Supporting Information for
information on local coordinate systems, refined parameters and
residual electron density maps of all multipolar models employed in this
paper). For better comparability of the results the construction of
atomic electron densities \rhoat\ and scattering factors was in all
cases based on atomic wave functions for neutral reference atoms as
implemented in the standard database of
\texttt{JANA2006}/\texttt{JANA2020}.\cite{Petricek14}

IAM refinements in this paper feature atomic scattering factors of
neutral magnesium and boron atoms. The atomic scattering factor of
magnesium was calculated using a configuration of
[$1s^2$,$2s^2$,$2p^6$]($3s^2$) with [$1s^2$,$2s^2$,$2p^6$] defining
the core and ($3s^2$) the valence.  For boron a [$1s^2$]($2s^1$,$2p^2$)
configuration was assumed.

To account for the expected charge transfer, in the Standard
Hansen-Coppens (SHC) multipolar refinements the $\zeta$-exponents for the valence radial function on magnesium
was calculated from the $s/p$ ratio of an [$1s^2$]($2s^2$,$2p^6$) configuration. At the same time, the
according $s/p$ ratio of boron was still calculated from an [$1s^2$]($2s^1$,$2p^2$) configuration.
Starting values of the valence population parameters $P_v$ were fixed
to 8 for magnesium and 4 for boron to account for an ionic starting
model composed of \Mgp\ and \Bm.  Aspherical density deformations were
described using one set of deformation functions centered at each of
the atoms in the asymmetric unit of \Mg\ and employing maximum
multipolar orders of $l_\mathrm{max} = 2$ (magnesium) and
$l_\mathrm{max} = 3$ (boron).  In contrast to earlier
attempts,\cite{Tsirelson03} our multipolar model allows for a free and
simultaneous least-squares refinement of its parameters without the
need for a manual variation of $P_v$(Mg) and $P_v$(B). This increases
the likelihood of identifying the global refinement minimum and
results in a good fit to the available reflection intensities up to
\sthlmax~$=$~1.3~\iA.

Extended Hansen-Coppens (EHC) multipolar refinements were employed in a detailed investigation
of the presence or absence of magnesium vacancies in our sample. Model
parameters were determined on the basis of static theoretical
structure factors for \Mg\ with a stoichiometric composition of
1~Mg~:~2~B and experimental single-crystal XRD data collected at
$T =$~100(2)~K (\sthlmax~$= 1.6$~\iA).  The spherical contributions to
\rhoat\ of the magnesium and boron atoms were represented by three
shells with electronic configurations of ($1s^2$)~/ ($2s^2$, $2p^6$)~/
($3s^2$) and ($1s^2$)~/ ($2s^1$)~/ ($2p^2$), respectively.
Additionally, one set of aspherical deformation functions with maximum
multipolar order $l_\mathrm{max} = 4$ was employed for each of the
atoms in the asymmetric unit of \Mg. In the refinements of theoretical
structure factors and experimental data, the number of varied
parameters was increased in a stepwise manner: In a first step, the
multipolar parameters were refined, while the magnesium occupation
factor was kept at a fixed value of 1.0. Thereby, the core multipolar
parameters were only varied in case of theoretical structure factors
(T-EHCM1). In case of experimental data, the final core multipolar
parameter values from T-EHCM1 were adopted, but the scale factor and
the anisotropic (harmonic) ADPs of the magnesium and boron atom were
varied (E-EHCM1). Then, the magnesium site occupation factor, the
scale factor, the anisotropic ADPs, and the core and valence
multipolar parameters (theoretical structure factors; T-EHCM2) or the
valence multipolar parameters (experimental data; E-EHCM2) were
relaxed in a joint refinement.

\subsection{DFT calculations}
\label{sec:methods-dft}

Density functional theory (DFT) calculations for structure factor
generation have been performed using the \texttt{Wien2k} suite of
programs.\cite{blaha_wien2k_2020,blaha_wien2k_2018} LAPW wave function
calculations employing the PBE functional \cite{Perdew_Burke_Ernzerhof_1996,Perdew_Burke_Ernzerhof_1997}, a $k$-point sampling mesh
of size 25$\times$25$\times$19 and a $R_{mt}K_{max}$ parameter of 10.0
were performed using the lattice parameters and fractional coordinates
obtained from the x-ray diffraction study at room temperature (see
above). Static structure factors up to a resolution limit of
6~\AA$^{-1}$ have been calculated with the \texttt{lapw3} module.
Lattice dynamic calculations via the finite-difference approach based
on a 5$\times$5$\times$3 supercell have been performed employing
\texttt{phonopy}\cite{phonopy} in combination with the
\texttt{VASP}\cite{kresse_efficiency_1996,kresse_efficient_1996,kresse_ab_1994,kresse_ab_1993}
code as force calculator. The PBE functional, a $k$-mesh sampling of
9$\times$9$\times$7 and an energy cut-off of 600 eV have been used
throughout. The $q$-mesh sampling for the calculation of thermal
displacement parameters was done on a 17$\times$17$\times$15 grid of
points.

The calculation of dynamic structure factors was done with a locally
modified version of the \texttt{DENPROP}
code\cite{volkov_basis-set_2009}, details of the implementation are
given in Appendix~C. Wave function data was taken from the
\texttt{Wien2k} calculations described above. The numerical
integration of the Stockholder atoms\cite{Hirshfeld_1977} was done on
a Lebedev-Laikov grid\cite{Lebedev_Laikov_1999} with 590 angular
points while using 923 and 961 radial points between 0.0001 and
16~a.u. from the atomic position for Mg and B,
respectively. Contributions to the Stockholder weights from an atomic
cluster consisting of 2013 and 2031 atoms were included in the
calculation for Mg and B, respectively. The numerical integration
errors employing these parameters were of the order of
10$^{-4}$~\textit{e} for the $F(000)$, the (non-iterated) stockholder charges
obtained from the partitioning of the total electron density
distribution were +0.2778~\textit{e} and $-$0.1389~\textit{e} for Mg
and B, respectively. Structure factors for different temperatures were
calculated employing the according experimentally determined unit cell
parameters and calculated ADP parameters (see Tab.~\ref{tab:DFT-ADPs} 
of the Supporting Information), respectively, while the stockholder 
atomic electron densities were kept fixed.

\section{Results and discussion} \label{sec:results}

\subsection{Magnesium site occupancy}
\label{sec:occupancy}

Model-free approaches for the analysis of x-ray diffraction (XRD) data
like the maximum entropy method (MEM) do not require an explicit
consideration of sample defects and provide a crystal-averaged
electron density distribution. To perform successful multipolar
refinements, however, one needs to consider deviations from the ideal
sample stoichiometry -- a frequent defect type which might also occur 
in other superconducting compounds, \textit{e.g.}
\YBCO.\cite{Jorgensen87,Jorgensen88,Dharwadkar87,Beno87,Manthiram87}
Such non-stoichiometric deviations can be determined by least-squares refinements  
of the respective atomic site occupation factors. Otherwise neglection of sample defects may
result in false model parameters as
demonstrated in the case study of B/C occupational disorder
in ScB$_2$C$_2$ by Haas \textit{et al.}\cite{Haas19} Therefore, we
focus on this aspect first.

For \Mg\ there has been a long-lasting debate about the existence or
absence of magnesium vacancies and how they control the physical and
especially superconducting properties. Whereas a large number of
authors argued in favor of the presence of magnesium vacancies up to
approx. 5~\% even in nominally stoichiometric samples
\cite{Serquis01,Chen01,Zhao01,Mori02b,Tsirelson03,Lee03a,Chen08,Zhigadlo10},
others questioned any significant deviations from the ideal
composition 1~Mg~:~2~B.\cite{Hinks02} Interestingly, most reports of
magnesium vacancies relied on the refinement of magnesium site
occupation factors using the independent-atom-model (IAM) to fit
powder or single-crystal XRD data.\cite{Serquis01,Mori02b,Tsirelson03,
  Lee03a,Zhigadlo10} By contrast, magnesium site occupation factors
derived from powder neutron diffraction experiments \cite{Hinks02}
showed no significant deviation from unity. This discrepancy may be
due to the fact that the IAM approach does not account for charge
transfer between ions, the presence of aspherical density deformations and
deformations of the atomic core densities resulting from the latter
two effects\cite{Fischer11,Batke13,Scherer14,Fischer21} which in turn
affect the determination of precise sample compositions.\cite{Haas19}
In case of the pseudo-Zintl phase \Mg, strong valence charge transfer
from the magnesium cations to the covalently bonded anionic boron
network \cite{Faessler,Kortus01,An01,Harima02,Mori02b,
  Tsirelson03,Wu04,Merz14} may hamper the determination of magnesium
occupation factors by standard XRD refinement techniques.\cite{Wu04}

To study this possible pitfall in more detail we performed IAM
structural refinements using neutral-atom scattering factors (\textit{i.e.} explicitly and erroneously \textit{not} taking charge transfer effects into account) to model
static theoretical structure factors of \Mg. These were derived from
periodic DFT calculations for stoichiometric \Mg. Accordingly, they
are not biased by the presence of vacancies, impurity atoms like carbon or other experimental
errors and thus provide an idealized test set (see Methods
section). In analogy to the treatment of experimental XRD data, the
scale factor, the anisotropic atomic displacement parameters (ADPs)
for magnesium and boron and the occupation factor of the magnesium
site were refined.  Notably, this refinement strategy is not only chosen for consistency with experiment. Due to the structural simplicity of \Mg\ with only two atoms in the asymmetric unit the aforementioned parameters are highly interdependent, so that their simultaneous refinement is essential for the following results. Furthermore, the magnesium occupation factor and the scale factor are correlated via the the number of electrons per asymmetric unit. This requires us to focus 
our study on one type of sample defect (in our case the Mg site occupancy), as a simultaneous refinement of B/Mg site occupation factors and the scale factor under a electroneutrality constraint is not feasible for \Mg. 

\begin{figure}[htbp]
  \centering
  \includegraphics[width=0.5\textwidth]{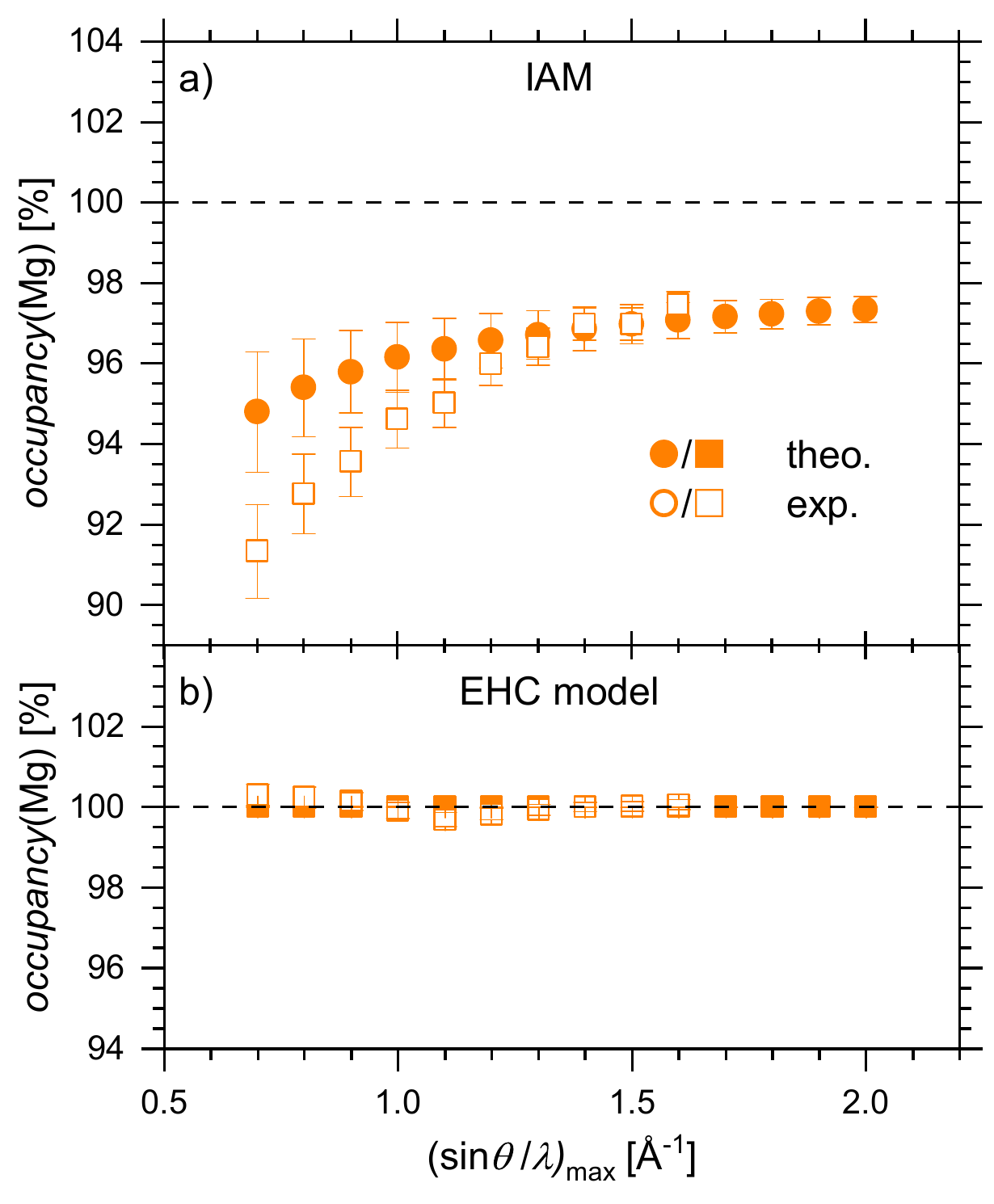}
  \caption{Variation of the magnesium site occupancy with increasing
    resolution \sthlmax\ for different scenarios: using a)~an IAM
    based on scattering factors of neutral atoms (circles) or
    b)~an EHC model based on multipolar parameters (squares) to fit
    theoretical structure factors (filled symbols) or experimental
    single-crystal XRD data collected at $T =$~100~K (empty symbols)
    at a resolution \sthlmax\ = 1.6~\iA. In each case, the magnesium
    site occupation factor was refined simultaneously with the scale
    factor and the ADPs of the magnesium and boron atoms.}
  \label{fig:theo-occupation-resden}
\end{figure}

Fig.~\ref{fig:theo-occupation-resden}a reveals the
dependency of the obtained magnesium site occupancy (represented by
filled circles) on the maximum reciprocal-space resolution \sthlmax\
of the employed data set. The resolution-dependent development of all
remaining refined parameters is available in
Fig.~\ref{fig:theo-IAM-scale-Uaniso-ai-param-overview} of the
Supporting Information.

It can be recognized from Fig.~\ref{fig:theo-occupation-resden}a that
IAM refinements result in false magnesium site occupancies especially
in case of low reciprocal-space resolution. 
Accordingly, in the resolution range
\sthlmax~$\leq 0.8$~\iA\ that is usually employed for standard XRD
experiments magnesium site occupancies of approx. 95~\% are
obtained. We note that this value is remarkably close to IAM-based site
occupancies reported in the
literature\cite{Serquis01,Lee03a,Mori02b,Tsirelson03} suggesting that
these results might also be affected by insufficient data resolution
and the choice of an inflexible IAM to model the data. Fig.~\ref{fig:theo-occupation-resden} also reveals that at
significantly higher resolutions \sthlmax, the magnesium site
occupancy increases and approaches a maximum value of 98~\%, while the
correct value of 100~\% is never reached. An analogous asymptotic
behavior of the refined magnesium site occupancy with increasing
resolution can be observed for experimental single-crystal XRD data
collected up to \sthlmax~$= 1.6$~\iA\ (crystal~1; $T =$~100(2)~K; open
circles in Fig.~\ref{fig:theo-occupation-resden}; for the resolution
dependence of all refined parameters see
Fig.~\ref{fig:exp-IAM-scale-Uaniso-ai-param-overview} of the
Supporting Information). Hence, our analysis suggests that false
atomic site occupancies in compounds characterized by polar bonds
and/or charged atoms are a natural consequence of using an independent
atom model.

Multipolar models applied to high-resolution XRD data may overcome the
inflexibility of the IAM by properly taking into account
(\textit{i})~charge transfer effects, (\textit{ii})~aspherical density
deformations due to chemical bonding, and (\textit{iii})~the
contraction or expansion of atomic shells (see Appendix~A).  The
success of the multipolar approach in the determination of sample
compositions was demonstrated by Haas \textit{et al.}\cite{Haas19} In
that study the coloring problem due to the mixed occupation of
carbon/boron atomic sites in the borocarbide ScB$_2$C$_2$
could be resolved using high-resolution XRD data and a Standard
Hansen-Coppens (SHC) multipolar model.\cite{Hansen78} At ultra-high
data resolutions (\sthl~$\gtrapprox$~1.4~\iA), however, also the
charge-transfer between core and valence shells and
core polarization effects due to chemical-bonding effects may need to
be considered.\cite{Bentley_Stewart_1976,Batke13} Otherwise, neglecting these core
contraction/expansion or polarization effects in the SHC
multipolar model can lead to drastically increased residual electron
density features in the core-density region of atoms and may hamper
the precise determination of ADPs\cite{Svendsen_2010,Fischer11,Scherer14} and/or atomic
positions via so-called core-asphericity shifts.\cite{Fischer21} In
the Extended Hansen Coppens (EHC) multipolar model proposed for example
by Batke \textit{et
  al.}\cite{Batke13} and Fischer \textit{et
  al.}\cite{Fischer11,Fischer21} the frozen-core approximation of the SHC model is lifted, 
  so that ultra-high-resolution XRD data can be
fitted with good accuracy (see Appendix~A).

In the following, we demonstrate the applicability of the multipolar
approach to the problem of magnesium site occupancies in \Mg. Similar
to our evaluation of the IAM approach, we start with static structure
factors from periodic DFT calculations for ideal \Mg\ lacking any
magnesium vacancies (see Methods section). Thus, a magnesium site
occupancy close to 100~\% should be obtained in case of a successful
modelling. In the least-squares refinement of multipolar models we
used reflection data featuring the maximum achieved resolution
\sthlmax\ = 1.6~\iA\ in our XRD experiments on \Mg\ (see Methods
section). Core polarization
effects present at \sthlmax\ = 1.6~\iA\ (see the resolution dependence
of the residual electron density at the magnesium position for IAM,
SHC and EHC models in Fig.~\ref{fig:resolution-dep-resden-Mg-pos} of
the Supporting Information) were taken into account by choosing a
multipolar model at the EHC level. In a final step of the model
refinement core and valence multipolar parameters, the scale factor,
the anisotropic ADPs and the magnesium site occupation factor were
varied jointly (denoted T-EHCM2 model; for more details, see the Methods section and the
Supporting Information). The obtained magnesium site occupancy of
100.152(14)~\% is very close to 100~\%, while the values of the scale
factor (1.01007(19)) and the ADPs
($U_{11}$(Mg)~$=$~0.000078(4)~\AA$^2$;
$U_{33}$(Mg)~$=$~0.000081(4)~\AA$^2$;
$U_{11}$(B)~$=$~$-$0.000085(5)~\AA$^2$;
$U_{33}$(B)~$=$~$-$0.000091(5)~\AA$^2$) show only minor deviations
with respect to their expected values of unity and zero, respectively.
Only a small dependency on the data resolution was observed hinting
for a high robustness of the results (filled squares in
Fig.~\ref{fig:theo-occupation-resden}b; resolution dependence of other
refined parameters in
Fig.~\ref{fig:theo-ehc-scale-Uaniso-ai-param-overview} of the
Supporting Information). Thus, the capability of the EHC approach to
provide correct atomic site occupancies even in the presence of polar
bonds or charged atoms appears drastically improved with respect to
the IAM approach.

Next, we check for the presence or absence of magnesium vacancies in
our \Mg\ sample by performing an EHC multipolar refinement on
experimental single-crystal XRD data collected at $T =$~100(2)~K
(crystal~1; \sthlmax~$= 1.6$~\iA). Thereby, core multipolar parameters
were fixed at their values obtained in the EHC model refinement of
theoretical structure factors (\sthlmax~$= 1.6$~\iA).  Again, a
magnesium site occupancy close to 100~\% was obtained (101.5(3)~\% at
\sthlmax~$= 1.6$~\iA; E-EHCM2) with only a minor effect of changes in
the data resolution (open rectangles in
Fig.~\ref{fig:theo-occupation-resden}b; resolution dependence of other
refined parameters in
Fig.~\ref{fig:exp-ehc-scale-Uaniso-ai-param-overview} of the
Supporting Information).  Thus, our combined experimental and
theoretical EHC refinements rule out any significant evidence for the
presence of magnesium vacancies in the investigated \Mg\
samples. Earlier reports on magnesium vacancies in the compound may
thus be mainly attributed to shortcomings of the employed IAM approach 
and/or usage of low-resolution XRD data of \Mg. Hence, this case study stresses
the importance of using (\textit{i})~a sufficiently high data
resolution and (\textit{ii})~appropriate multipolar models in the data
analysis, when reliable and accurate site occupancies need to be
derived from x-ray diffraction data.

\subsection{Temperature-dependent changes in the electron density distribution}
\label{sec:temp-dep-density}

\begin{figure}[htbp]
  \centering
  \includegraphics[width=0.8\textwidth]{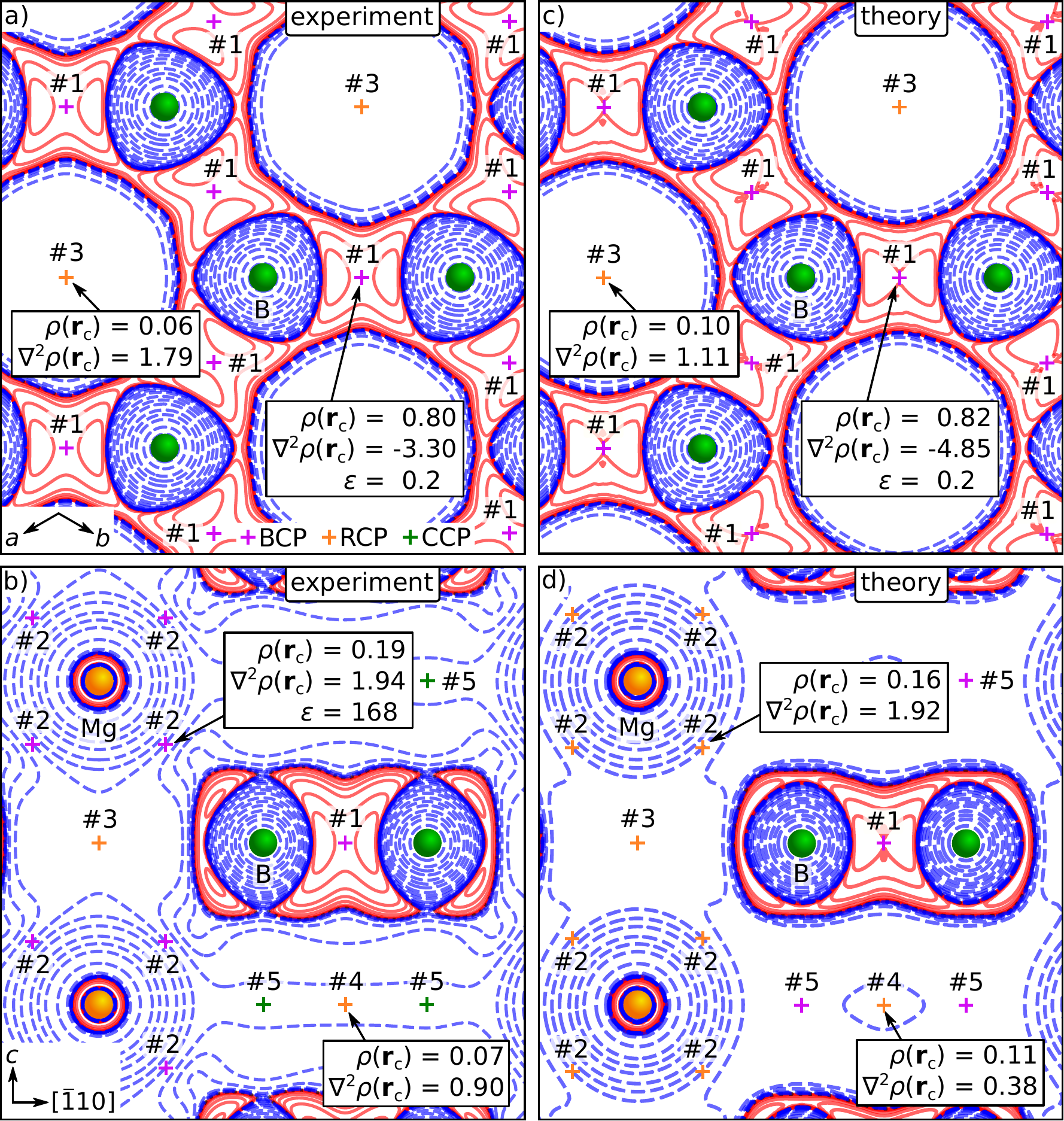}
  \caption{Maps of the Laplacian of the electron density,
    $\nabla^2 \rho(\mathbf{r})$, in the boron network parallel to the
    $a$-$b$ plane (a,c), and in a plane parallel to the c-axis 
    enclosing the magnesium and boron atoms (b,d). The static $\rho(\mathbf{r})$ distributions were obtained from
    EHC model E-EHCM1 using experimental single-crystal XRD data
    (a,c; $T =$~100~K) and from periodic DFT calculations (b,d; see
    text).  Contour lines are drawn for negative (solid red lines) and
    positive values of $\nabla^2 \rho(\mathbf{r})$ (dashed blue lines)
    at levels of $\pm 2 \cdot 10^n$, $\pm 4 \cdot 10^n$, and
    $\pm 8 \cdot 10^n$ with $n \in \{-2, \dots, 3\}$;  
    \rhoc\ values (in \eA) and Laplacian \Lc\ values (in \eAA).
    The bond ellipticity $\epsilon$ values are specified
    at the locations of critical points (indicated by crosses). Their
    numbering refers to Tab.~\ref{tab:qtaim-analysis} in the
    Supporting Information.}
  \label{fig:lapl-ehc-100k-crystal1}
\end{figure}

We first focus on topological analyses of static electron density 
distributions as derived from an EHC multipolar refinement of XRD 
data collected at $T =$~100(2)~K (\rhoEHC; \sthlmax~$= 1.6$~\iA; 
model E-EHCM1) and periodic DFT calculations using the Quantum 
Theory of Atoms in Molecules (QTAIM)\cite{Bader,Popelier} before 
we discuss potential changes induced by the onset of superconductivity.  
Maps of the Laplacian of the
electron density, $\nabla^2\rho(\mathbf{r})$, in planes showing
nearest-neighbor boron-boron and boron-magnesium contacts are given in
Fig.~\ref{fig:lapl-ehc-100k-crystal1} (topological characteristics and
Laplacian maps for the other multipolar models in this paper are
available in Tab.~\ref{tab:qtaim-analysis},
Fig.~\ref{fig:lapl-crystal1-b-plane} and
Fig.~\ref{fig:lapl-crystal1-mg-b-plane} of the Supporting
Information).  In agreement with a description of \Mg\ as a
pseudo-Zintl phase\cite{Faessler} and previous experimental
\cite{Mori02b, Tsirelson03,Wu04,Merz14} and theoretical
results,\cite{Kortus01,An01,Harima02} only the bond-critical point
(BCP) at the midpoint of the B--B contact (\#1 in
Fig.~\ref{fig:lapl-ehc-100k-crystal1}) shows a large \rhoc\ of
0.80~\eA\ [DFT: 0.82~\eA] and a strongly negative \Lc\ value of
$-$3.30~\eAA\ [DFT: $-$4.85~\eAA] in line with the presence of
covalent interactions.  The magnesium-boron contact (CP \#2 in
Fig.~\ref{fig:lapl-ehc-100k-crystal1}b), by contrast, is characterized
by rather electrostatic closed-shell interactions indicated by a small \rhoc\ value
of 0.19~\eA\ [DFT: 0.16~\eA] and a positive \Lc\ value of
$+$1.94~\eAA\ [DFT: $+$1.92~\eAA].\footnote{In analogy to the results
  of the charge-density study by Tsirelson \textit{et
    al.}\cite{Tsirelson03} some of the Hessian eigenvalues of critical
  points \#2 and \#5 are close to zero (Fig.~\ref{fig:lapl-ehc-100k-crystal1}). Thus, the corresponding
  signatures of the critical points are topologically unstable (see also the Supporting Information).} 
  This is also
evident from the respective atomic charges of $+$1.6~\textit{e} and
$-$0.8~\textit{e} [DFT: $+$1.6~\textit{e}/$-$0.8~\textit{e}] as a consequence of the
pronounced charge transfer from the electropositive magnesium atoms 
towards the boron atoms which form the
a graphene-type network characterized by delocalized $\pi$ electrons.

On the basis of MEM analyses of powder XRD data Nishibori \textit{et
  al.}\cite{Nishibori01} suggested that the chemical bonding situation in \Mg\
changes significantly upon cooling from room temperature to 15~K,
\textit{i.e.} a temperature below the superconducting \Tc.  According to the 
authors these electronic structure changes are clearly indicated by an increased density 
at the B--B BCP (1.0~\eA\ at 15~K compared to 0.9~\eA\ at RT) which results from 
a charge transfer from delocalized $\pi$ to localized $\sigma$ bonds in the hexagonal boron
layers\cite{Nishibori01} (see the
structural model in Fig.~\ref{fig:mgb2-structure}).  However, the MEM
analyses employed by Nishibori \textit{et al.}\cite{Nishibori01} 
solely provide density distributions \rhomem\ that are compatible
with observed structure factor amplitudes and maximize information
entropy.\cite{Collins82,Gatti-guided-tour,Coppens-MEM} As such they
only show the combined effect of density deformations due to chemical bonding as well as 
dynamic and static atomic displacements\cite{Gatti-guided-tour}. Hence, a definitive
attribution of the observed electron density increase to one of these
factors is not possible. To identify the origin and nature of the temperature-dependent
density variation at the B-B BCPs we performed high-resolution charge density studies
employing a single crystal (crystal~1) at various temperatures above and below the
\Tc. The according $T_\mathrm{c} \approx$~37.5~K was determined from magnetic
susceptibility studies of a single crystal (crystal~2) from the same synthesis
batch. Our measurements relied on a low-temperature diffractometer
setup featuring a closed-cycle helium cryocooler to reach sub-nitrogen
cryogenic temperatures (further information in the Methods section and
the Supporting Information). The employed cooling equipment puts
restrictions on the possible goniometer settings, so that the
available data resolution is reduced from \sthlmax~$=$~1.6~\iA\ to
1.3~\iA. Therefore, Standard Hansen-Coppens (SHC) multipolar
models\cite{Hansen78} with a fixed magnesium site occupancy of 100~\%
were refined against the low-temperature XRD data to maintain a high
data-to-parameter ratio and to reduce potential parameter correlations
(see Appendix~A, Methods section and Supporting Information).

\begin{figure}[p]
  \centering
  \includegraphics[width=0.8\textwidth]{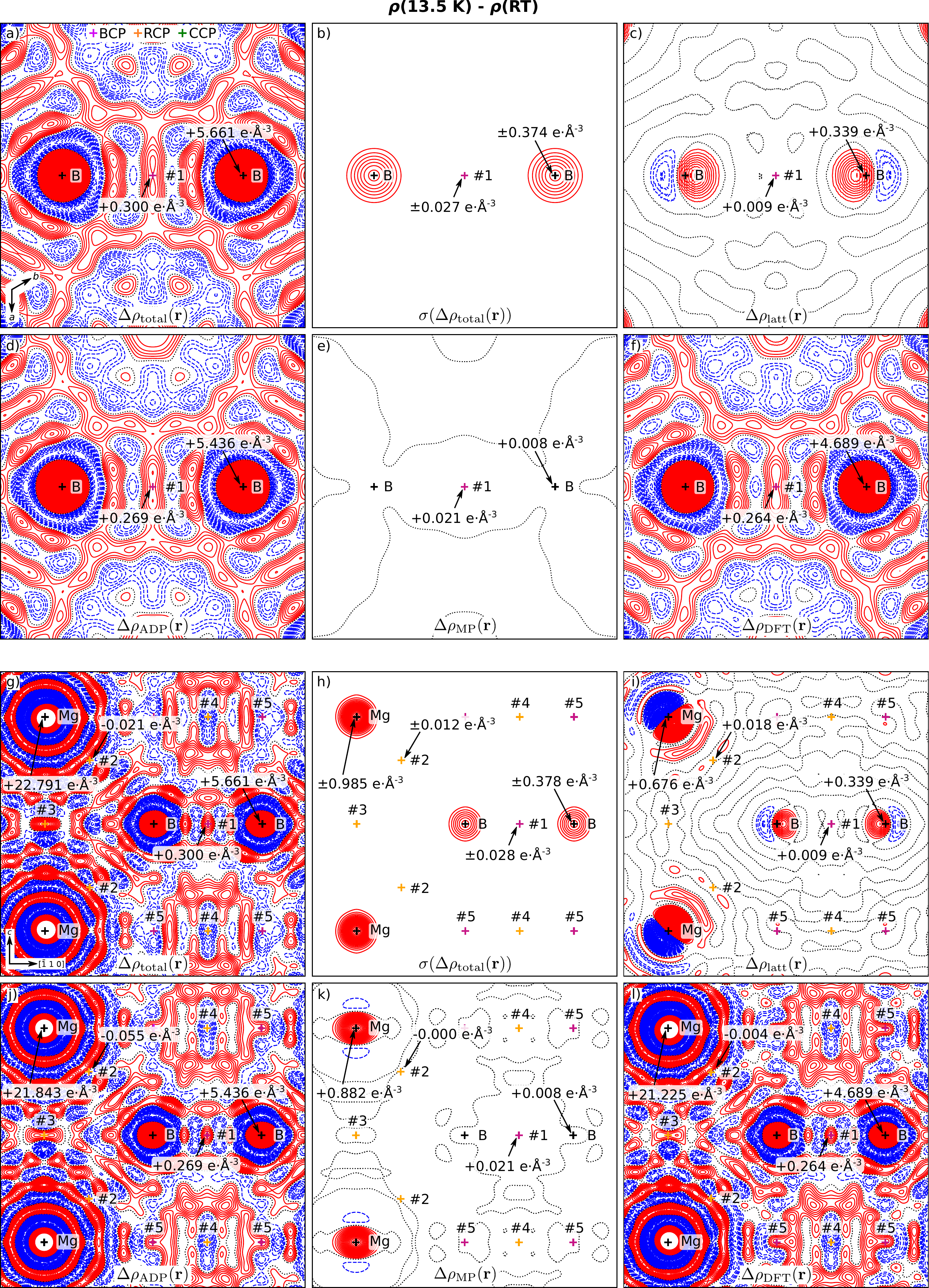}
  \caption{Iso-contour maps of differences between electron density
    distributions at 13.5~K and RT in planes parallel (a-f) and
    perpendicular to the hexagonal boron layers in \Mg\ (g-l). The
    standard deviation $\sigma$(\Drhototal) of the total difference 
    density \Drhototal\ (a,g) has been specified in b) and h). It has
    been further decomposed into contributions from changes in the lattice
    parameters (c,i), the ADPs (d,j), and the multipolar parameters
    (MP; e,k). Standard DFT techniques can reproduce \Drhototal\
    without consideration of the superconducting transition (f,l; see
    text).  Contour lines at positive (red; solid), zero (black;
    dotted) and negative values (blue; dashed) are equally spaced by
    increments of $\pm$0.05~\eA. Crosses indicate the location of the
    QTAIM critical points introduced in
    Fig.~\ref{fig:lapl-ehc-100k-crystal1}.}
  \label{fig:rho-diff-decomposition-13p5K-RT}
\end{figure}

As a starting point, we compare the
total temperature-dependent changes of the electron density
distribution between 13.5~K and room-temperature (RT) as obtained by
our SHC refinements with the \rhomem\ changes
between 15~K and RT as observed by Nishibori \textit{et
  al.}\cite{Nishibori01} To directly visualize the
\textit{modification} of the electron density distribution, we
consider the total temperature-dependent difference electron density,
\Drhototal, between 13.5~K and RT
(Fig.~\ref{fig:rho-diff-decomposition-13p5K-RT}a and
Fig.~\ref{fig:rho-diff-decomposition-13p5K-RT}g) obtained by
subtracting densities $\rho_\mathrm{o}(\mathbf{r}\text{, 13.5~K})$ and
$\rho_\mathrm{o}(\mathbf{r}\text{, RT})$ from each other.  Each of the
individual densities \rhoobsT\
is calculated by Fourier summation over
the observed structure factor amplitudes at the respective temperature
$T$ (see Appendix~B). Similar to \rhomem, but in contrast to the previously
discussed model density \rhoEHC, each density \rhoobsT\ reflects the
\textit{joint} effect of chemical bonding as well as static and
dynamic atomic displacements on the electron density distribution at a
specific temperature.

Fig.~\ref{fig:rho-diff-decomposition-13p5K-RT}a and
Fig.~\ref{fig:rho-diff-decomposition-13p5K-RT}g show iso-contour maps
of \Drhototal\ in planes parallel and perpendicular to the hexagonal
boron layers in \Mg. Complementary maps of the standard deviation of
\Drhototal\ (see Appendix~B for details on their generation) are given
in Fig.~\ref{fig:rho-diff-decomposition-13p5K-RT}b and
Fig.~\ref{fig:rho-diff-decomposition-13p5K-RT}h to ease the
differentiation between significant and non-significant features.  In fact,
inspection of Fig.~\ref{fig:rho-diff-decomposition-13p5K-RT}a and
Fig.~\ref{fig:rho-diff-decomposition-13p5K-RT}g reveals an increase of
the electron density at the B--B BCP (\#1 in
Fig.~\ref{fig:rho-diff-decomposition-13p5K-RT}) by 0.300~\eA\ and a
decrease by 0.021~\eA\ at the B$\cdots$Mg BCP (\#2 in
Fig.~\ref{fig:rho-diff-decomposition-13p5K-RT}) upon cooling from RT
to 13.5~K. The identified density accumulation at the B--B BCP is in
qualitative agreement with the increase of \rhomem\ by 0.1~\eA\
between RT and 15~K reported by Nishibori \textit{et
  al.}\cite{Nishibori01} Yet, these are not the strongest
temperature-dependent changes pointed out by \Drhototal: The maps in
Fig.~\ref{fig:rho-diff-decomposition-13p5K-RT}a and
Fig.~\ref{fig:rho-diff-decomposition-13p5K-RT}g are clearly dominated
by atom-centered spherical features with a radial alternation of
sign. At the atomic positions, \Drhototal\ is found to be strongly
positive with values of 5.661~\eA\ for boron and 22.791~\eA\ for
magnesium.

To proceed beyond the MEM level and to investigate the origin of the
observed features we decompose \Drhototal\ into components according
to

\begin{align}
  \Delta \rho_\mathrm{total}(\mathbf{r}) = \Delta \rho_\mathrm{latt}(\mathbf{r}) +
  \Delta \rho_\mathrm{ADP}(\mathbf{r}) + \Delta \rho_\mathrm{MP}(\mathbf{r}) +
  \Delta \rho_\mathrm{res}(\mathbf{r})
  \label{eq:Drhototal-decomposition}
\end{align}

\noindent using Fourier summations over \textit{observed} and/or
\textit{calculated} structure factor amplitudes from our multipolar
models (see Appendix~B). This procedure enables the visualization of
the individual difference density contributions of changes in
(\textit{i})~the lattice parameters (\Drholatt;
Fig.~\ref{fig:rho-diff-decomposition-13p5K-RT}c and
Fig.~\ref{fig:rho-diff-decomposition-13p5K-RT}i), ($ii$)~the atomic
displacement parameters (ADPs) (\DrhoADP;
Fig.~\ref{fig:rho-diff-decomposition-13p5K-RT}d and
Fig.~\ref{fig:rho-diff-decomposition-13p5K-RT}j), $(iii)$~the
multipolar parameters (\DrhoMP;
Fig.~\ref{fig:rho-diff-decomposition-13p5K-RT}e and
Fig.~\ref{fig:rho-diff-decomposition-13p5K-RT}k), and $(iv)$~residual
changes (\Drhores) not captured by the employed multipolar models. It
becomes evident that the lattice contraction between RT and 13.5~K
leads to dipolar patterns centered at the atomic positions in
\Drholatt\ (Fig.~\ref{fig:rho-diff-decomposition-13p5K-RT}c and
Fig.~\ref{fig:rho-diff-decomposition-13p5K-RT}i). But at the same
time, the effect of a shrinkage in lattice parameters $a$ and $c$ on
the density between the atomic positions is minimal (0.009~\eA\ at the
B--B BCP and 0.018~\eA\ at the B$\cdots$Mg BCP). Density changes
\DrhoMP\ (Fig.~\ref{fig:rho-diff-decomposition-13p5K-RT}e and
Fig.~\ref{fig:rho-diff-decomposition-13p5K-RT}k) are connected to a
variation of multipolar parameters (MP) and -- if present -- may
indicate a modified chemical bonding situation in \Mg\ as proposed by
Nishibori \textit{et al}.\cite{Nishibori01} The contribution of
\DrhoMP\ to \Drhototal\ is, however, even smaller than that of
\Drholatt\ (0.021~\eA\ at the B--B BCP and $-$0.000~\eA\ at the
B$\cdots$Mg BCP). The only notable feature in maps of \DrhoMP\ are
quadrupole-like patterns accumulating density at the positions of the
magnesium atoms (Fig.~\ref{fig:rho-diff-decomposition-13p5K-RT}k).
Finally, maps Fig.~\ref{fig:rho-diff-decomposition-13p5K-RT}d and
Fig.~\ref{fig:rho-diff-decomposition-13p5K-RT}j show density changes
\DrhoADP\ related to a temperature-dependent variation of ADPs.  Their
striking similarity to maps of \Drhototal\
(Fig.~\ref{fig:rho-diff-decomposition-13p5K-RT}a and
Fig.~\ref{fig:rho-diff-decomposition-13p5K-RT}g) emphasizes that the
large temperature difference between RT and 13.5~K renders the reduction
of thermal smearing the most important source of temperature-dependent
density changes (0.269~\eA\ at the B--B BCP and $-$0.055~\eA\ at the
B$\cdots$Mg BCP).  Not only the spherically alternating patterns at
the atomic positions in \Drhototal\ are reproduced by \DrhoADP, but
also most of the features in the inter-atomic region. Accordingly, the
increase of \Drhototal\ at the B--B BCP by 0.300~\eA\ can be
decomposed into contributions of 0.009~\eA\ ($\hat{=}$~3.0~\%) from
\Drholatt, 0.021~\eA\ ($\hat{=}$~7.0~\%) from \DrhoMP, and 0.269~\eA\
($\hat{=}$~89.7~\%) from \DrhoADP.  Small remaining contributions to
\Drhototal\ that cannot be captured by the employed multipolar models
and cannot be assigned to variations in lattice, MP or ADP parameters
are contained in the residual difference density, \Drhores\
($\hat{=}$~0.3~\%; corresponding iso-contour maps are available in
Fig.~\ref{fig:rho-diff-decomposition-13p5K-RT-supp} of the Supporting
Information).

Further evidence for the dominant role of thermal smearing is provided
by the fact that standard DFT calculations without consideration of
the superconducting transition can reproduce the maps of \Drhototal\
in Fig.~\ref{fig:rho-diff-decomposition-13p5K-RT}a and
Fig.~\ref{fig:rho-diff-decomposition-13p5K-RT}g to a large extent: The
maps of \DrhoDFT\ in Fig.~\ref{fig:rho-diff-decomposition-13p5K-RT}f
and Fig.~\ref{fig:rho-diff-decomposition-13p5K-RT}l were generated
ab-initio from \textit{dynamic} theoretical structure factors
(\sthlmax~=~1.3~\iA; see the Methods section and Appendix~C for details on the procedure)
using lattice parameters at 13.5~K and RT as the only experimental
input. At the location of the B--B BCP, \DrhoDFT\ amounts to
0.264~\eA\ comparing to 0.300~\eA\ in \Drhototal\ and 0.269~\eA\ in
\DrhoADP.

\begin{figure}[p]
  \centering
  \includegraphics[width=0.8\textwidth]{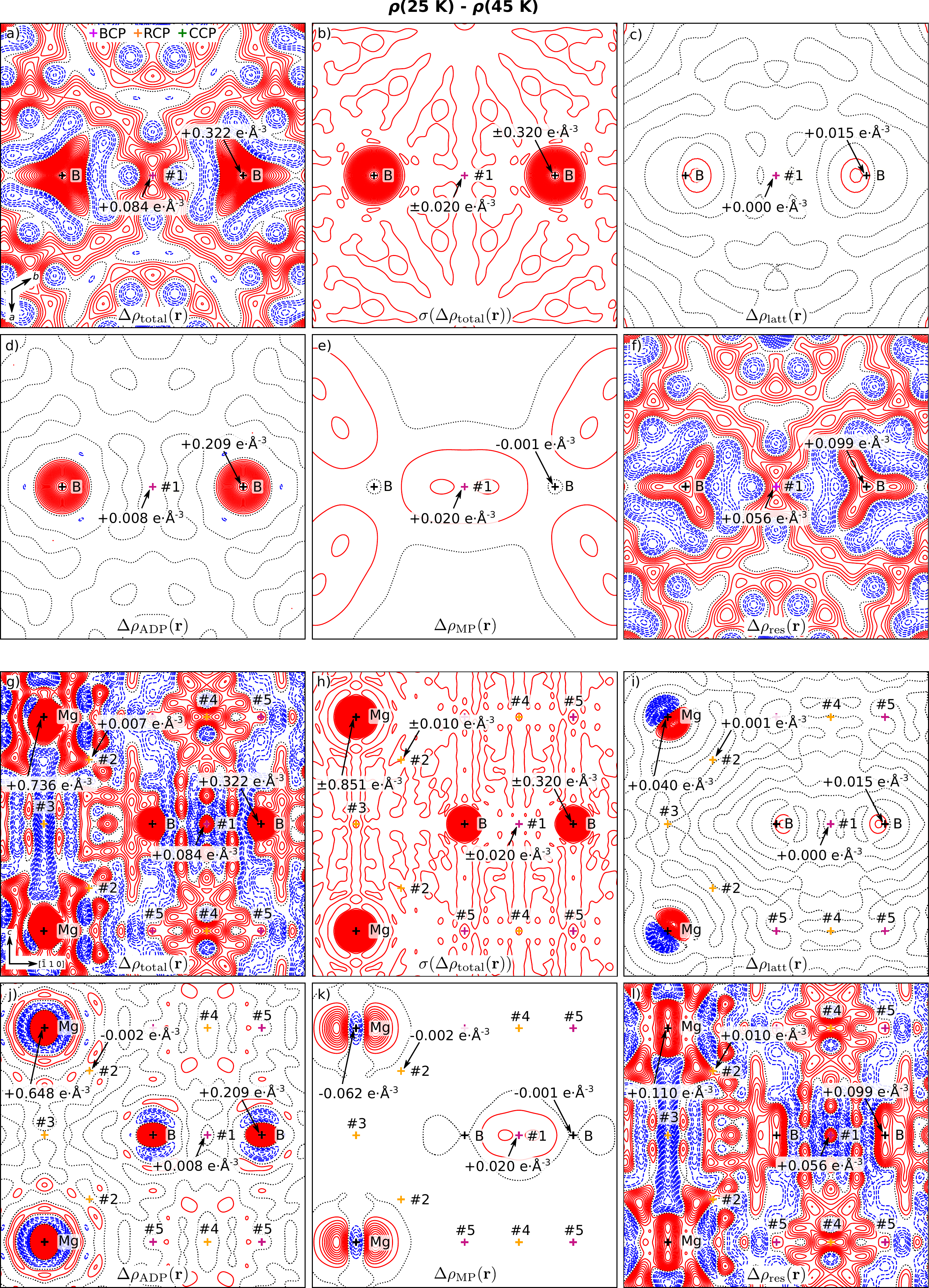}
  \caption{Iso-contour maps of differences between electron density
    distributions at 25~K and 45~K in planes parallel (a-f) and
    perpendicular to the hexagonal boron layers in \Mg\ (g-l). The
    standard
    deviation $\sigma$(\Drhototal) of the total difference density
    \Drhototal\ (a,g) has been specified in b) and h). Electron density 
    differences have
    been decomposed into contributions from changes in the lattice
    parameters (c,i), the ADPs (d,j), and the multipolar parameters
    (MP; e,k) as well as residual changes not captured by the employed
    multipolar model (f,l; see text).  Contour lines at positive (red;
    solid), zero (black; dotted) and negative values (blue; dashed)
    are equally spaced by increments of $\pm$0.01~\eA. Crosses
    indicate the location of the QTAIM critical points introduced in
    Fig.~\ref{fig:lapl-ehc-100k-crystal1}.}
  \label{fig:rho-diff-decomposition-25K-45K}
\end{figure}

To reduce the impact of thermal smearing and lattice shrinkage on
\Drhototal\ and to increase the sensitivity for potential other
changes during the superconducting transition, we performed additional
single-crystal XRD experiments closely above (45~K) and below (25~K)
the $T_\mathrm{c} \approx$ 37.5~K of our \Mg\ sample. Resulting maps
of \Drhototal\ with a reduced spacing of contour values
($\pm$0.01~\eA\ instead of $\pm$0.05~\eA) are given in
Fig.~\ref{fig:rho-diff-decomposition-25K-45K}. It can be recognized
that in spite of the reduced temperature window between 45~K and 25~K
\Drhototal\ remains positive at the B--B BCP (\#1 in
Fig.~\ref{fig:rho-diff-decomposition-25K-45K}).  The absolute value of
\Drhototal\ at this position (0.084~\eA;
Fig.~\ref{fig:rho-diff-decomposition-25K-45K}a), however, is significantly reduced
with respect to the larger temperature window
between RT and 13.5~K (0.300~\eA;
Fig.~\ref{fig:rho-diff-decomposition-13p5K-RT}a). In accordance with
the closer spacing of measuring temperatures, the decomposition of
\Drhototal\ at the B--B BCP yields only minor contributions from
temperature-dependent changes in lattice parameters as described by
\Drholatt\ (0.000~\eA\ $\hat{=}$~0.0~\%;
Fig.~\ref{fig:rho-diff-decomposition-25K-45K}c) and thermal smearing
as described by \DrhoADP\ (0.008~\eA\ $\hat{=}$~9.5~\%;
Fig.~\ref{fig:rho-diff-decomposition-25K-45K}d). Nevertheless,
inspection of the according map of \DrhoMP\ in
Fig.~\ref{fig:rho-diff-decomposition-25K-45K}e reveals that again the
largest part of the difference density \Drhototal\ at the B--B BCP
position cannot be attributed to changes in bonding-induced
density deformations (0.020~\eA\ $\hat{=}$~23.8~\%). Instead, most of
\Drhototal\ at the B--B BCP is absorbed by the residual difference
density \Drhores\ (0.056~\eA\ $\hat{=}$~66.7~\%;
Fig.~\ref{fig:rho-diff-decomposition-25K-45K}f) for which an
assignment to a specific origin is not possible by means of the
employed multipolar models. We note however, that the obtained value of \Drhores\ at the
B--B BCP is below the three-fold standard deviation $\pm$0.060~\eA\ of
the difference density \Drhototal\ at the same position
(Fig.~\ref{fig:rho-diff-decomposition-25K-45K}b) and therefore close
to the limit of detectability.

Prevalent features in the maps of \Drhototal\
between 25~K and 45~K (Fig.~\ref{fig:rho-diff-decomposition-25K-45K}a
and Fig.~\ref{fig:rho-diff-decomposition-25K-45K}g) are 
centered at the positions of magnesium and boron atoms. 
These can partly be modelled by a reduction of the (harmonic) ADPs in
the multipolar models between 45~K and 25~K (see \DrhoADP\ in
Fig.~\ref{fig:rho-diff-decomposition-25K-45K}d and
Fig.~\ref{fig:rho-diff-decomposition-25K-45K}j). But non-trivial
difference density patterns around the boron atom in the $a$-$b$ plane
and the magnesium atom in the $[\overline{1} 1 0]$-$c$ plane remain
undescribed (see maps of \Drhores\ in
Fig.~\ref{fig:rho-diff-decomposition-25K-45K}f and
Fig.~\ref{fig:rho-diff-decomposition-25K-45K}l). This and a notable
discrepancy between maps of \Drhototal\
(Fig.~\ref{fig:rho-diff-decomposition-25K-45K}a and
Fig.~\ref{fig:rho-diff-decomposition-25K-45K}g) and \DrhoDFT\
(Fig.~\ref{fig:rho-diff-decomposition-25K-45K-supp}a and
Fig.~\ref{fig:rho-diff-decomposition-25K-45K-supp}b) derived from DFT
calculations considering only changes in harmonic thermal motion
between 25~K and 45~K (see
Fig.~\ref{fig:rho-diff-decomposition-25K-45K-supp} of the Supporting
Information) might indicate a minor modification of potential
anharmonic contributions to the atomic displacements in \Mg.  Notably,
anharmonicity of phonons in \Mg\ has been revealed in DFT studies by
Liu \textit{et al.}\cite{Liu01} and Yildirim \textit{et
  al.},\cite{Yildirim01} although the importance of this effect for
the emergence of superconductivity in the compound remains
controversial.\cite{Shukla03,Baron07,Blumberg07,Calandra07,
  DAstuto07,Mialitsin07} The anharmonic effects in our sample,
however, are too small to be captured by anharmonic ADPs. Their
inclusion into our multipolar models resulted in only insignificant
Gram-Charlier coefficients up to fourth order.

Hence, in our single-crystal XRD studies we could identify the
accumulation of additional electron density at the B--B BCP upon
cooling across the superconducting \Tc\ of \Mg.  This is in
qualitative agreement with the temperature-dependent behavior of the
density \rhomem\ obtained in previous MEM analyses of powder XRD data
by Nishibori \textit{et al.}\cite{Nishibori01} At the same time, our
results do not support their hypothesis of a reorganization of
bonding-induced density deformations as a primary source of the
observed temperature-dependent density changes. Multipolar refinements
of our XRD data in combination with DFT studies instead link density
increases at the BCP positions in \Mg\ to a reduction of thermal
smearing as a natural consequence of sample cooling.

\section{Conclusion}
\label{sec:conclusion}

To conclude, in our single-crystal x-ray diffraction (XRD) experiments
we could not find evidence for a significant modification of the
one-electron density distribution in \Mg\ directly related to the
onset of superconductivity. With the help of multipolar refinements
any marked electron density differences obtained from XRD experiments
above and below the transition temperature \Tc\ can be traced back to
changes in thermal smearing. Their magnitude varies with the size of the temperature
interval between the compared measurements and is correctly predicted
by standard DFT techniques without the need for a consideration of
superconductivity in \Mg. Yet, the precise effect of a modified
thermal smearing on the electron density distribution is by no means
trivial. Namely, not only the electron density distribution close to
the atomic positions of magnesium and boron is affected. Electron
density may also be accumulated or depleted in the inter-atomic region
away from the atomic nuclei.  If XRD measurements at different
temperatures are compared in model-free electron density studies,
changes in thermal smearing may therefore be misinterpreted in terms
of changes in chemical bonding.

But also simple model-based approaches turn out to be error prone,
when applied to the pseudo-Zintl phase \Mg. If electron density shifts
due to chemical bonding and sizable charge transfer between magnesium
and boron remain unmodeled in refinements of XRD data on the IAM
level, wrong magnesium vacancy concentrations in the range of
approx. 5~\% may be obtained. By applying highly flexible Extended
Hansen Coppens multipolar
models\cite{Fischer11,Batke13,Haas19,Fischer21} to high-resolution
single-crystal XRD data we could demonstrate that our samples are
fully stoichiometric with an insignificant vacancy concentration at
the magnesium site. These findings have potential implications for the
discussion about chemical control parameters of the superconductivity
in \Mg. Alternative explanations for systematic trends in the \Tc\
with the weighed-in sample stoichiometry should be considered,
\textit{e.g.} the solution behavior of impurities in the starting
materials magnesium and boron.\cite{Hinks02}

\section{Appendix A}
\label{sec:appendixA}

In this paper, multipolar models at the Standard Hansen Coppens
(SHC)\cite{Stewart77,Hansen78} and the Extended Hansen Coppens
(EHC)\cite{Fischer11,Batke13,Fischer21} level are employed.
Generally, the flexibility of multipolar models is increased with
respect to the IAM, as each atomic density contribution \rhoat\ to
the model density 

\begin{align}
  \rho_\mathrm{model}(\mathbf{r}) = \sum_i{\rho_{\mathrm{at},i}(\mathbf{r} - \mathbf{r}_i)}
  \label{eq:rho-model}
\end{align}

\noindent is decomposed into a sum over density contributions from a
spherical core, a spherical valence and an aspherical valence
deformation as

\begin{align}
  \rho_\mathrm{at}(\mathbf{r}) & = \underbrace{P_c \rho_c(r)}_\mathrm{(frozen)~spherical~core} +
  \underbrace{P_v \kappa_v^3 \rho_v(\kappa_v,r)}_\mathrm{spherical~valence} \nonumber\\
  & + \underbrace{\sum_{l = 0}^{l_\mathrm{max}}{(\kappa'_{v,l})^3 R_l(\kappa'_{v,l},r)
    \sum_{m = -l}^l{P_{lm}d_{lm}(\theta,\phi)}}}_\mathrm{aspherical~valence~deformation}
  \label{eq:rho-at-SHC}
\end{align}

\noindent Thereby, the population parameters $P_c$ and $P_v$ control
the number of electrons attributed to the core and valence region.  In
multipolar refinements at the SHC level, the $P_c$ parameter is
usually kept at a fixed value, so that the spherical core density is
effectively frozen. The spherical valence density, by contrast, is not
fixed in order to account for charge transfer between the individual
pseudo atoms via the $P_v$ parameter and furthermore for contraction or
expansion in radial direction by means of the $\kappa_v$
parameter. Likewise, the radial part $R_l(\kappa'_{v,l},r)$ of the
aspherical valence deformation functions is contracted/expanded by
means of the $\kappa'_v$ parameter, while the asphericity of the
valence density due to chemical bonding effects is controlled by
the population parameters
$P_{lm}$ of the multipolar deformation density functions
$d_{lm}(\theta,\phi)$. 

The Extended Hansen-Coppens (EHC) approach systematically improves the
SHC multipolar model by increasing its flexibility. An enhancement of
the density fit can be accomplished by splitting the spherical core
and spherical valence contribution to \rhoat\ into different core and
valence shells whose occupation and radial extent can be varied
individually. This leads to the EHC model as for example proposed by Fischer
\textit{et al.} \cite{Fischer11}, in which the pseudo-atom densities are written as

\begin{align}
  \rho_\mathrm{at}(\mathbf{r}) & = \underbrace{P_{1,c} \kappa_{1,c}^3
            \rho_{1,c}(\kappa_{1,c},r) + P_{2,c} \kappa_{2,c}^3
            \rho_{2,c}(\kappa_{2,c},r) + \cdots}_\text{spherical core shells} \nonumber\\
          & + \underbrace{P_{1,v} \kappa_{1,v}^3
            \rho_{1,v}(\kappa_{1,v},r) + P_{2,v} \kappa_{2,v}^3
            \rho_{2,v}(\kappa_{2,v},r) + \cdots}_\text{spherical valence shells} \nonumber\\
          &  + \underbrace{\sum_{l = 0}^{l_\mathrm{max}}{(\kappa'_{v})^3 R_l(\kappa'_{v},r) 
            \sum_{m = -l}^l{P_{v,lm} d_{lm}(\theta,\phi)}}}_\text{aspherical valence deformation}
  \label{eq:rho-at-EHC}
\end{align}

A further improvement of model flexibility beyond the currently discussed level 
may be achieved by taking into account aspherical core polarizations\cite{Bentley_Stewart_1976}, as 
has been demonstrated for example by Fischer \textit{et al.} in a recent experimental 
x-ray charge-density study of $\alpha$-boron.\cite{Fischer21}

\section{Appendix B}
\label{sec:appendixB}

In our analysis of temperature-dependent changes to the electron
density distribution we focus on differences between observed
(\rhoobs) or calculated electron density distributions (\rhocalc).
This allows us to trace the origin of temperature-dependent changes by
decomposing the total density difference into contributions from
changes of (\textit{i})~lattice parameters, ($ii$)~atomic displacement
parameters (ADPs), $(iii)$~multipolar parameters and $(iv)$~residual
factors.  Both, \rhoobs\ and \rhocalc, are obtained from Fourier
summations over all measured reflections $\mathbf{h}$ and take into
account the limited experimental resolution. This, however, does not
lead to serious Fourier truncation errors, as only differences between
densities \rhoobs\ or \rhocalc\ are considered in the following. The
quantity

\begin{align}
  \rho_\mathrm{o}(\mathbf{r}) = \frac{s}{V} \sum_\mathbf{h}{|F_\mathbf{h}^\mathrm{o}|
  \exp\left(-2 \pi i \mathbf{h} \cdot \mathbf{r} + i \varphi_\mathbf{h}^\mathrm{c}\right)}
  \label{eq:rhoo}
\end{align}

\noindent ($V$:~unit cell volume; $s$:~scale factor) is computed from
observed structure factor amplitudes $|F_\mathbf{h}^\mathrm{o}|$,
\textit{i.e.} the square roots of the measured reflection intensities.
We note that due to the lack of observed structure factor phases
$\varphi_\mathbf{h}^\mathrm{o}$ calculated phases
$\varphi_\mathbf{h}^\mathrm{c}$ have to be used in the computation of
both \rhoobs\ and \rhocalc.  The usage of
$\varphi_\mathbf{h}^\mathrm{c}$ in Eq.~(\ref{eq:rhoo}), however,
represents no approximation, since ($i$)~the structure of \Mg\ is
centro-symmetric. Hence, the phase angles are limited to values of 0
and $\pi$, which again ($ii$)~are controlled by the fixed fractional
coordinates of magnesium and boron atoms on special positions in the
\Mg\ unit cell.

Similar to the electron density distribution from a MEM analysis,
\rhoobs\ contains the joint effects of chemical bonding, and static or
dynamic atomic displacements on the electron density distribution.  It
is also affected by experimental noise. The quantity

\begin{align}
  \rho_\mathrm{c}(\mathbf{r}) = \frac{1}{V} \sum_\mathbf{h}{|F_\mathbf{h}^\mathrm{c}|
  \exp\left(-2 \pi i \mathbf{h} \cdot \mathbf{r} + i \varphi_\mathbf{h}^\mathrm{c}\right)}
  \label{eq:rhoc}
\end{align}
  
\noindent is computed from structure factor amplitudes
$|F_\mathbf{h}^\mathrm{c}|$ calculated in our case on the basis of the
respective multipolar model. The individual model parameters determine
to which degree \rhocalc\ is affected by charge transfer, chemical
density deformations as well as static and dynamic atomic
displacements. Accordingly, we can systematically study 
the influence of temperature-dependent changes in (\textit{i}) lattice, 
(\textit{ii}) thermal and (\textit{iii}) multipolar parameters on the electronic structure
my mapping the
 \textit{difference} between electron density distributions
observed at different temperatures.

To simplify the following discussion, we introduce a shorthand
notation to indicate the employed parameters in the generation of
\rhoobs\ or \rhocalc\ maps. Our analysis showed that the \rhoobs\ maps
depend not only on the structure factor amplitudes
$|F_\mathbf{h}^\mathrm{o}|$, but also the lattice parameters (latt) of
\Mg\ at a specific temperature. The temperature, at which 
${|F_\mathbf{h}^\mathrm{o}|}$ and lattice parameters
were determined, will from now on be indicated by
specific superscripts in front of and behind the $\rho$ symbol:

\begin{align}
  \tensor*[^{|F_\mathbf{h}^\mathrm{o}|}]{\rho}{^{\mathrm{latt}}_{\mathrm{o}}}
\end{align}

On the other hand, the \rhocalc\ maps depend on the ADPs, the
multipolar parameters (MP) and the lattice parameters in the
underlying multipolar model at a specific temperature. The multipolar
models, from which the individual parameters were derived, are
indicated by superscripts or subscripts grouped around the $\rho$
symbol:

\begin{align}
  \tensor*[^{\mathrm{ADP}}_{\mathrm{MP}}]{\rho}{^{\mathrm{latt}}_{\mathrm{c}}}
\end{align}

\noindent A specific multipolar model is thereby characterized by the
measuring temperature of its underlying XRD data set. We emphasize
that parameters from different data sets or models may be combined.
For example, a \rhocalc\ map may be based on lattice and ADP
parameters pertaining to a temperature of 13.5~K and multipolar
parameters pertaining to room temperature.

To obtain the total electron density difference between two measuring
temperatures $T_1$ and $T_2$ ($T_1 < T_2$), we generate observed
density maps \rhoobs\ from XRD data sets collected at $T_1$ and $T_2$.
Thereby, the lattice parameters and structure factor amplitudes
$|F_\mathbf{h}^\mathrm{o}|$ at the respective temperatures are used.
The resulting maps are subtracted from each other:

\begin{align}
  \Delta \rho_\mathrm{total}(\mathbf{r}) =
  \tensor*[^{T_1}]{\rho}{^{T_1}_{\mathrm{o}}}(\mathbf{r}) -
  \tensor*[^{T_2}]{\rho}{^{T_2}_{\mathrm{o}}}(\mathbf{r})
  \label{eq:Drhototal}
\end{align}

Changes solely due to a variation of lattice parameters are also accessible from
the differences between two \rhoobs\ maps. In that case the difference maps are
generated using
the same $|F_\mathbf{h}^\mathrm{o}|$ values corresponding to the
higher temperature $T_2$, but different lattice parameters
corresponding to the lower temperature $T_1$ and the higher
temperature $T_2$, respectively:

\begin{align}
  \Delta \rho_\mathrm{latt}(\mathbf{r}) =
  \tensor*[^{T_2}]{\rho}{^{T_1}_{\mathrm{o}}}(\mathbf{r}) -
  \tensor*[^{T_2}]{\rho}{^{T_2}_{\mathrm{o}}}(\mathbf{r})
  \label{eq:Drholatt}
\end{align}

\noindent It is important to note that the choice of origin in the
comparison of \rhoobs\ or \rhocalc\ maps with differing lattice
parameters is non-trivial. To ensure consistent results the position
of a B--B BCP was used as a reference point in the generation of
difference density maps.

All remaining control parameters which influence the density
distribution in \Mg\ can be studied by analysis of \rhocalc\ as
obtained from our multipolar models. We thereby fix the lattice
parameters to their values at the lower temperature $T_1$, since the
effects of changes in the unit cell dimensions on the density already
have been extracted via Eq.~(\ref{eq:Drholatt}).  In that case, the
contribution of changes in the ADP parameters \DrhoADP\ becomes
visible, when \rhocalc\ maps for ($i$)~a model with multipolar (MP)
parameters corresponding to $T_2$ and ADPs corresponding to $T_1$ and
($ii$)~a model with MP parameters and ADPs corresponding to $T_2$ are
subtracted from each other:

\begin{align}
  \Delta \rho_\mathrm{ADP}(\mathbf{r}) =
  \tensor*[^{T_1}_{T_2}]{\rho}{^{T_1}_{\mathrm{c}}}(\mathbf{r}) -
  \tensor*[^{T_2}_{T_2}]{\rho}{^{T_1}_{\mathrm{c}}}(\mathbf{r})
  \label{eq:DrhoADP}
\end{align}

The contribution of changes in the MP parameters \DrhoMP\ is then
revealed by subtracting \rhocalc\ maps for ($i$)~a model with ADPs and
MP parameters corresponding to $T_1$ and ($ii$)~a model with ADPs
corresponding to $T_1$ and MP parameters corresponding to $T_2$ from
each other.

\begin{align}
  \Delta \rho_\mathrm{MP}(\mathbf{r}) =
  \tensor*[^{T_1}_{T_1}]{\rho}{^{T_1}_{\mathrm{c}}}(\mathbf{r}) -
  \tensor*[^{T_1}_{T_2}]{\rho}{^{T_1}_{\mathrm{c}}}(\mathbf{r})
  \label{eq:DrhoMP}
\end{align}

All residual contributions to \Drhototal\ not captured by our
multipolar models are contained in

\begin{align}
  \Delta \rho_\mathrm{res}(\mathbf{r}) = \Delta \rho_\mathrm{total}(\mathbf{r}) -
  \Delta \rho_\mathrm{latt}(\mathbf{r}) -
  \Delta \rho_\mathrm{ADP}(\mathbf{r}) - \Delta \rho_\mathrm{MP}(\mathbf{r})
  \label{eq:Drhores}
\end{align}

To assess the significance of features in \Drhototal\ or its
underlying contributions, the underlying error needs to be
known. Therefore, we generated maps of the standard deviation
$\sigma$(\Drhototal) over two thousand \Drhototal\ maps generated with
pseudo random numbers from the random module in NumPy\cite{Harris20}
added to the values of the structure factor amplitude
$|F_\mathbf{h}^\mathrm{o}|$ and scale factor $s$. Employed pseudo
random numbers $x$ follow normal distributions, whereby the
probability density functions

\begin{align}
  p(x) = \frac{1}{\sqrt{2\pi \sigma^2}} \exp{\left[-\frac{\left(x-\mu\right)^2}{2\sigma^2}\right]}
\end{align}

\noindent are determined individually for each
$|F_\mathbf{h}^\mathrm{o}|$ or $s$ using a mean $\mu$ of 0 and a
standard deviation $\sigma$ equal to the respective estimated standard
deviation.

\section{Appendix C}
\label{sec:appendixC}

The generation of the \DrhoDFT\ maps requires the calculation of
dynamic structure factors from DFT calculations. The folding of
a calculated static scattering factor $F_{\mathbf{h}}$ with a
Debye-Waller factor obtained from an 
atomic ADP is a trivial step for one-atomic structures like
Diamond. In contrast, compounds with more than one atom in the asymmetric unit such as \Mg\ require the decomposition
of the total crystal electron density $\rho_\mathrm{total,stat}(\mathbf{r})$
into atomic contributions $\rho_i(\mathbf{r})$ prior to the folding with the probability
density function $P(\mathbf{u})$ according to

\begin{align}
  \rho_{i,\mathrm{dyn}}(\mathbf{r})=\int_{-\infty}^{\infty} \rho_{i,\mathrm{stat}}(\mathbf{r-u})P(\mathbf{u})
\end{align}

\noindent where $P(\mathbf{u})$ is the probability of atom $i$ being
displaced by \textbf{u} from its rest position and the
Fourier-Transform of $P(\mathbf{u})$ is the Debye Waller factor
$Q(\mathbf{h})$.

We have implemented this folding process into a locally modified
version of the \texttt{DENPROP} code\cite{volkov_basis-set_2009},
which for this purpose has been interfaced to solid-state
programs such as
\texttt{WIEN2K}.\cite{blaha_wien2k_2020,blaha_wien2k_2018} Some
details of this implementation are outlined in the following.

\texttt{DENPROP} can be used to calculate static x-ray structure
factors from molecular wave functions represented in a Gaussian-type
(GTO) or Slater-type (STO) basis set. In the former case the according
Fourier-Transforms can be calculated analytically, whereas the
calculation needs to be performed numerically in the latter case.
But while the Fourier-Transform of an atom centered GTO basis function is thus readily calculated, the
separation of the two-center terms into atomic contributions needs to
be addressed properly (for an example, see the according implementation
in the \texttt{CRYSTAL}
code\cite{dovesi_crystal17_2017,dovesi_quantum-mechanical_2018}).

The numerical approach implemented in \texttt{DENPROP} for the STO basis sets does not suffer from
such ambiguities, but on the other hand requires sufficiently accurate integration
grids. These are obtained by an atomic partitioning of the total electron
density, for example via the (iterative or non-iterative) stockholder approach. The resulting atomic densities
$\rho_{i,\mathrm{stat}}(\mathbf{r})$ can then be accurately integrated
on standard Lebedev-Laikov grids, thus yielding static atomic scattering
factor contributions $f_{i,\mathrm{stat}}(\mathbf{h})$ which are finally
summed up to the $F_\mathrm{DFT,stat}(\mathbf{h})$. 

As this procedure
intrinsically produces atomic densities and subsequently atomic
scattering factors of the stockholder atoms from a total electron density distribution, we
added a routine for the folding of the
$f_{i,\mathrm{stat}}(\mathbf{h})$ with the according $Q(\mathbf{h})$
calculated from user provided isotropic or anisotropic ADPs. To generalize the
data-input format and thus allow for calculation of dynamical
structure factors also for solid state compounds represented for
example in a plane wave basis, we further generalized the routine
providing $\rho_\mathrm{total,stat}(\mathbf{r})$. In a first step all points of the atomic
radial and angular integration grids are generated in Cartesian
(x,y,z) format. This list of points can subsequently be used as
input for any code capable of providing the total electron density of
the system under investigation at a given point in space. Instead of
being calculated solely from databases of atomic wave functions in GTO
or STO format, the neutral atom reference electron density for the
stockholder partitioning can now also be directly supplied by the
user. For the present study we employed the atomic radial electron
densities from which the \texttt{WIEN2k} code constructs the
first-guess crystal density for the SCF calculation, while the total
electron density of \Mg\ was calculated from the \texttt{Wien2k} LAPW
wave function at the pre-generated list of points by the \texttt{CRITIC}
code.\cite{otero-de-la-roza_critic:_2009,otero-de-la-roza_critic2_2014}
Contrary to the molecular case, were the number of atoms which need to
be considered in the final integration step is simply defined by the
number of atoms in the molecule (when assuming that the chosen pseudo unit cell is
large enough for intermolecular interactions to be negligible), in solids the numerical accuracy critically depends on
the definition of an sufficiently large atomic cluster surrounding the
individual atom to be integrated. Our tests have shown, that a cut-off
radius of 15~\AA\ is usually sufficient for the generation of this
cluster.

In summary, our generalized implementation allows for the generation
of dynamic structure factors for a broad variety of use-cases, ranging
for example\cite{Batke_2017} from fully relativistic four-component calculations on molecules
with the \texttt{DIRAC} code\cite{DIRAC22} to all-electron solid state
calculations employing full-potential LAPW codes such as \texttt{Wien2k} or \texttt{ELK}.\cite{elk}

\FloatBarrier

\begin{acknowledgement}
Work done at Ames Laboratory (MYX and PCC, growth of single crystalline \Mg) was supported by the U.S. Department of Energy, Office of Basic Energy Science, Division of Materials Sciences and Engineering. Ames Laboratory is operated for the U.S. Department of Energy by Iowa State University under Contract No. DE-AC02-07CH11358.

\end{acknowledgement}

\begin{suppinfo}

The following files are available free of charge.
\begin{itemize}
\item supporting.pdf: Supporting Information with details on the
  synthesis and crystal growth, investigated samples, the collection and processing of XRD data and
  the refined SHC and EHC multipolar models as well as
  a table of salient ab-initio ADP values and additional maps
  illustrating the decomposition of temperature-dependent changes in
  the electron density distribution.
\end{itemize}

\end{suppinfo}

\bibliography{manuscript}

\end{document}


\captionsetup[longtable]{labelfont=rm,textfont=rm}


\section{Crystal growth}
\label{sec:crystal-growth}
Single crystals of \Mg\ were grown by a high pressure method from excess solution of magnesium. Elemental Mg and B were packed inside a 1 cm diameter boron nitride (BN)
crucible in the ratio of Mg:B = 1:0.7. A pressed Mg-pellet and B powder were placed in the crucible and the crucible was tightly packed with BN powder to prevent any Mg leakage at high pressure and temperature. Using a cubic anvil furnace, a pressure of 3.5~GPa was applied to the crucible at room temperature, and it was then heated to 1430~\degr C and held at this temperature for 1~hr. Afterwards
the temperature was decreased to the 650~\degr C melting point of Mg over a time of
6 hrs. After completion of the crystal growth, the furnace heating was turned off to
achieve a faster cooling and then the pressure was released. This procedure yields \Mg\
single crystals embedded inside the solidified Mg-melt. Individual single crystals were
extracted by distilling the excess Mg at 750~\degr C inside an evacuated quartz tube. Given that the heater used in the high temperature/high pressure furnace is a graphite tube, a slight C-doping is possible, reducing \Tc\ slightly from the reported value for pure \Mg.

\newpage

\section{Investigated samples}
\label{sec:investigated-samples}

\subsection{Crystal 1}
\label{sec:photo-crystal-1}

\begin{figure}[htb]
  \centering
  \includegraphics[height=6.5cm]{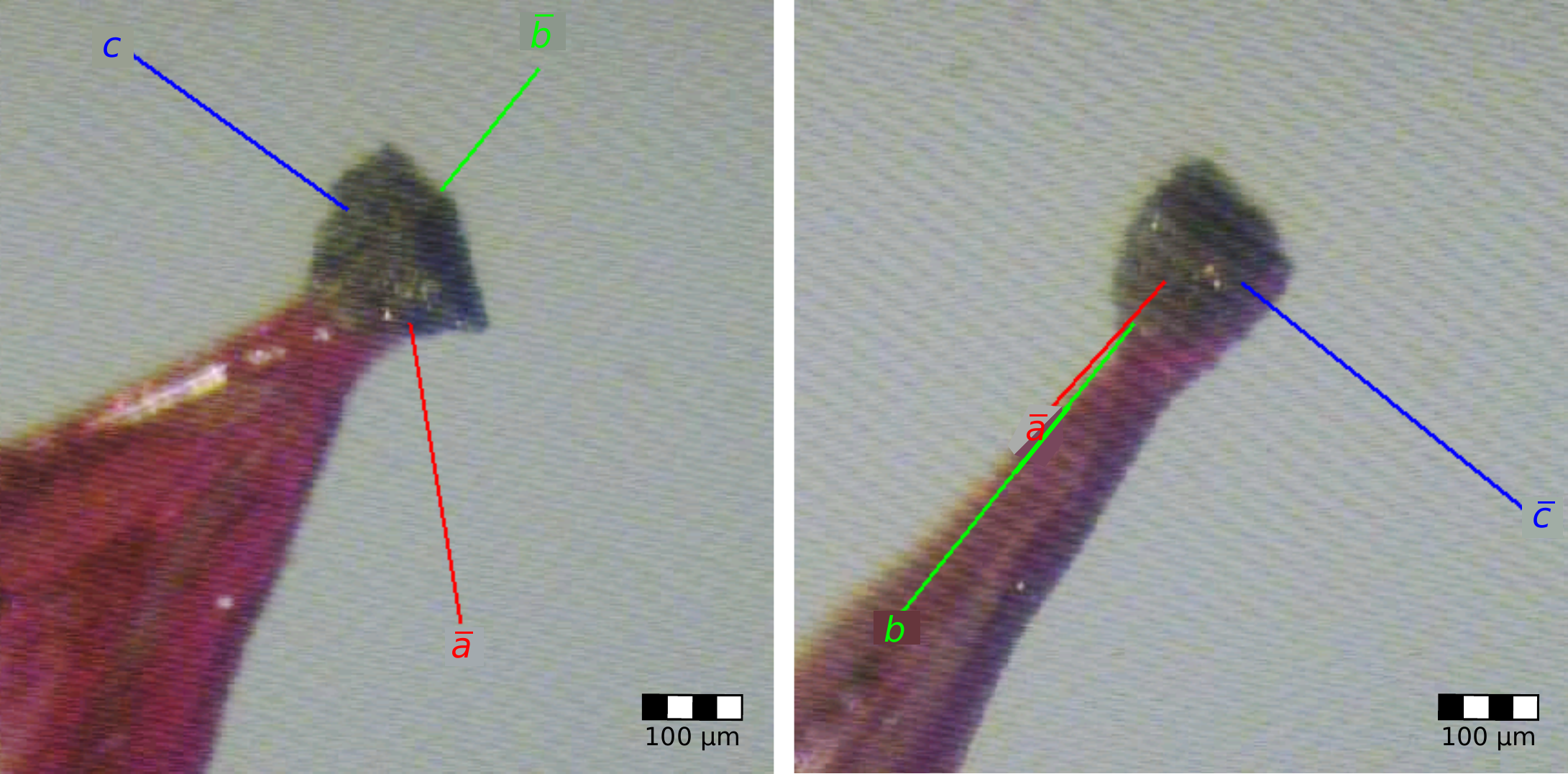}
  \caption{Photographic image of \Mg\ single crystal 1 used in the XRD
    experiments. Crystal axes $a$, $b$ and $c$ referring to the
    hexagonal unit cell are indicated by colored lines.}
  \label{fig:photo-crystal-1}
\end{figure}

\subsection{Crystal 2}
\label{sec:photo-crystal-2}

\begin{figure}[htb]
  \centering
  \includegraphics[height=6.5cm]{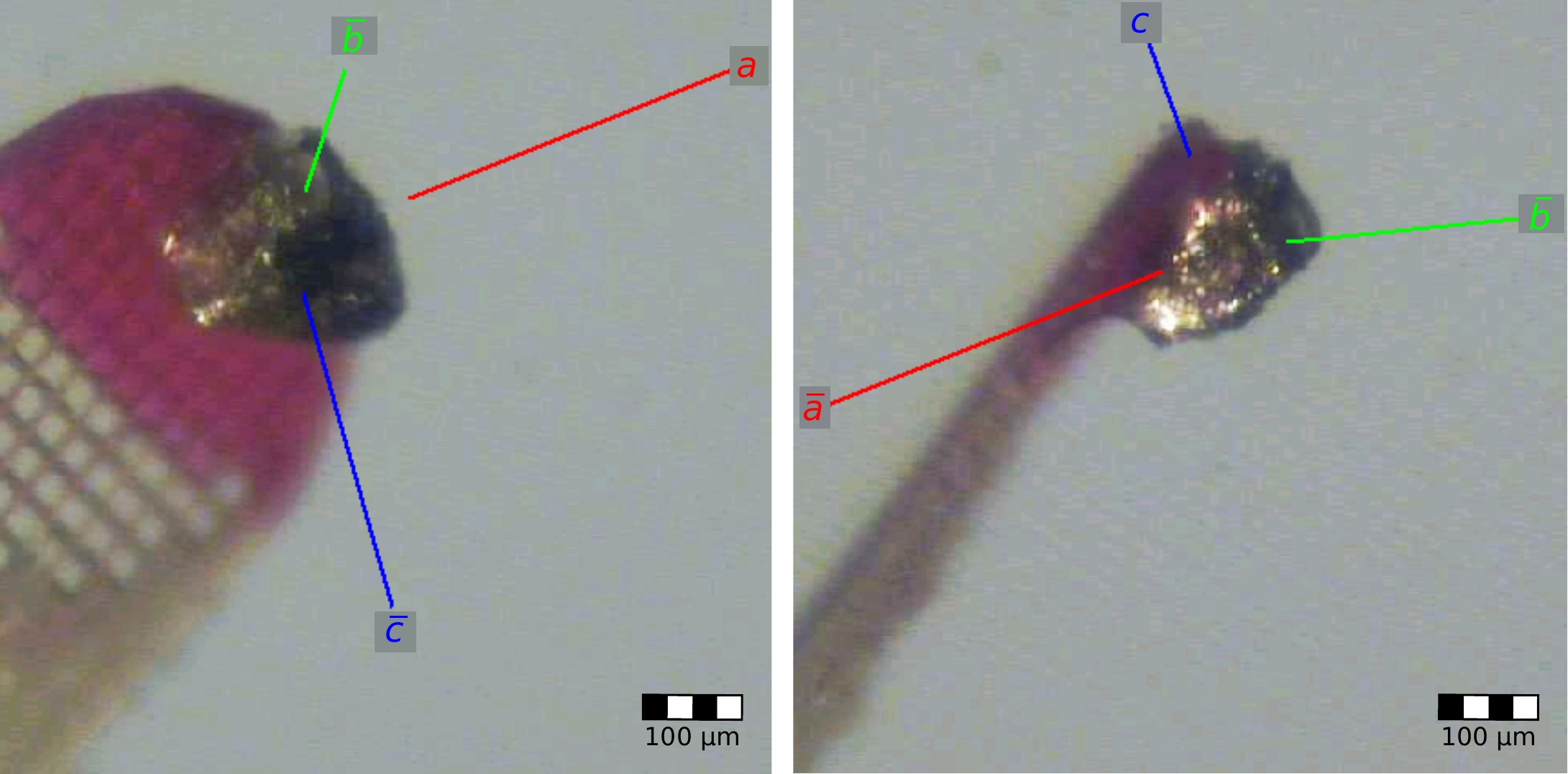}
  \caption{Photographic image of \Mg\ single crystal 2 used in the
    magnetization measurements. Crystal axes $a$, $b$ and $c$
    referring to the hexagonal unit cell are indicated by colored
    lines.}
  \label{fig:photo-crystal-2}
\end{figure}

\begin{figure}[htb]
  \centering
  \includegraphics[height=9cm]{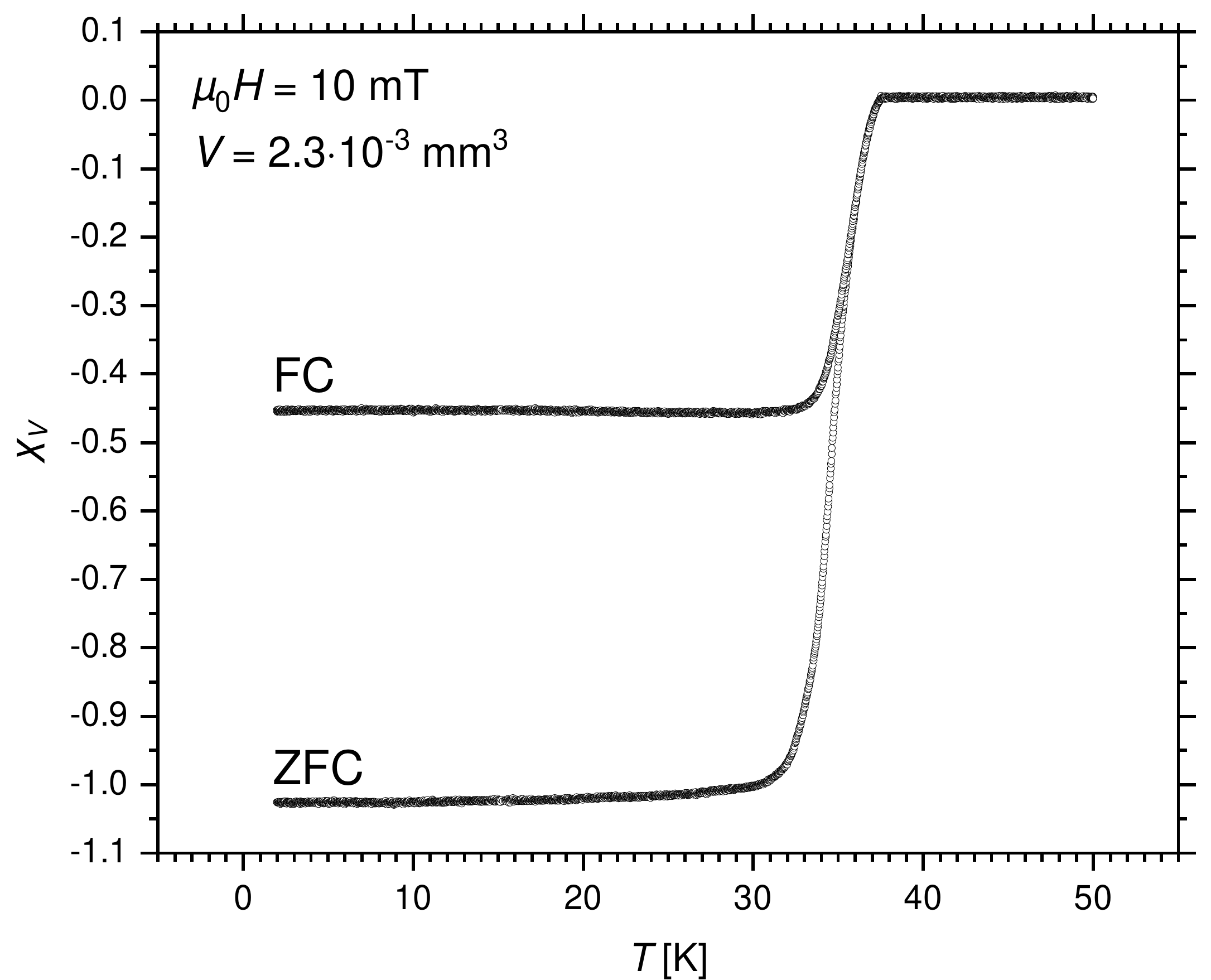}
  \caption{Temperature-dependent DC magnetization of crystal~2
    measured under zero-field cooling (ZFC) and field-cooling (FC)
    conditions in an applied field $\mu_0H =$~10~mT.}
  \label{fig:zfc-fc-crystal-2}
\end{figure}

\FloatBarrier
\hspace{1em}
\newpage

\section{Single-crystal XRD experiments}
\label{sec:experiment-overview}

\subsection{Overview}

\begin{longtable}[c]{p{0.25\textwidth}>{\centering}p{0.2\textwidth}
    >{\centering\arraybackslash}p{0.2\textwidth}}
  \hline
    \centering \textbf{experiment} & \textbf{1} & \textbf{2}\\
    \hline\hline\endhead
    sample & crystal~1 & crystal~1\\
    sample dim. [\textmu m$^3$] & 132$\times$138$\times$180 & 132$\times$138$\times$180\\
    sample photo & Fig.~\ref{fig:photo-crystal-1} & Fig.~\ref{fig:photo-crystal-1}\\[0.3cm]
    diffractometer & 1 & 2\\[0.3cm]
    \multirow{4}{*}{temperature(s) [K]} & \multirow{4}{*}{100(2)} & 13.5(5)\\[-0.3cm]
    && 25.086(5)\\[-0.3cm]
    && 45.009(7)\\[-0.3cm]
    && 298.3(1)\\
    sample cooling & open-flow N$_2$ & closed-cycle He\\
    vacuum chamber & -- & beryllium domes\\
    \hline
    \caption{Overview of the single-crystal XRD experiments on
      \Mg. The numbering of diffractometer setups is explained in the
      Methods Section of the main paper.}
  \label{tab:single-crystal-xrd-experiments}
\end{longtable}

\FloatBarrier
\newpage

\subsection{Details}

\begin{longtable}[c]{p{0.3\textwidth}>{\centering\arraybackslash}p{0.23\textwidth}}
  \hline
  experiment number & \multirow{2}{*}{1} \\[-0.3cm]
  in Tab.~\ref{tab:single-crystal-xrd-experiments} &\\
    temperature & 100(2)~K\\
    \multirow{3}{*}{unit cell dimensions} & $a =$~3.0849(1)~\AA\\
            & $c =$~3.5113(2)~\AA\\
            & $V =$~28.939(3)~\AA$^3$\\
            calculated density & 2.6213~g$\cdot$cm$^{-3}$\\
    sample & crystal~1\\
    crystal size & 132$\times$138$\times$180~\mumm\\
    wave length & 0.56087~\AA\\
    absorption correction & multi-scan\\
    transm. ratio (max/min) & 0.9582 / 0.9283\\
    absorption coefficient & 0.322~mm$^{-1}$\\
    $F(000)$ &  22\\
    $\theta$ range & 5\degr\ to 63\degr\\
    range in $hkl$ &  -9/9, -9/8, -11/10\\
    total no. reflections & 15642\\
    independent reflections & 237 ($R_\mathrm{int} =$~0.0245)\\
    reflections with $I \geq 3\sigma(I)$ & 237\\
    \hline
    \caption{Experimental parameters of the high-resolution
      single-crystal XRD data set collected for crystal 1 at a
      temperature of 100(2)~K (experiment~1 in
      Tab.~\ref{tab:single-crystal-xrd-experiments}).}
  \label{tab:crystal-1-100k-dataset}
\end{longtable}

\newpage

\begin{longtable}[c]{p{0.3\textwidth}>{\centering}p{0.23\textwidth}
    >{\centering\arraybackslash}p{0.23\textwidth}}
    \hline
    \centering $T$~[K] & 298.3(1) & 45.009(7)\\
    \hline
    experiment number & \multirow{2}{*}{2} & \multirow{2}{*}{2}\\[-0.3cm]
    in Tab.~\ref{tab:single-crystal-xrd-experiments} &&\\
    \multirow{3}{*}{unit cell dimensions} & $a =$~3.09413(12)~\AA & $a =$~3.08861(4)~\AA\\
            & $c =$~3.52625(15)~\AA & $c =$~3.51561(6)~\AA\\
            & $V =$~29.236(2)~\AA$^3$ & $V =$~29.044(1)~\AA$^3$\\
    calculated density & 2.6085~g$\cdot$cm$^{-3}$ & 2.6257~g$\cdot$cm$^{-3}$\\
    crystal size & \multicolumn{2}{c}{132$\times$138$\times$180~\mumm}\\
    wave length & \multicolumn{2}{c}{0.56087~\AA}\\
    absorption correction & \multicolumn{2}{c}{multi-scan}\\
    transm. ratio (max/min) & 0.9824 / 0.9354 & 0.9825 / 0.9323\\
    absorption coefficient & 0.321~mm$^{-1}$ & 0.323~mm$^{-1}$\\
    $F(000)$ &  \multicolumn{2}{c}{22}\\
    $\theta$ range & \multicolumn{2}{c}{5\degr\ to 48\degr}\\
    range in $hkl$ & -4/7, -8/4, -9/9 & -4/7, -8/4, -9/9\\
    total no. reflections & 3387 & 3389\\
    independent reflections & 148 ($R_\mathrm{int} =$~0.004) &
    148 ($R_\mathrm{int} =$~0.004)\\
    reflections with $I \geq 3\sigma(I)$ & 148 & 148\\
    \hline
    \caption{Experimental parameters of the low-temperature XRD data
      sets collected for crystal 1 at temperatures of 298.3(1)~K and
      45.009(7)~K (experiment~2 in
      Tab.~\ref{tab:single-crystal-xrd-experiments}).}
  \label{tab:crystal-1-lt-datasets-1}
\end{longtable}

\newpage

\begin{longtable}[c]{p{0.3\textwidth}>{\centering}p{0.23\textwidth}
    >{\centering\arraybackslash}p{0.23\textwidth}}
    \hline
    \centering $T$~[K] & 25.086(5)~K & 13.5(5)~K\\
    \hline
    experiment number & \multirow{2}{*}{2} & \multirow{2}{*}{2}\\[-0.3cm]
    in Tab.~\ref{tab:single-crystal-xrd-experiments} &&\\
    \multirow{3}{*}{unit cell dimensions} & $a =$~3.08836(3)~\AA & $a =$~3.08934(3)~\AA\\
            & $c =$~3.51515(6)~\AA & $c =$~3.51629(5)~\AA\\
            & $V =$~29.036(1)~\AA$^3$ & $V =$~29.063(1)~\AA$^3$\\
    calculated density & 2.6263~g$\cdot$cm$^{-3}$ & 2.6240~g$\cdot$cm$^{-3}$\\
    crystal size & \multicolumn{2}{c}{132$\times$138$\times$180~\mumm}\\
    wave length & \multicolumn{2}{c}{0.56087~\AA}\\
    absorption correction & \multicolumn{2}{c}{multi-scan}\\
    transm. ratio (max/min) & 0.9824 / 0.9310 & 0.9824 / 0.9421\\
    absorption coefficient & 0.323~mm$^{-1}$ & 0.323~mm$^{-1}$\\
    $F(000)$ &  \multicolumn{2}{c}{22}\\
    $\theta$ range & \multicolumn{2}{c}{5\degr\ to 48\degr}\\
    range in $hkl$ & -4/7, -8/4, -9/9 & -4/8, -8/4, -9/9\\
    total no. reflections & 3387 & 3387\\
    independent reflections & 148 ($R_\mathrm{int} =$~0.0038) &
    148 ($R_\mathrm{int} =$~0.0036)\\
    reflections with $I \geq 3\sigma(I)$ & 148 & 148\\
    \hline
    \caption{Experimental parameters of the low-temperature XRD data
      sets collected for crystal 1 at temperatures of 25.086(5)~K and
      13.5(5)~K (experiment~2 in
      Tab.~\ref{tab:single-crystal-xrd-experiments}).}
  \label{tab:crystal-1-lt-datasets-2}
\end{longtable}

\FloatBarrier
\newpage

\subsection{Comparability of temperature-dependent XRD
  experiments}
\label{sec:lt-comp-temp-points}

Several measures were taken to ensure comparability between
single-crystal XRD data sets collected at different temperatures
(experiment~2 in Tab.~\ref{tab:single-crystal-xrd-experiments}; data
set parameters in Tab.~\ref{tab:crystal-1-lt-datasets-1}):

\begin{enumerate}
\item All temperature-dependent measurements on a sample were taken in
  direct sequence without re-orientation and without interruption of
  cooling.
\item Integration and reduction of all data sets were performed with
  identical parameters in EVAL14\cite{Duisenberg03} and
  SADABS,\cite{Krause15} respectively.
\item Only reflections which have been collected at all four temperature points were
  read into SADABS\cite{Krause15} for data reduction to limit the data sets to the same number of hkl reflections. Intensities
  of symmetry-equivalent reflections were merged.
\end{enumerate}

\FloatBarrier
\newpage

\section{Resolution dependence of the magnesium site occupation
  factor}
\label{sec:res-dependence-mg}

\subsection{IAM using static theoretical structure factors}

\begin{figure}[htb]
  \centering
  \includegraphics[height=11cm]{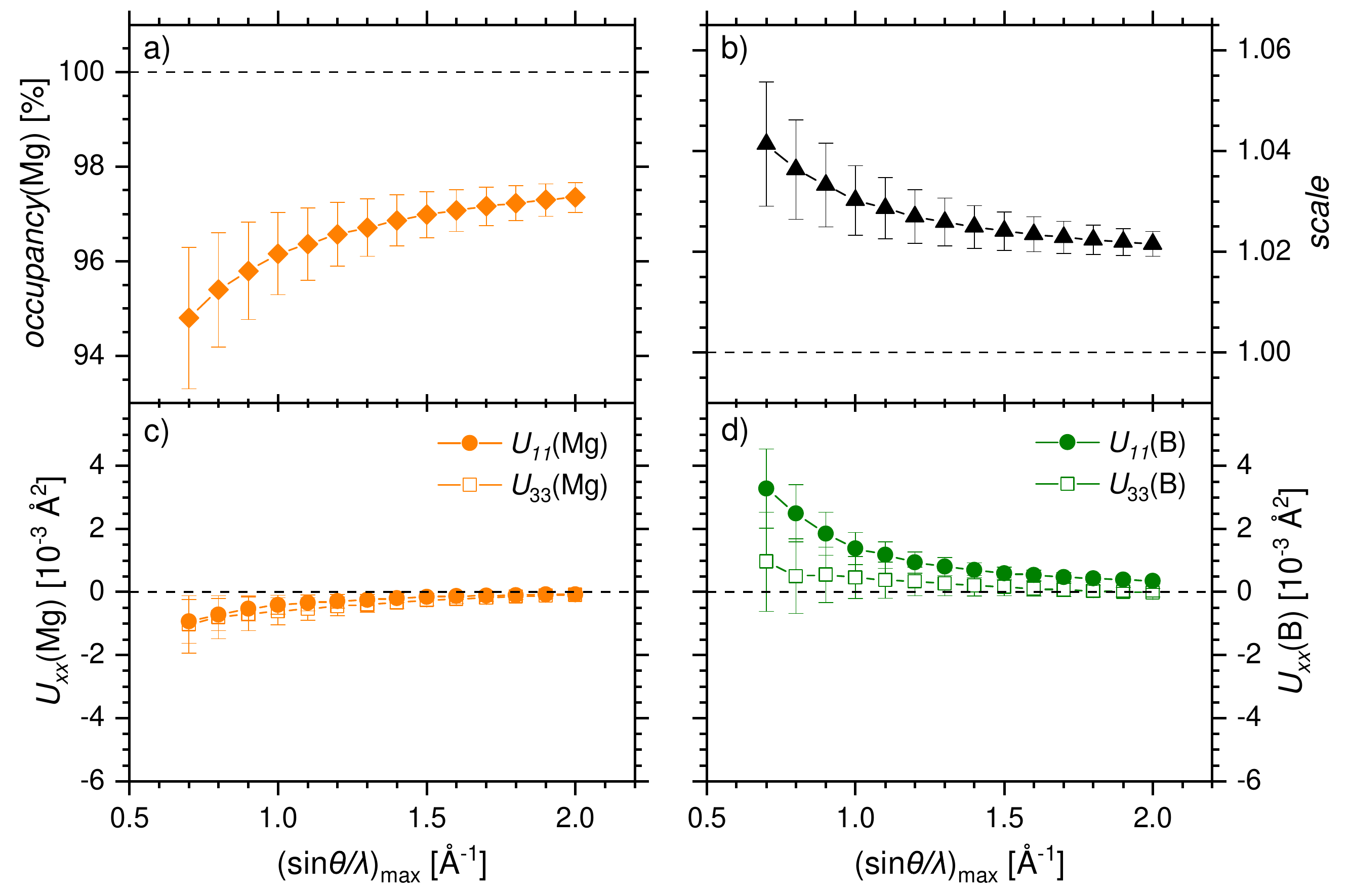}
  \caption{Variation of a)~the magnesium site occupancy, b)~the scale
    factor, and the anisotropic ADPs $U_{11}$ and $U_{33}$ for c)~the
    magnesium atom and d)~the boron atom with increasing resolution
    $(\sin\theta/\lambda)_{max}$ of static theoretical structure
    factors.  All given parameters were refined simultaneously employing IAMs based on scattering factors for neutral
    atoms. Expected parameter values are marked by dashed horizontal
    lines, while error bars denote the estimated standard deviation 
    ($1 \times \sigma$) of the 
    parameter values as obtained from least-squares refinement 
    using unit weights for all reflection intensities.}
  \label{fig:theo-IAM-scale-Uaniso-ai-param-overview}
\end{figure}

\newpage

\subsection{IAM using experimental XRD data}

\begin{figure}[htb]
  \centering
  \includegraphics[height=11cm]{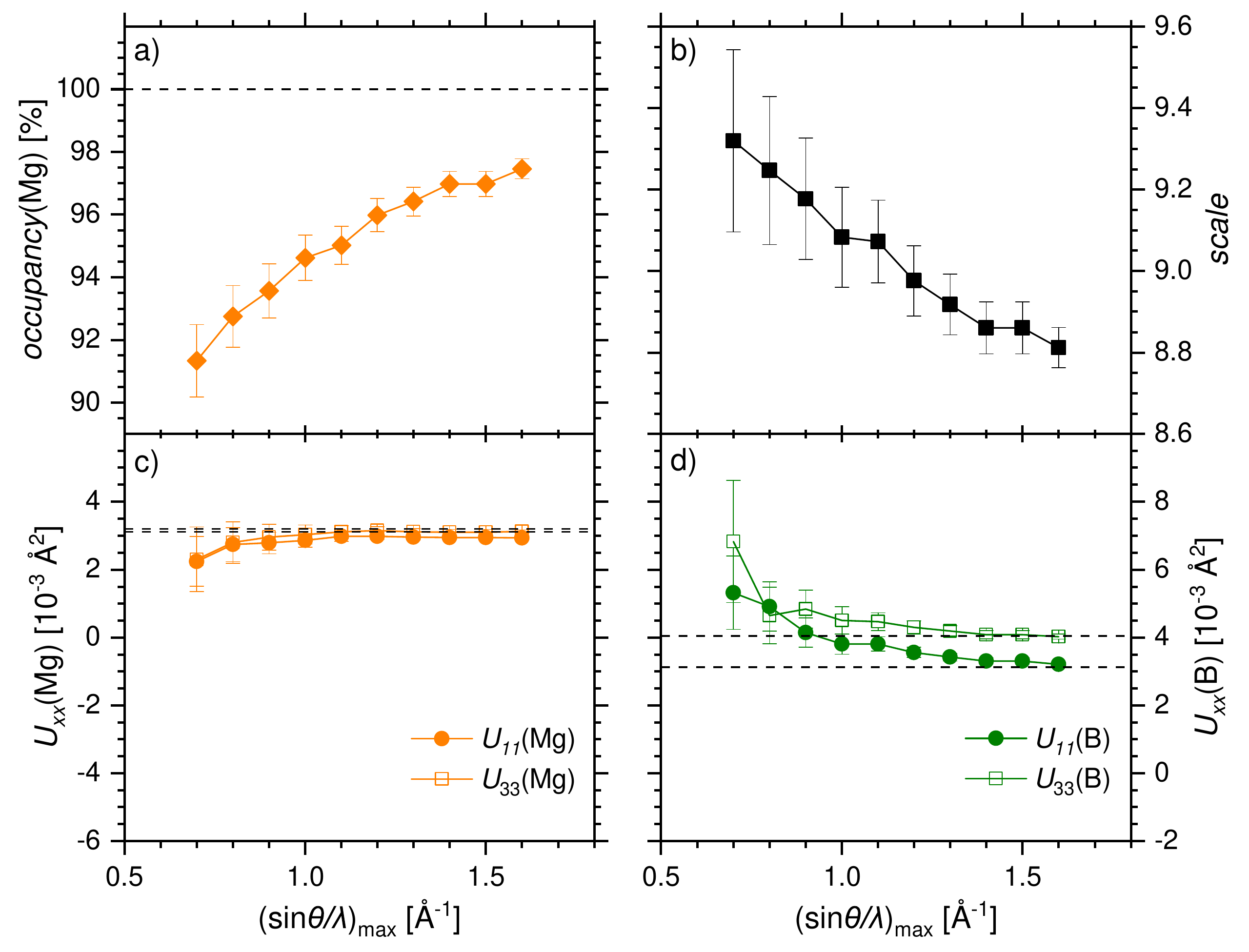}
  \caption{Variation of a)~the magnesium site occupation factor,
    b)~the scale factor, and the anisotropic ADPs $U_{11}$ and
    $U_{33}$ for c)~the magnesium atom and d)~the boron atom with
    increasing resolution $(\sin\theta/\lambda)_{max}$ of experimental
    single-crystal XRD data (experiment 1 in
    Tab.~\ref{tab:single-crystal-xrd-experiments}; data set parameters
    in Tab.~\ref{tab:crystal-1-100k-dataset}).  All given parameters
    were refined simultaneously employing IAMs based on
    scattering factors for neutral atoms. Expected parameter values
    are marked by dashed horizontal lines. In case of c) and d) these
    are the ADP values for $T =$~100~K derived from DFT calculations.
    Error bars refer to the estimated standard deviation  
    ($1 \times \sigma$) of the parameter values 
    as obtained from least-squares refinement.}
  \label{fig:exp-IAM-scale-Uaniso-ai-param-overview}
\end{figure}

\newpage

\subsection{EHC multipolar model using static theoretical
  structure factors}

\begin{figure}[htb]
  \centering
  \includegraphics[height=11cm]{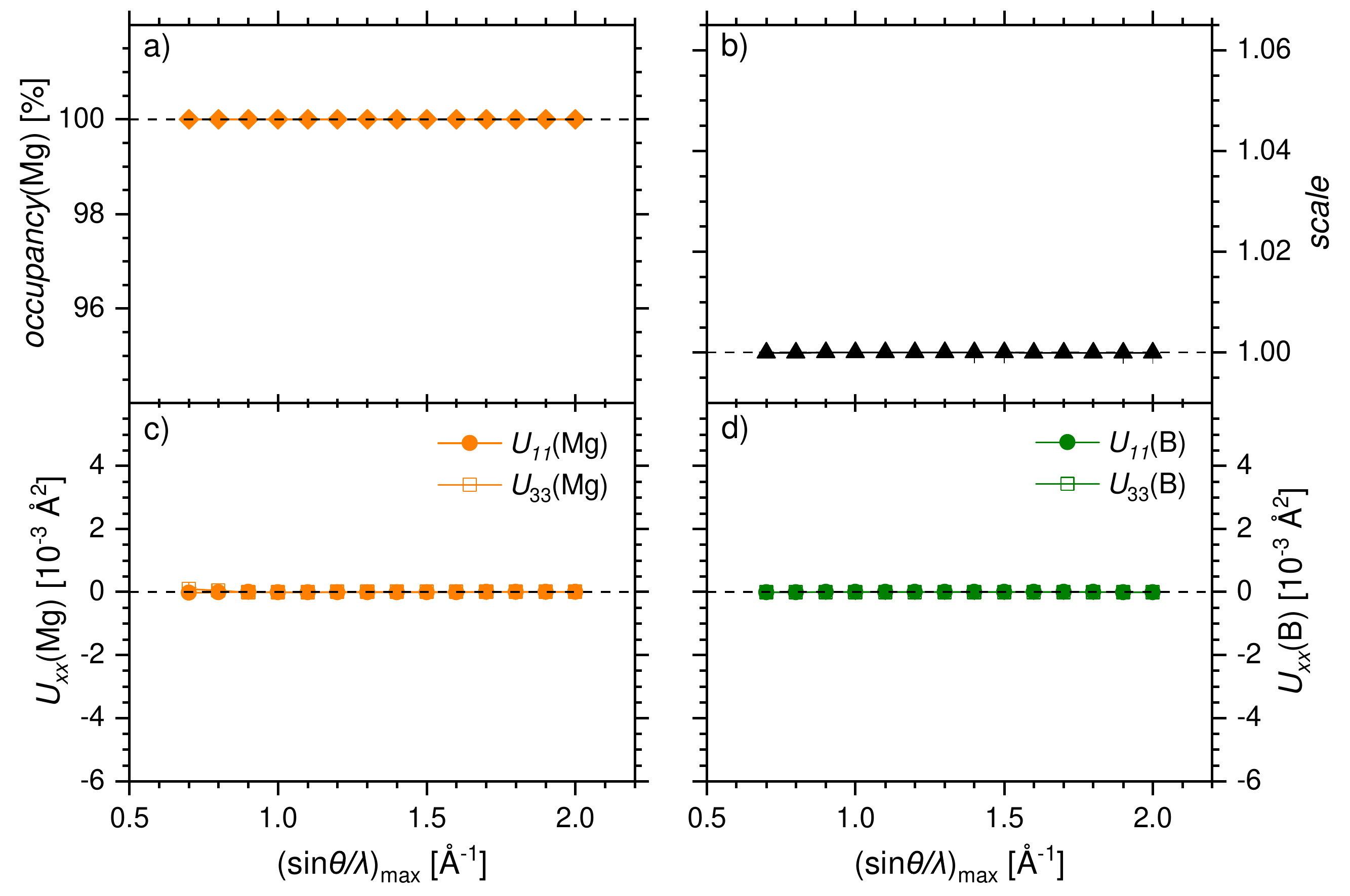}
  \caption{Variation of a)~the magnesium site occupancy, b)~the scale
    factor, and the anisotropic ADPs $U_{11}$ and $U_{33}$ for c)~the
    magnesium atom and d)~the boron atom with increasing resolution
    $(\sin\theta/\lambda)_{max}$ of static theoretical structure
    factors.  All given parameters were refined simultaneously employing EHC models with multipolar parameters fixed to their
    values in T-EHCM1 (Tab.~\ref{tab:theo-ehc}). Expected parameter
    values are marked by dashed horizontal lines, while error bars
    denote the estimated standard deviation ($1 \times \sigma$) of the 
    parameter values as obtained from least-squares refinement 
    using unit weights for all reflection intensities.}
  \label{fig:theo-ehc-scale-Uaniso-ai-param-overview}
\end{figure}

\newpage

\subsection{EHC multipolar model using experimental XRD data}

\begin{figure}[htb]
  \centering
  \includegraphics[height=11cm]{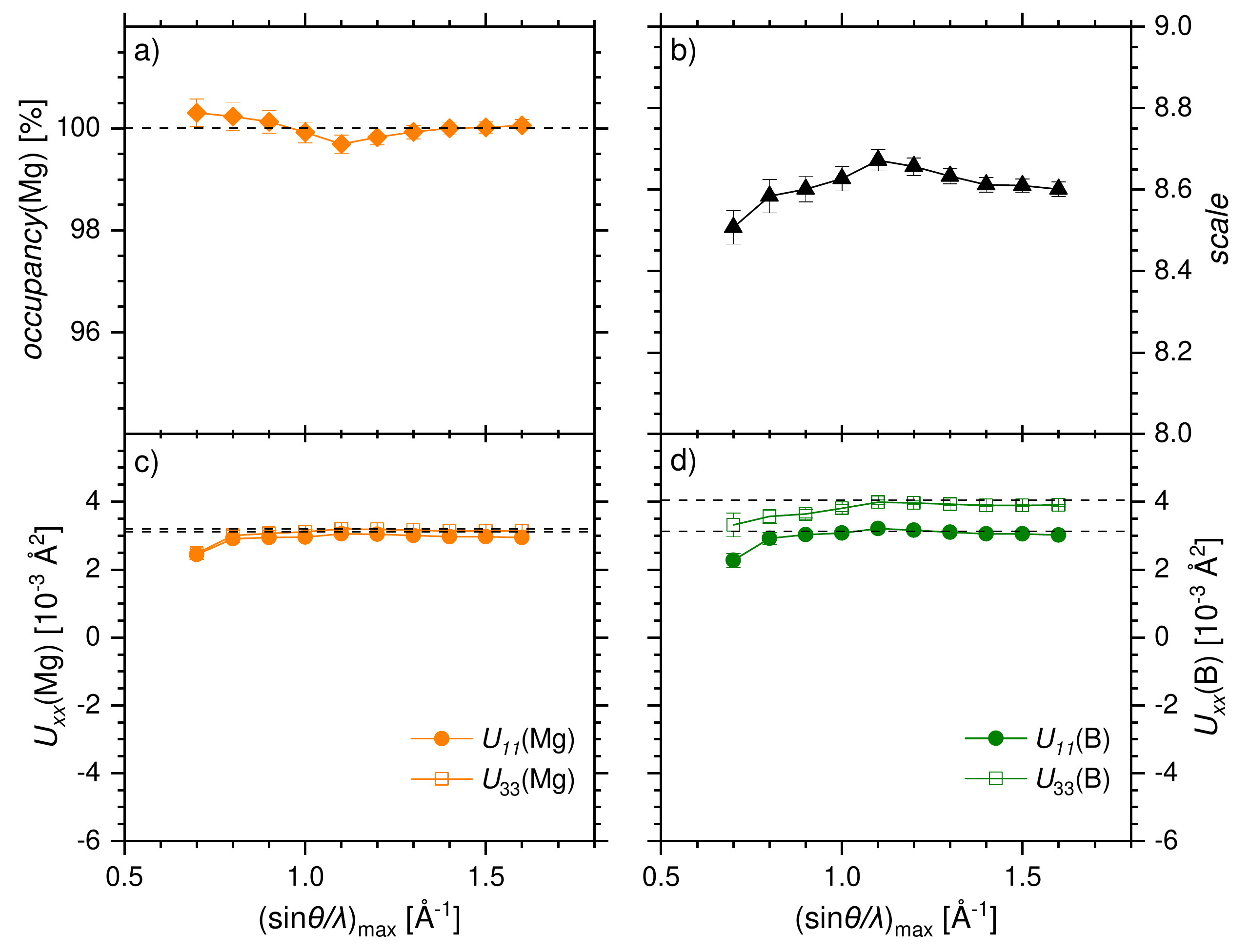}
  \caption{Variation of a)~the magnesium site occupation factor,
    b)~the scale factor, and the anisotropic ADPs $U_{11}$ and
    $U_{33}$ for c)~the magnesium atom and d)~the boron atom with
    increasing resolution $(\sin\theta/\lambda)_{max}$ of experimental
    single-crystal XRD data (experiment 1 in
    Tab.~\ref{tab:single-crystal-xrd-experiments}; data set parameters
    in Tab.~\ref{tab:crystal-1-100k-dataset}). All given parameters
    were refined simultaneously employing EHC models with
    multipolar parameters fixed to their values in E-EHCM1
    (Tab.~\ref{tab:exp-ehc}). Expected parameter values are marked by
    dashed horizontal lines. In case of c) and d) this is the ADP
    values for $T =$~100~K derived from DFT calculations.
    Error bars refer to the estimated standard deviation  
    ($1 \times \sigma$) of the parameter values 
    as obtained from least-squares refinement.}
  \label{fig:exp-ehc-scale-Uaniso-ai-param-overview}
\end{figure}

\FloatBarrier
\newpage

\section{Multipolar models}
\label{sec:mp-models}

\subsection{Local coordinate systems}

All multipolar models in this paper refer to the same local
coordinate systems for the two symmetry-independent atoms in the \Mg\
structure, \textit{i.e.} Mg and B. In both cases, the local $x$ and
$z$ axes are oriented parallel to the $a$ and $c$ axes of the unit
cell, respectively. The $y$ axes are chosen to complete the
right-handed orthonormal coordinate systems.

\vspace{1cm}

\begin{figure}[htb]
  \centering
  \includegraphics[width=0.8\textwidth]{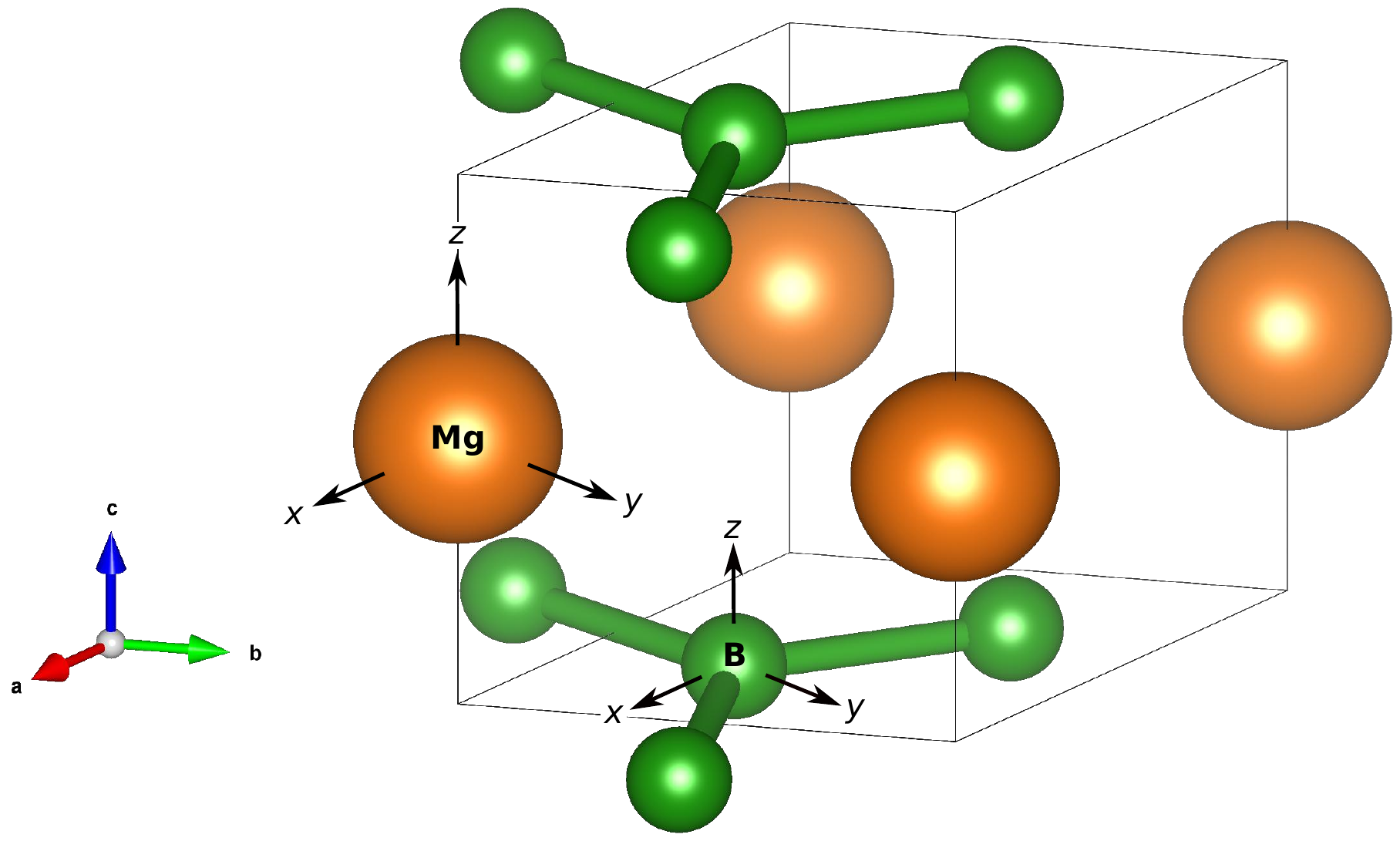}
  \caption{Definition of local coordinate systems for the two
    symmetry-independent atoms Mg and B.}
  \label{fig:local-cs}
\end{figure}

\FloatBarrier
\newpage

\subsection{Model parameters}

\subsubsection{EHC models using static theoretical structure factors}

\begin{longtable*}[c]{>{\centering}p{0.15\textwidth}>{\centering}p{0.1\textwidth}
    >{\centering}p{0.19\textwidth}>{\centering\arraybackslash}p{0.19\textwidth}}
    \hline
    && \textbf{T-EHCM1} & \textbf{T-EHCM2}\\
    \hline\hline\endhead
    & $scale$ & 1.0$^\ast$ & 1.01007(19)\\
    \hline
    Mg1 & $occ.$ & 1.0$^\ast$ & 1.00152(14)\\
    {[~]($1s^2$)} & $x$ & \multicolumn{2}{c}{0$^\ast$}\\
    & $y$ & \multicolumn{2}{c}{0$^\ast$}\\
    & $z$ & \multicolumn{2}{c}{0.5$^\ast$}\\
    & $U_{11}$ [\AA$^2$] & 0.0$^\ast$ & 0.000078(4)\\
    & $U_{33}$ [\AA$^2$] & 0.0$^\ast$ & 0.000081(4)\\
    & $P_c$ & \multicolumn{2}{c}{0.0$^\ast$}\\
    & $P_v$ & 1.9781(7) & 1.9671(7)\\
    & $\kappa$ & 1.0148(7) & 1.0363(12)\\
    \hline
    Mg2 & $occ.$ & 1.0$^\ast$ & 1.00152$^\ast$\\
    {[~]($2s^2$, $2p^6$)} & $x$ & \multicolumn{2}{c}{0$^\ast$}\\
    & $y$ & \multicolumn{2}{c}{0$^\ast$}\\
    & $z$ & \multicolumn{2}{c}{0.5$^\ast$}\\
    & $U_{11}$ [\AA$^2$] & 0.0$^\ast$ & 0.000078$^\ast$\\
    & $U_{33}$ [\AA$^2$] & 0.0$^\ast$ & 0.000081$^\ast$\\
    & $P_c$ & \multicolumn{2}{c}{0.0$^\ast$}\\
    & $P_v$ & 7.968863$^\ast$ & 7.916841$^\ast$\\
    & $\kappa$ & 1.01006(9) & 1.01055(10)\\
    \hline
\end{longtable*}

\newpage

\begin{longtable*}[c]{>{\centering}p{0.15\textwidth}>{\centering}p{0.1\textwidth}
    >{\centering}p{0.19\textwidth}>{\centering\arraybackslash}p{0.19\textwidth}}
    \hline
    && \textbf{T-EHCM1} & \textbf{T-EHCM2}\\
    \hline\hline\endhead
    Mg3 & $occ.$ & 1.0$^\ast$ & 1.00152$^\ast$\\
    {[~]($3s^2$)} & $x$ & \multicolumn{2}{c}{0$^\ast$}\\
    & $y$ & \multicolumn{2}{c}{0$^\ast$}\\
    & $z$ & \multicolumn{2}{c}{0.5$^\ast$}\\
    & $U_{11}$ [\AA$^2$] & 0.0$^\ast$ & 0.000078$^\ast$\\
    & $U_{33}$ [\AA$^2$] & 0.0$^\ast$ & 0.000081$^\ast$\\
    & $P_c$ & \multicolumn{2}{c}{0.0$^\ast$}\\
    & $P_v$ & 1.9142(4) & 1.9399(15)\\
    & $\kappa$ & 0.7939(17) & 0.7981(16)\\
    & $\kappa'$ & 1.363(7) & 1.366(6)\\
    & $P_{20}$ & 0.0025(6) & 0.0016(6)\\
    & $P_{40}$ & -0.1556(17) & -0.1555(17)\\
    \hline
    B1 & $occ.$ & 1.0$^\ast$ & 1.0$^\ast$\\
    {[~]($1s^2$)} & $x$ & \multicolumn{2}{c}{$\frac{2}{3}^\ast$}\\
    & $y$ & \multicolumn{2}{c}{$\frac{1}{3}^\ast$}\\
    & $z$ & \multicolumn{2}{c}{0$^\ast$}\\
    & $U_{11}$ [\AA$^2$] & 0.0$^\ast$ & -0.000085(5)\\
    & $U_{33}$ [\AA$^2$] & 0.0$^\ast$ & -0.000091(5)\\
    & $P_c$ & \multicolumn{2}{c}{0.0$^\ast$}\\
    & $P_v$ & 1.9777(4) & 1.9860(4)\\
    & $\kappa$ & 1.00554(19) & 0.9922(4)\\
    \hline
\end{longtable*}

\newpage

\begin{longtable*}[c]{>{\centering}p{0.15\textwidth}>{\centering}p{0.1\textwidth}
    >{\centering}p{0.19\textwidth}>{\centering\arraybackslash}p{0.19\textwidth}}
    \hline
    && \textbf{T-EHCM1} & \textbf{T-EHCM2}\\
    \hline\hline\endhead
    B2 & $occ.$ & 1.0$^\ast$ & 1.0$^\ast$\\
    {[~]($2s^1$)} & $x$ & \multicolumn{2}{c}{$\frac{2}{3}^\ast$}\\
    & $y$ & \multicolumn{2}{c}{$\frac{1}{3}^\ast$}\\
    & $z$ & \multicolumn{2}{c}{0$^\ast$}\\
    & $U_{11}$ [\AA$^2$] & 0.0$^\ast$ & -0.000085$^\ast$\\
    & $U_{33}$ [\AA$^2$] & 0.0$^\ast$ & -0.000091$^\ast$\\
    & $P_c$ & \multicolumn{2}{c}{0.0$^\ast$}\\
    & $P_v$ & 1.5874(7) & 1.5888(7)\\
    & $\kappa$ & 0.9161(9) & 0.9468(8)\\
    & $\kappa'$ & 0.9586(18) & 0.9562(18)\\
    & $P_{20}$ & -0.0727(3) & -0.0727(3)\\
    & $P_{33-}$ & 0.0816(4) & 0.0816(4)\\
    & $P_{40}$ & 0.0019(3) & 0.0019(3)\\
    \hline
\end{longtable*}

\newpage

\begin{longtable}[c]{>{\centering}p{0.15\textwidth}>{\centering}p{0.1\textwidth}
    >{\centering}p{0.19\textwidth}>{\centering\arraybackslash}p{0.19\textwidth}}
    \hline
    && \textbf{T-EHCM1} & \textbf{T-EHCM2}\\
    \hline\hline\endhead
    B3 & $occ.$ & 1.0$^\ast$ & 1.0$^\ast$\\
    {[~]($2p^2$)} & $x$ & \multicolumn{2}{c}{$\frac{2}{3}^\ast$}\\
    & $y$ & \multicolumn{2}{c}{$\frac{1}{3}^\ast$}\\
    & $z$ & \multicolumn{2}{c}{0$^\ast$}\\
    & $U_{11}$ [\AA$^2$] & 0.0$^\ast$ & -0.000085$^\ast$\\
    & $U_{33}$ [\AA$^2$] & 0.0$^\ast$ & -0.000091$^\ast$\\
    & $P_c$ & \multicolumn{2}{c}{0.0$^\ast$}\\
    & $P_v$ & 1.5044(8) & 1.5133(8)\\
    & $\kappa$ & 1.0744(13) & 1.0040(14)\\
    & $\kappa'$ & 0.9586$^\ast$ & 0.9562$^\ast$\\
    & $P_{20}$ & -0.0727$^\ast$ & -0.0727$^\ast$\\
    & $P_{33-}$ & 0.0816$^\ast$ & 0.0816$^\ast$\\
    & $P_{40}$ & 0.0019$^\ast$ & 0.0019$^\ast$\\
    \hline\hline
    \multicolumn{2}{c}{independent reflections} & \multicolumn{2}{c}{238}\\
    \multicolumn{2}{c}{data / parameters} & 13.22 & 9.92\\
    \multicolumn{2}{c}{goodness-of-fit on $F^2$} & 0.01 & 0.01\\
    \multicolumn{2}{c}{$R$ (all data)} & 0.03 & 0.02\\
    \multicolumn{2}{c}{$wR$ (all data)} & 0.03 & 0.03\\
    \multicolumn{2}{c}{largest diff. peak} & \multirow{2}{*}{0.02/-0.02} &
    \multirow{2}{*}{0.01/-0.01}\\[-0.2cm]
    \multicolumn{2}{c}{and hole [\eA]} &&\\
    \hline
    \caption{Model parameters refined using static theoretical
      structure factors up to a resolution of
      $(\sin\theta/\lambda)_{max} = 1.6$~\AA$^{-1}$. Parameter values
      marked by an asterisk ($\ast$) were not refined or subject to
      constraints. In T-EHCM1 scale factor, thermal parameters
      $U_{11}$ and $U_{33}$ for Mg and B, and the occupation factor for
      Mg were fixed. In T-EHCM2, these parameters were refined
      simultaneously with the multipolar parameters.}
  \label{tab:theo-ehc}
\end{longtable}

\newpage

\subsubsection{EHC models using experimental XRD data}

\begin{longtable*}[c]{>{\centering}p{0.15\textwidth}>{\centering}p{0.1\textwidth}
    >{\centering}p{0.19\textwidth}>{\centering\arraybackslash}p{0.19\textwidth}}
    \hline
    && \textbf{E-EHCM1} & \textbf{E-EHCM2}\\
    \hline\hline\endhead
    & $scale$ & 74.1(2) & 71.9(5)\\
    \hline
    Mg1 & $occ.$ & 1.0$^\ast$ & 1.015(3)\\
    {[~]($1s^2$)} & $x$ & \multicolumn{2}{c}{0$^\ast$}\\
    & $y$ & \multicolumn{2}{c}{0$^\ast$}\\
    & $z$ & \multicolumn{2}{c}{0.5$^\ast$}\\
    & $U_{11}$ [\AA$^2$] & 0.002950(14) & 0.002947(14)\\
    & $U_{33}$ [\AA$^2$] & 0.003143(17) & 0.003140(17)\\
    & $P_c$ & \multicolumn{2}{c}{0.0$^\ast$}\\
    & $P_v$ & \multicolumn{2}{c}{1.978130$^\ast$}\\
    & $\kappa$ & \multicolumn{2}{c}{1.014839$^\ast$}\\
    \hline
    Mg2 & $occ.$ & 1.0$^\ast$ & 1.015$^\ast$\\
    {[~]($2s^2$, $2p^6$)} & $x$ & \multicolumn{2}{c}{0$^\ast$}\\
    & $y$ & \multicolumn{2}{c}{0$^\ast$}\\
    & $z$ & \multicolumn{2}{c}{0.5$^\ast$}\\
    & $U_{11}$ [\AA$^2$] & 0.002950$^\ast$ & 0.002947$^\ast$\\
    & $U_{33}$ [\AA$^2$] & 0.003143$^\ast$ & 0.003140$^\ast$\\
    & $P_c$ & \multicolumn{2}{c}{0.0$^\ast$}\\
    & $P_v$ & \multicolumn{2}{c}{7.968863$^\ast$}\\
    & $\kappa$ & \multicolumn{2}{c}{1.010059$^\ast$}\\
    \hline
\end{longtable*}

\newpage

\begin{longtable*}[c]{>{\centering}p{0.15\textwidth}>{\centering}p{0.1\textwidth}
    >{\centering}p{0.19\textwidth}>{\centering\arraybackslash}p{0.19\textwidth}}
    \hline
    && \textbf{E-EHCM1} & \textbf{E-EHCM2}\\
    \hline\hline\endhead
    Mg3 & $occ.$ & 1.0$^\ast$ & 1.015$^\ast$\\
    {[~]($3s^2$)} & $x$ & \multicolumn{2}{c}{0$^\ast$}\\
    & $y$ & \multicolumn{2}{c}{0$^\ast$}\\
    & $z$ & \multicolumn{2}{c}{0.5$^\ast$}\\
    & $U_{11}$ [\AA$^2$] & 0.002950$^\ast$ & 0.002947$^\ast$\\
    & $U_{33}$ [\AA$^2$] & 0.003143$^\ast$ & 0.003140$^\ast$\\
    & $P_c$ & \multicolumn{2}{c}{0.0$^\ast$}\\
    & $P_v$ & 1.6(3) & 1.5(3)\\
    & $\kappa$ & 0.81(8) & 0.82(8)\\
    & $\kappa'$ & \multicolumn{2}{c}{1.362923$^\ast$}\\
    & $P_{20}$ & -0.04(5) & -0.04(5)\\
    & $P_{40}$ & -0.46(11) & -0.42(11)\\
    \hline
    B1 & $occ.$ & 1.0$^\ast$ & 1.0$^\ast$\\
    {[~]($1s^2$)} & $x$ & \multicolumn{2}{c}{$\frac{2}{3}^\ast$}\\
    & $y$ & \multicolumn{2}{c}{$\frac{1}{3}^\ast$}\\
    & $z$ & \multicolumn{2}{c}{0$^\ast$}\\
    & $U_{11}$ [\AA$^2$] & 0.00303(3) & 0.00298(3)\\
    & $U_{33}$ [\AA$^2$] & 0.00391(4) & 0.00387(4)\\
    & $P_c$ & \multicolumn{2}{c}{0.0$^\ast$}\\
    & $P_v$ & \multicolumn{2}{c}{1.977650$^\ast$}\\
    & $\kappa$ & \multicolumn{2}{c}{1.005537$^\ast$}\\
    \hline
\end{longtable*}

\newpage

\begin{longtable*}[c]{>{\centering}p{0.15\textwidth}>{\centering}p{0.1\textwidth}
    >{\centering}p{0.19\textwidth}>{\centering\arraybackslash}p{0.19\textwidth}}
    \hline
    && \textbf{E-EHCM1} & \textbf{E-EHCM2}\\
    \hline\hline\endhead
    B2 & $occ.$ & 1.0$^\ast$ & 1.0$^\ast$\\
    {[~]($2s^1$)} & $x$ & \multicolumn{2}{c}{$\frac{2}{3}^\ast$}\\
    & $y$ & \multicolumn{2}{c}{$\frac{1}{3}^\ast$}\\
    & $z$ & \multicolumn{2}{c}{0$^\ast$}\\
    & $U_{11}$ [\AA$^2$] & 0.00303$^\ast$ & 0.00298$^\ast$\\
    & $U_{33}$ [\AA$^2$] & 0.00391$^\ast$ & 0.00387$^\ast$\\
    & $P_c$ & \multicolumn{2}{c}{0.0$^\ast$}\\
    & $P_v$ & 1.60(12) & 2.330861$^\ast$\\
    & $\kappa$ & 0.90(2) & 0.852(16)\\
    & $\kappa'$ & 1.11(11) & 1.16(11)\\
    & $P_{20}$ & -0.063(18) & -0.058(16)\\
    & $P_{33-}$ & 0.051(16) & 0.046(15)\\
    & $P_{40}$ & -0.007(13) & -0.002(11)\\
    \hline
\end{longtable*}

\newpage

\begin{longtable}[c]{>{\centering}p{0.15\textwidth}>{\centering}p{0.1\textwidth}
    >{\centering}p{0.19\textwidth}>{\centering\arraybackslash}p{0.19\textwidth}}
    \hline
    && \textbf{E-EHCM1} & \textbf{E-EHCM2}\\
    \hline\hline\endhead
    B3 & $occ.$ & 1.0$^\ast$ & 1.0$^\ast$\\
    {[~]($2p^2$)} & $x$ & \multicolumn{2}{c}{$\frac{2}{3}^\ast$}\\
    & $y$ & \multicolumn{2}{c}{$\frac{1}{3}^\ast$}\\
    & $z$ & \multicolumn{2}{c}{0$^\ast$}\\
    & $U_{11}$ [\AA$^2$] & 0.00303$^\ast$ & 0.00298$^\ast$\\
    & $U_{33}$ [\AA$^2$] & 0.00391$^\ast$ & 0.00387$^\ast$\\
    & $P_c$ & \multicolumn{2}{c}{0.0$^\ast$}\\
    & $P_v$ & 1.665498$^\ast$ & 0.97(17)\\
    & $\kappa$ & 0.98(5) & 1.24(9)\\
    & $\kappa'$ & 1.11$^\ast$ & 1.16$^\ast$\\
    & $P_{20}$ & -0.063$^\ast$ & -0.058$^\ast$\\
    & $P_{33-}$ & 0.051$^\ast$ & 0.046$^\ast$\\
    & $P_{40}$ & -0.007$^\ast$ & -0.002$^\ast$\\
    \hline\hline
    \multicolumn{2}{c}{independent reflections} & \multicolumn{2}{c}{237}\\
    \multicolumn{2}{c}{data / parameters} & 14.81 & 13.94\\
    \multicolumn{2}{c}{goodness-of-fit on $F^2$} & 0.92 & 0.91\\
    \multicolumn{2}{c}{$R$ (all data)} & 0.69 & 0.69\\
    \multicolumn{2}{c}{$wR$ (all data)} & 1.96 & 1.95\\
    \multicolumn{2}{c}{largest diff. peak} & \multirow{2}{*}{0.2/-0.22} &
    \multirow{2}{*}{0.2/-0.22}\\[-0.2cm]
    \multicolumn{2}{c}{and hole [\eA]} &&\\
    \hline
    \caption{Model parameters refined using experimental x-ray
      diffraction data (resolution
      $(\sin\theta/\lambda)_{max} = 1.6$~\AA$^{-1}$; experiment~1 in
      Tab.~\ref{tab:single-crystal-xrd-experiments}; data set
      parameters in Tab.~\ref{tab:crystal-1-100k-dataset}) collected
      at $T =$~100(2)~K. Parameter values marked by an asterisk
      ($\ast$) were not refined or subject to constraints. In E-EHCM1
      the occupation factor for Mg was fixed, while it was refined in
      E-EHCM2.}
  \label{tab:exp-ehc}
\end{longtable}

\newpage

\subsubsection{SHC models using temperature-dependent experimental XRD
  data}

\enlargethispage{1em}

\begin{longtable*}[c]{>{\centering}p{0.2\textwidth}>{\centering}p{0.1\textwidth}
    >{\centering}p{0.19\textwidth}
    >{\centering\arraybackslash}p{0.19\textwidth}}
    \hline
    && \textbf{298.3~K} & \textbf{45.009~K}\\
    \hline\hline\endhead
    & $scale$ & 8.72(4) & 8.56(4)\\
    \hline
    Mg & $occ.$ & \multicolumn{2}{c}{1.0$^\ast$}\\
    {[$1s^2$]($2s^2$, $2p^6$)} & $x$ & \multicolumn{2}{c}{0$^\ast$}\\
    & $y$ & \multicolumn{2}{c}{0$^\ast$}\\
    & $z$ & \multicolumn{2}{c}{0.5$^\ast$}\\
    & $U_{11}$ [\AA$^2$] & 0.00543(6) & 0.00299(5)\\
    & $U_{33}$ [\AA$^2$] & 0.00585(7) & 0.00309(5)\\
    & $P_c$ & \multicolumn{2}{c}{2.0$^\ast$}\\
    & $P_v$ & 8.028831$^\ast$ & 8.054975$^\ast$\\
    & $\kappa$ & 1.004(4) & 1.011(2)\\
    & $\kappa'$ & \multicolumn{2}{c}{1.0$^\ast$}\\
    & $P_{20}$ & 0.023(14) & 0.020(9)\\
    \hline
\end{longtable*}

\newpage

\begin{longtable}[c]{>{\centering}p{0.2\textwidth}>{\centering}p{0.1\textwidth}
    >{\centering}p{0.19\textwidth}
    >{\centering\arraybackslash}p{0.19\textwidth}}
    \hline
    && \textbf{298.3~K} & \textbf{45.009~K}\\
    \hline\hline\endhead
    B & $occ.$ & \multicolumn{2}{c}{1.0$^\ast$}\\
    {[$1s^2$]($2s^1$, $2p^2$)} & $x$ & \multicolumn{2}{c}{$\frac{2}{3}^\ast$}\\
    & $y$ & \multicolumn{2}{c}{$\frac{1}{3}^\ast$}\\
    & $z$ & \multicolumn{2}{c}{0$^\ast$}\\
    & $U_{11}$ [\AA$^2$] & 0.00422(8) & 0.00303(6)\\
    & $U_{33}$ [\AA$^2$] & 0.00577(9) & 0.00372(7)\\
    & $P_c$ & \multicolumn{2}{c}{2.0$^\ast$}\\
    & $P_v$ & 3.99(3) & 3.97(3)\\
    & $\kappa$ & 0.82(3) & 0.83(3)\\
    & $\kappa'$ & 0.76(9) & 0.74(8)\\
    & $P_{20}$ & -0.22(9) & -0.21(9)\\
    & $P_{33-}$ & 0.39(12) & 0.41(11)\\
    \hline\hline
    \multicolumn{2}{c}{independent reflections} & 148 & 148\\
    \multicolumn{2}{c}{data / parameters} & 12.33 & 12.33\\
    \multicolumn{2}{c}{goodness-of-fit on $F^2$} & 1.74 & 1.18\\
    \multicolumn{2}{c}{$R$ (all data)} & 0.95 & 0.65\\
    \multicolumn{2}{c}{$wR$ (all data)} & 3.78 & 2.41\\
    \multicolumn{2}{c}{largest diff. peak} & \multirow{2}{*}{0.44/-0.33} & 
                   \multirow{2}{*}{0.13/-0.23}\\[-0.2cm]
    \multicolumn{2}{c}{and hole [\eA]} &&\\
    \hline
    \caption{Model parameters refined using experimental XRD data
      ($(\sin\theta/\lambda)_{max} = 1.3$~\AA$^{-1}$; experiment~2 in
      Tab.~\ref{tab:single-crystal-xrd-experiments}; data set
      parameters in Tab.~\ref{tab:crystal-1-lt-datasets-1}) collected
      at temperatures of 298.3~K and 45.009~K. Parameter values marked
      by an asterisk ($\ast$) were not refined or subject to
      constraints.}
  \label{tab:lt-shc-1}
\end{longtable}

\newpage

\begin{longtable*}[c]{>{\centering}p{0.2\textwidth}>{\centering}p{0.1\textwidth}
    >{\centering}p{0.19\textwidth}
    >{\centering\arraybackslash}p{0.19\textwidth}}
    \hline
    && \textbf{25.086~K} & \textbf{13.5~K}\\
    \hline\hline\endhead
    & $scale$ & 26.90(9) & 27.07(12)\\
    \hline
    Mg & $occ.$ & \multicolumn{2}{c}{1.0$^\ast$}\\
    {[$1s^2$]($2s^2$, $2p^6$)} & $x$ & \multicolumn{2}{c}{0$^\ast$}\\
    & $y$ & \multicolumn{2}{c}{0$^\ast$}\\
    & $z$ & \multicolumn{2}{c}{0.5$^\ast$}\\
    & $U_{11}$ [\AA$^2$] & 0.00292(4) & 0.00301(5)\\
    & $U_{33}$ [\AA$^2$] & 0.00305(4) & 0.00313(6)\\
    & $P_c$ & \multicolumn{2}{c}{2.0$^\ast$}\\
    & $P_v$ & 8.098066$^\ast$ & 8.029143$^\ast$\\
    & $\kappa$ & 1.008(2) & 1.010(3)\\
    & $\kappa'$ & \multicolumn{2}{c}{1.0$^\ast$}\\
    & $P_{20}$ & 0.009(9) & 0.00(1)\\
    \hline
\end{longtable*}

\newpage

\begin{longtable}[c]{>{\centering}p{0.2\textwidth}>{\centering}p{0.1\textwidth}
    >{\centering}p{0.19\textwidth}
    >{\centering\arraybackslash}p{0.19\textwidth}}
    \hline
    && \textbf{25.086~K} & \textbf{13.5~K}\\
    \hline\hline\endhead
    B & $occ.$ & \multicolumn{2}{c}{1.0$^\ast$}\\
    {[$1s^2$]($2s^1$, $2p^2$)} & $x$ & \multicolumn{2}{c}{$\frac{2}{3}^\ast$}\\
    & $y$ & \multicolumn{2}{c}{$\frac{1}{3}^\ast$}\\
    & $z$ & \multicolumn{2}{c}{0$^\ast$}\\
    & $U_{11}$ [\AA$^2$] & 0.00300(4) & 0.00304(6)\\
    & $U_{33}$ [\AA$^2$] & 0.00362(5) & 0.00375(7)\\
    & $P_c$ & \multicolumn{2}{c}{2.0$^\ast$}\\
    & $P_v$ & 3.95(2) & 3.99(3)\\
    & $\kappa$ & 0.83(2) & 0.83(3)\\
    & $\kappa'$ & 0.76(6) & 0.80(9)\\
    & $P_{20}$ & -0.23(6) & -0.20(9)\\
    & $P_{33-}$ & 0.41(8) & 0.37(10)\\
    \hline\hline
    \multicolumn{2}{c}{independent reflections} & 148 & 148\\
    \multicolumn{2}{c}{data / parameters} & 12.33 & 12.33\\
    \multicolumn{2}{c}{goodness-of-fit on $F^2$} & 1.15 & 1.26\\
    \multicolumn{2}{c}{$R$ (all data)} & 0.57 & 0.69\\
    \multicolumn{2}{c}{$wR$ (all data)} & 2.32 & 2.56\\
    \multicolumn{2}{c}{largest diff. peak} & \multirow{2}{*}{0.15/-0.18} &
                   \multirow{2}{*}{0.10/-0.22}\\[-0.2cm]
    \multicolumn{2}{c}{and hole [\eA]} &&\\
    \hline
    \caption{Model parameters refined using experimental x-ray
      diffraction data ($(\sin\theta/\lambda)_{max} = 1.3$~\AA$^{-1}$;
      experiment~2 in Tab.~\ref{tab:single-crystal-xrd-experiments};
      data set parameters in Tab.~\ref{tab:crystal-1-lt-datasets-2})
      collected at temperatures of 25.086~K and 13.5~K. Parameter
      values marked by an asterisk ($\ast$) were not refined or
      subject to constraints.}
  \label{tab:lt-shc-2}
\end{longtable}

\FloatBarrier
\newpage

\subsection{Residual electron density maps}

\subsubsection{Resolution dependence}

\begin{figure}
  \centering
  \includegraphics[width=1.0\textwidth]{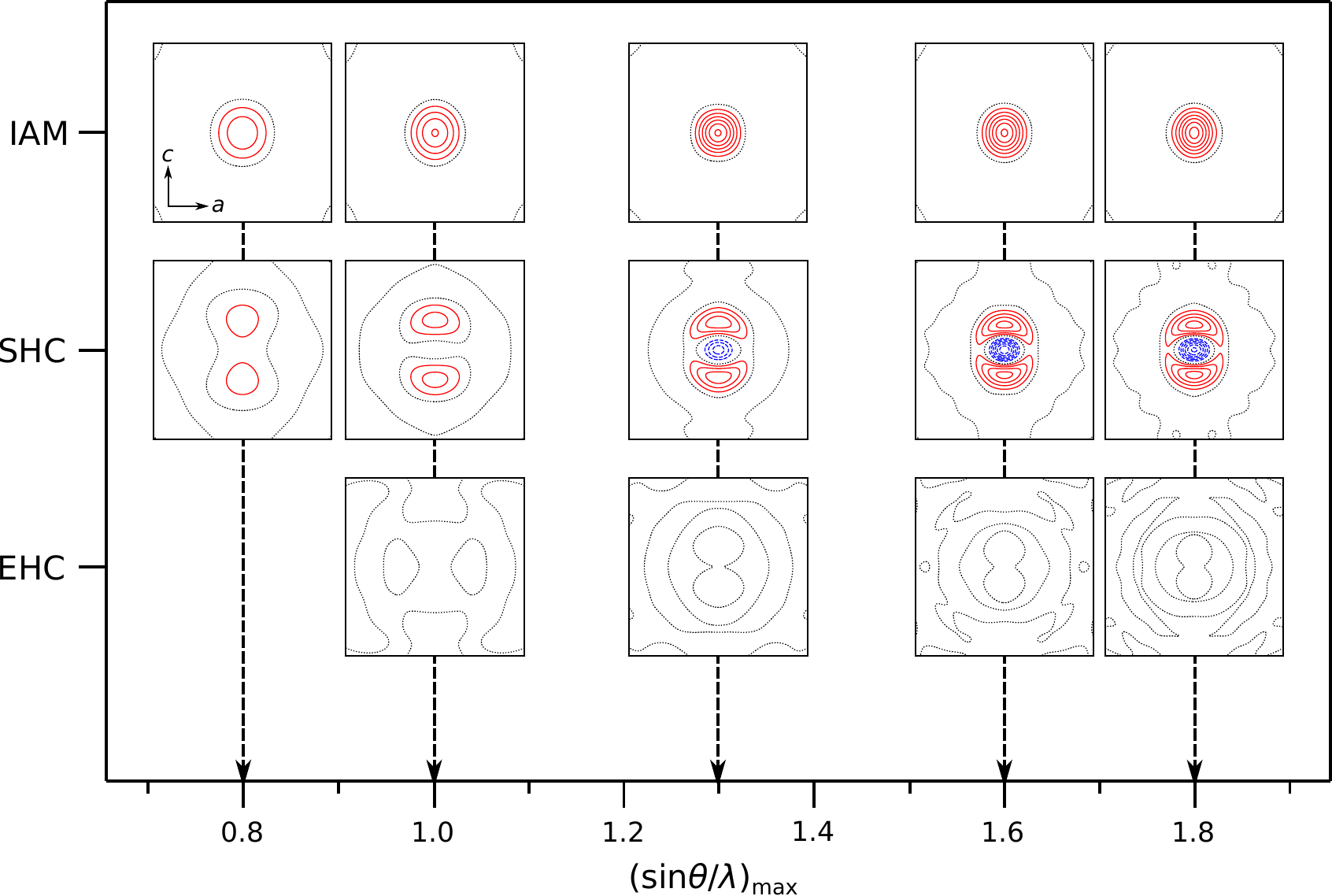}  
  \caption{Variation of the residual electron density centered at 
    the magnesium position in the $a$-$c$ plane with increasing 
    resolution \sthlmax\ of static theoretical structure factors
    for different scenarios: using a)~an IAM based on
    form factors of neutral atoms, b)~an SHC model in analogy to
    Tab.~\ref{tab:lt-shc-1} and Tab.~\ref{tab:lt-shc-2}, or c)~and EHC
    model. In case of b) and c) core and/or valence
    multipolar parameters were refined at each \sthlmax. Iso-contour
    lines are drawn at $\pm$0.05~\eA\ using solid red lines for
    positive, dotted black lines for zero and dashed blue lines for
    negative values of the residual electron density, respectively.}
  \label{fig:resolution-dep-resden-Mg-pos}
\end{figure}

\FloatBarrier
\newpage

\subsubsection{EHC models using static theoretical structure factors}

\begin{figure}
  \centering
  \includegraphics[width=0.84\textwidth]{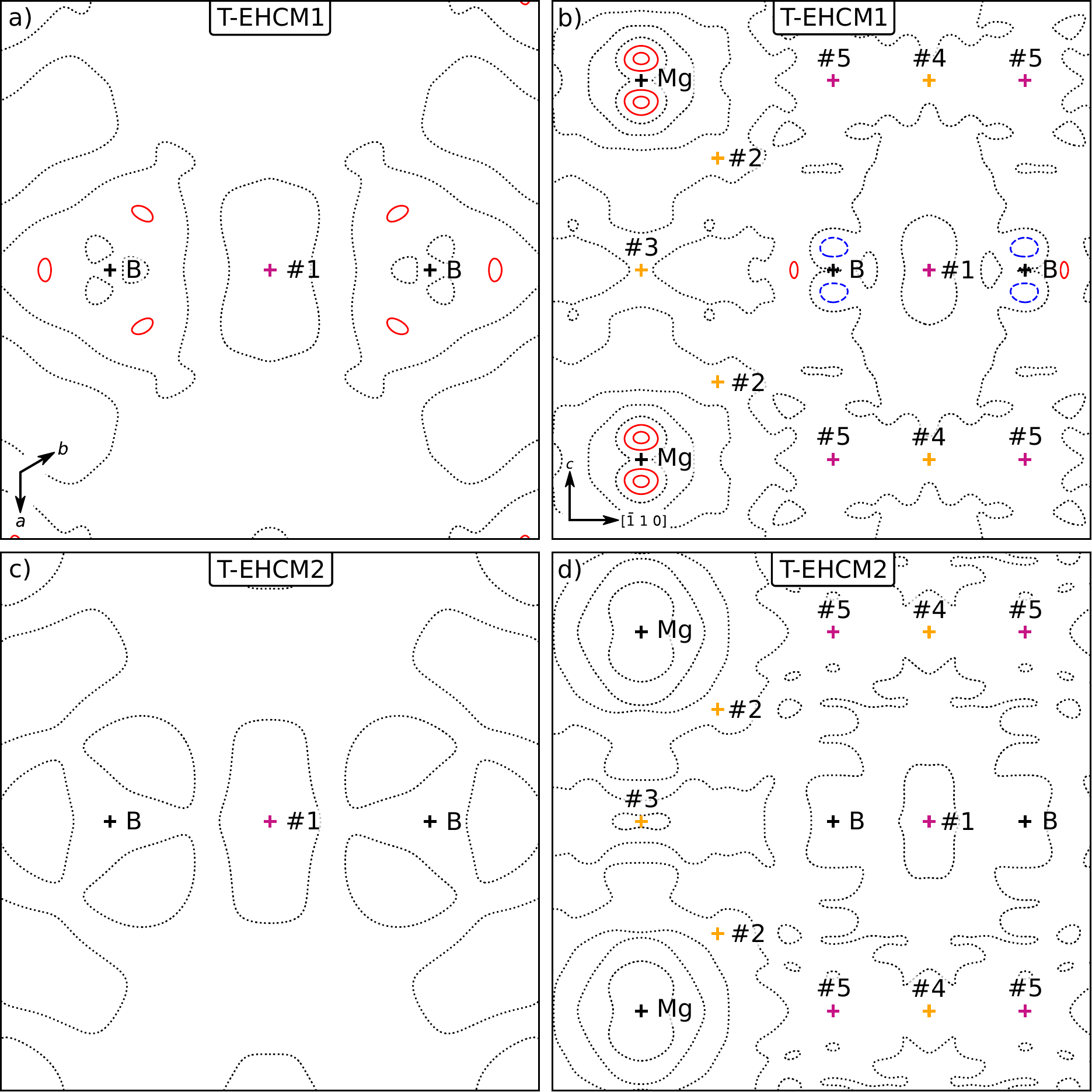}  
  \caption{Residual electron density maps for the Extendend
    Hansen-Coppens (EHC) models T-EHCM1 (a,b) and T-EHCM2 (c,d;
    Tab.~\ref{tab:theo-ehc}) refined using static theoretical
    structure factors.  Maps are oriented parallel (a,c) and
    perpendicular (b,d) to the hexagonal boron layers in
    \Mg. Iso-contour lines are drawn at $\pm$\textbf{0.01}~\eA\ using solid
    red lines for positive, dotted black lines for zero and dashed
    blue lines for negative values of the residual electron density,
    respectively.}
  \label{fig:resden-Mg-B-plane-theo-EHCM1}
\end{figure}

\FloatBarrier
\newpage

\subsubsection{EHC models using experimental XRD data}

\begin{figure}[htb]
  \centering
  \includegraphics[width=0.84\textwidth]{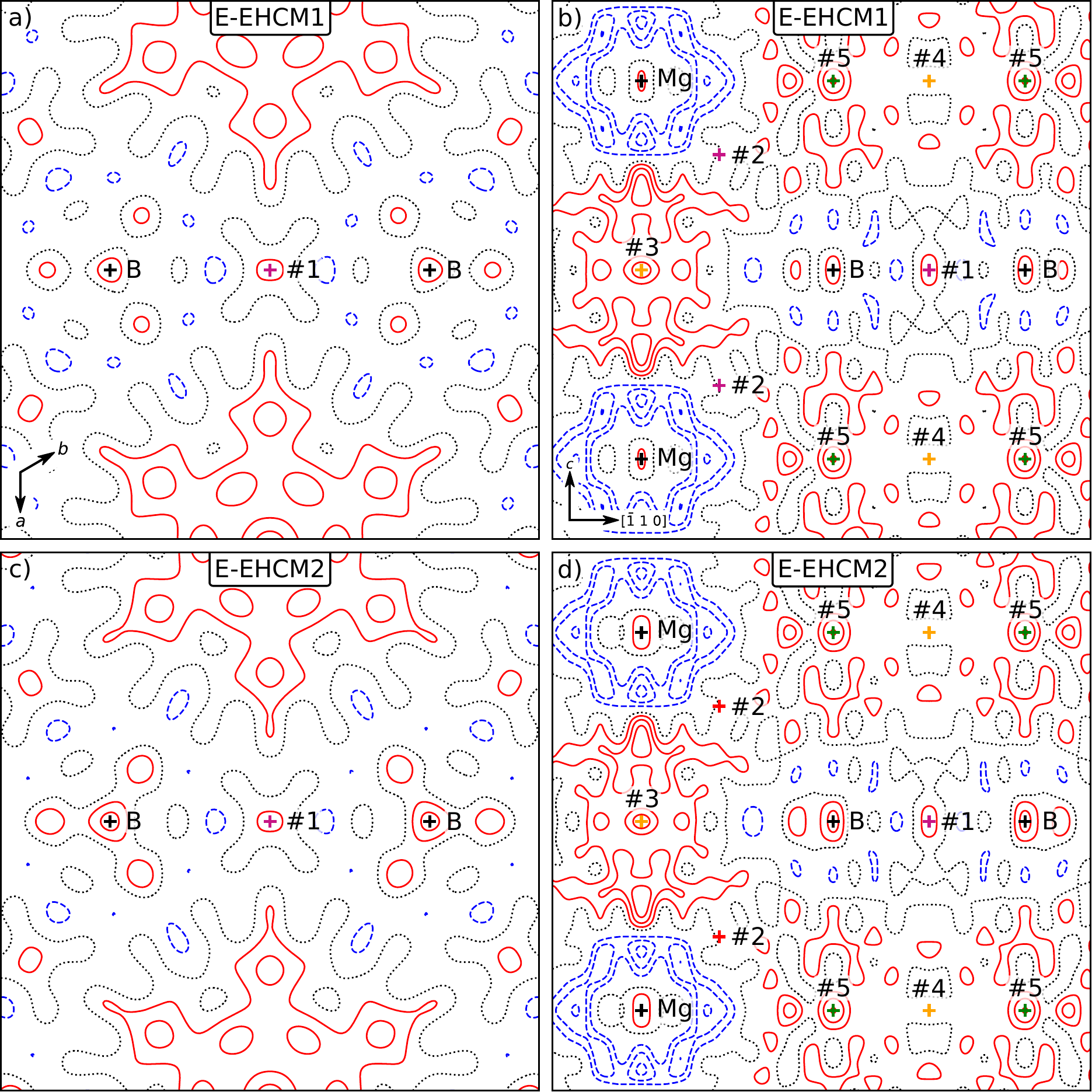}  
  \caption{Residual electron density maps for the Extendend
    Hansen-Coppens (EHC) models E-EHCM1 (a,b) and E-EHCM2 (c,d)
    (Tab.~\ref{tab:exp-ehc}) refined using experimental XRD data
    ($(\sin\theta/\lambda)_{max} = 1.6$~\AA$^{-1}$; experiment~1 in
    Tab.~\ref{tab:single-crystal-xrd-experiments}; data set parameters
    in Tab.~\ref{tab:crystal-1-100k-dataset}) collected at
    $T =$~100(2)~K. Maps are oriented parallel (a,c) and perpendicular
    (b,d) to the hexagonal boron layers in \Mg. Iso-contour lines are
    drawn at $\pm$0.05~\eA\ using solid red lines for positive,
    dotted black lines for zero and dashed blue lines for negative
    values of the residual electron density, respectively.}
  \label{fig:resden-Mg-B-plane-exp-EHCM1}
\end{figure}

\FloatBarrier
\hspace{1em}
\newpage

\subsubsection{SHC models using temperature-dependent experimental XRD
  data}

\begin{figure}
  \centering
  \includegraphics[width=0.84\textwidth]{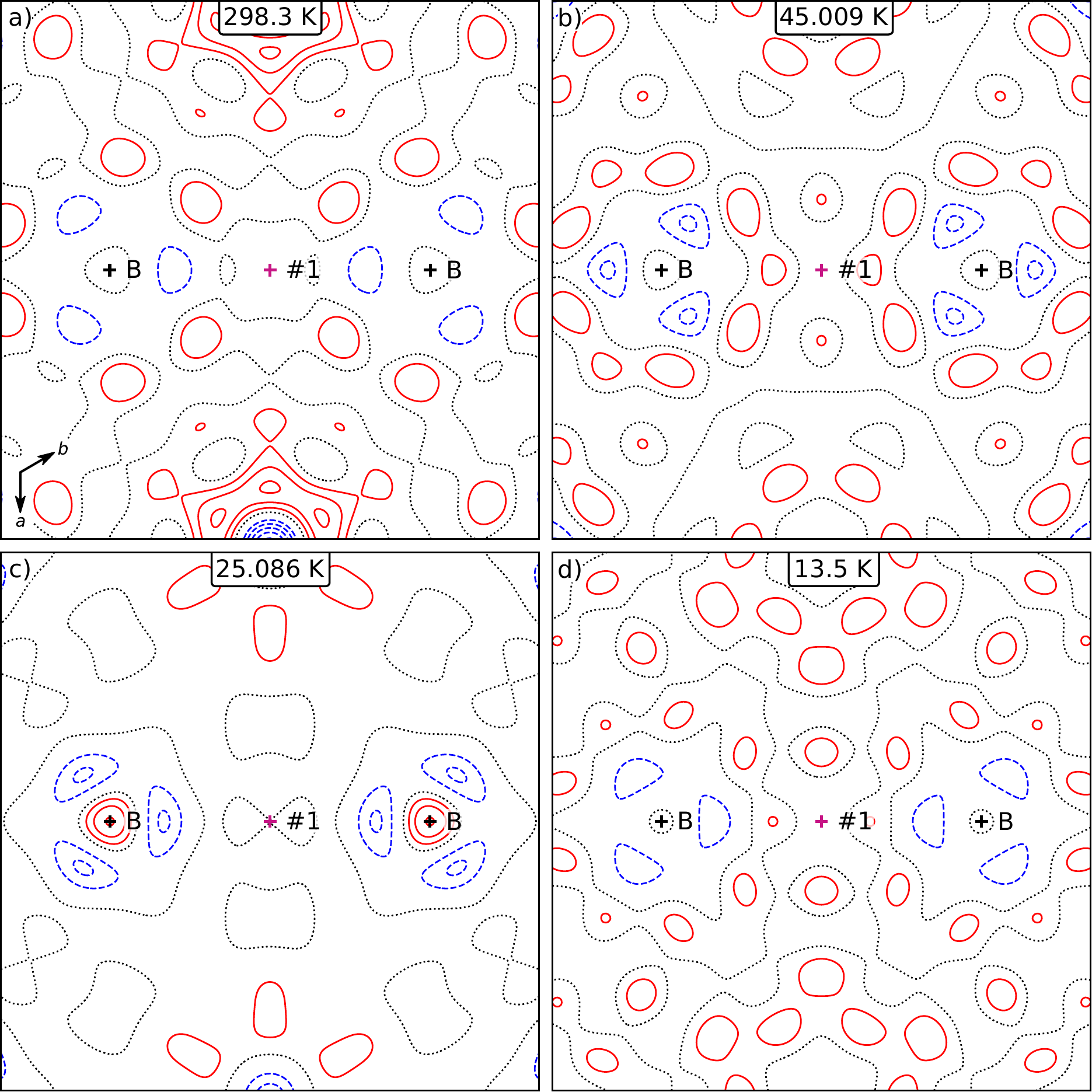}  
  \caption{Residual electron density maps for SHC multipolar models
    refined using experimental XRD data collected at temperatures
    between a)~298.3(1)~K and d)~13.5(5)~K (Tab.~\ref{tab:lt-shc-1}
    and Tab.~\ref{tab:lt-shc-2}).  Maps are oriented parallel to the
    hexagonal boron layers in \Mg.  Iso-contour lines are drawn at
    $\pm$0.05~\eA\ using solid red lines for positive, dotted black
    lines for zero and dashed blue lines for negative values of the
    residual electron density, respectively.}
  \label{fig:resden-B-plane-exp-SHCM}
\end{figure}

\begin{figure}
  \centering
  \includegraphics[width=0.84\textwidth]{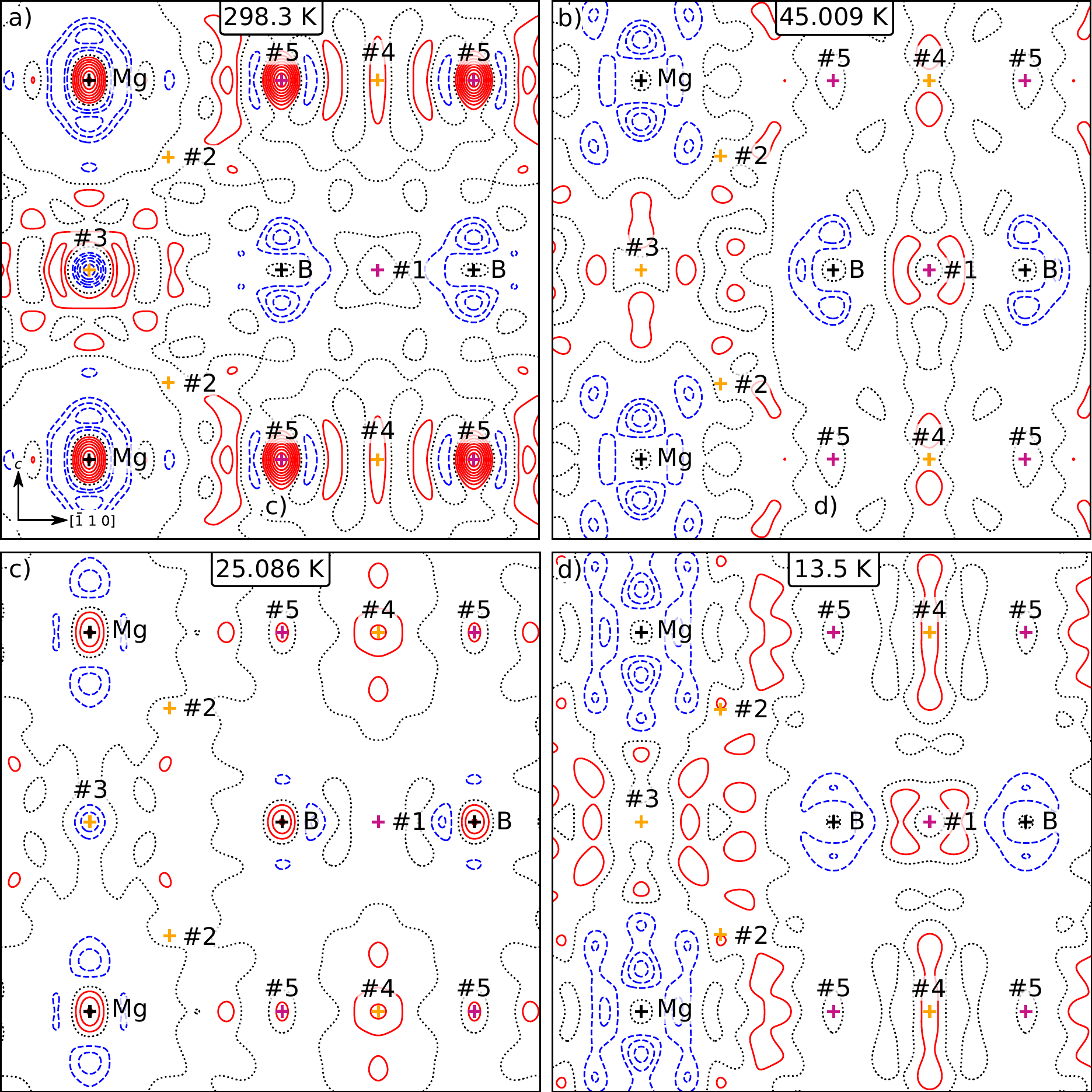}  
  \caption{Residual electron density maps for SHC multipolar models
    refined using experimental XRD data collected at temperatures
    between a)~298.3(1)~K and d)~13.5(5)~K (Tab.~\ref{tab:lt-shc-1}
    and Tab.~\ref{tab:lt-shc-2}).  Maps are oriented perpendicular to
    the hexagonal boron layers in \Mg.  Iso-contour lines are drawn at
    $\pm$0.05~\eA\ using solid red lines for positive, dotted black
    lines for zero and dashed blue lines for negative values of the
    residual electron density, respectively.}
  \label{fig:resden-Mg-B-plane-exp-SHCM}
\end{figure}

\begin{figure}
  \centering
  \includegraphics[width=0.9\textwidth]{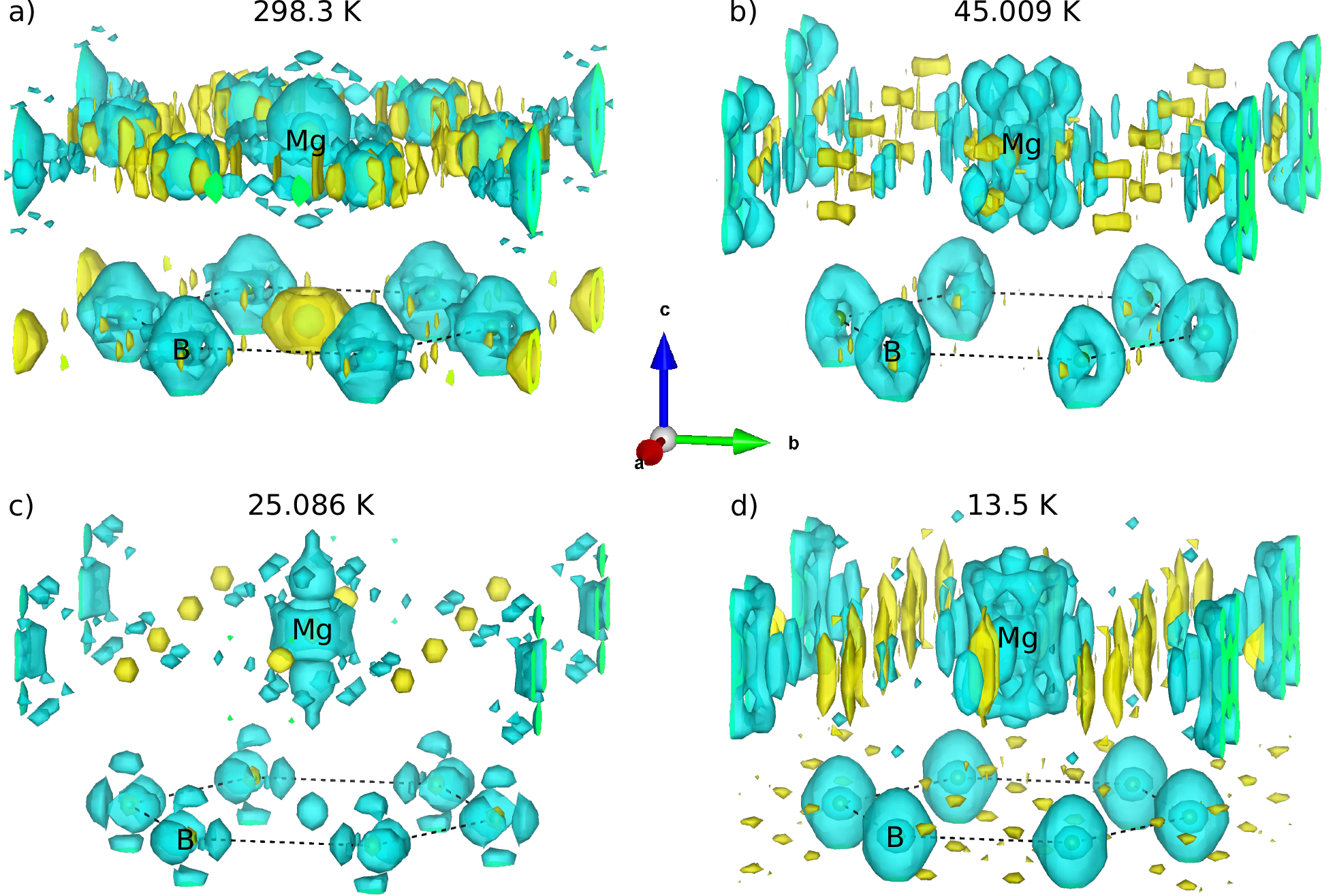} 
  \caption{Iso-surface plots of the residual electron density for SHC
    multipolar models refined using experimental XRD data collected at
    temperatures between a)~298.3(1)~K and d)~13.5(5)~K
    (Tab.~\ref{tab:lt-shc-1} and
    Tab.~\ref{tab:lt-shc-2}). Iso-surfaces are plotted at
    $\pm$0.08~\eA\ using yellow and turquoise colors for positive and
    negative values of the residual electron density, respectively.}
  \label{fig:resden-3d-exp-SHCM}
\end{figure}

\FloatBarrier

\section{QTAIM analyses}
\label{sec:qtaim}

\subsubsection{Topological characteristics}
\label{sec:topol-characteristics}

\begin{longtable*}[c]{>{\centering}p{0.04\textwidth}>{\centering}p{0.08\textwidth}
    >{\centering}p{0.1\textwidth}>{\centering}p{0.08\textwidth}
    >{\centering}p{0.08\textwidth}>{\centering}p{0.08\textwidth}
    >{\centering}p{0.08\textwidth}>{\centering}p{0.08\textwidth}
    >{\centering\arraybackslash}p{0.08\textwidth}}
  \hline &&&& \textbf{pos.} &&&& $H$ \\[-0.2cm]
  \textbf{CP} & \multirow{2}{*}{\textbf{model}} & $T$ & \multirow{2}{*}{\textbf{rank}}
  & $x$ & \rhoc & \Lc & \multirow{2}{*}{$\epsilon$} & $\lambda_1$\\[-0.2cm]
  \textbf{\#} && [K] && $y$ & [\eA] & [\eAA] && $\lambda_2$\\[-0.2cm]
  &&&& $z$ &&&& $\lambda_3$\\
  \hline\hline\endhead
  \multirow{18}{*}{1} &&&& 0.5 &&&& -3.02\\[-0.3cm]
  & EHC\textsuperscript{a} & 100(2) & $(3, -1)$ & 0.5 & 0.80 & -3.30 & 0.2 & -2.43\\[-0.3cm]
  &&&& 0 &&&& 2.15\\
  &&&& 0.5 &&&& -3.10\\[-0.3cm]
  & SHC & 298.3(1) & $(3, -1)$ & 0.5 & 0.85 & -5.41 & 0.3 & -2.47\\[-0.3cm]
  &&&& 0 &&&& 0.16\\
  &&&& 0.5 &&&& -3.05\\[-0.3cm]
  & SHC & 45.009(7) & $(3, -1)$ & 0.5 & 0.84 & -5.23 & 0.3 & -2.41\\[-0.3cm]
  &&&& 0 &&&& 0.23\\
  &&&& 0.5 &&&& -3.25\\[-0.3cm]
  & SHC & 25.086(5) & $(3, -1)$ & 0.5 & 0.86 & -5.56 & 0.3 & -2.57\\[-0.3cm]
  &&&& 0 &&&& 0.26\\
  &&&& 0.5 &&&& -3.40\\[-0.3cm]
  & SHC & 13.5(5) & $(3, -1)$ & 0.5 & 0.87 & -5.79 & 0.3 & -2.64\\[-0.3cm]
  &&&& 0 &&&& 0.25\\
  &&&& 0.5 &&&& -3.52\\[-0.3cm]
  & Ref.~\citenum{Tsirelson03} & RT & $(3, -1)$ & 0.5 & 0.81 & -4.63 & 0.3 & -2.70\\[-0.3cm]
  &&&& 0 &&&& 1.59\\
  &&&& 0.5 &&&&-3.30\\[-0.3cm]
  & DFT\textsuperscript{b} & -- & $(3, -1)$ & 0.5 & 0.82 & -4.93 & 0.2 & -2.87\\[-0.3cm]
  &&&& 0 &&&&1.25\\
  &&&& 0.5 &&&& -3.20\\[-0.3cm]
  & EHC\textsuperscript{c} & -- & $(3, -1)$ & 0.5 & 0.81 & -3.79 & 0.2 & -2.75\\[-0.3cm]
  &&&& 0 &&&& 2.16\\
  \hline
\end{longtable*}

\newpage

\begin{longtable*}[c]{>{\centering}p{0.04\textwidth}>{\centering}p{0.08\textwidth}
    >{\centering}p{0.1\textwidth}>{\centering}p{0.08\textwidth}
    >{\centering}p{0.08\textwidth}>{\centering}p{0.08\textwidth}
    >{\centering}p{0.08\textwidth}>{\centering}p{0.08\textwidth}
    >{\centering\arraybackslash}p{0.08\textwidth}}
  \hline &&&& \textbf{pos.} &&&& $H$ \\[-0.2cm]
  \textbf{CP} & \multirow{2}{*}{\textbf{model}} & $T$ & \multirow{2}{*}{\textbf{rank}}
  & $x$ & \rhoc & \Lc & \multirow{2}{*}{$\epsilon$} & $\lambda_1$\\[-0.2cm]
  \textbf{\#} && [K] && $y$ & [\eA] & [\eAA] && $\lambda_2$\\[-0.2cm]
  &&&& $z$ &&&& $\lambda_3$\\
  \hline\hline\endhead
  \multirow{18}{*}{2} &&&& 0.865 &&&& -0.74\\[-0.3cm]
  & EHC\textsuperscript{a} & 100(2) &
  $(3, -1)$ & 0.135 & 0.19 & 1.94 & 168 & -0.01\\[-0.3cm]
  &&&& 0.305 &&&& 2.68\\
  &&&& 0.863 &&&& -0.40\\[-0.3cm]
  & SHC & 298.3(1) & $(3, +1)$ & 0.137 & 0.15 & 2.21 & -- & 0.03\\[-0.3cm]
  &&&& 0.297 &&&& 2.58\\
  &&&& 0.862 &&&& -0.41\\[-0.3cm]
  & SHC & 45.009(7) & $(3, +1)$ & 0.138 & 0.15 & 2.27 & -- & 0.02\\[-0.3cm]
  &&&& 0.301 &&&& 2.66\\
  &&&& 0.862 &&&& -0.41\\[-0.3cm]
  & SHC & 25.086(5) & $(3, +1)$ & 0.139 & 0.14 & 2.24 & -- & 0.03\\[-0.3cm]
  &&&& 0.300 &&&& 2.62\\
  &&&& 0.863 &&&& -0.39\\[-0.3cm]
  & SHC & 13.5(5) & $(3, +1)$ & 0.137 & 0.15 & 2.17 & -- & 0.03\\[-0.3cm]
  &&&& 0.297 &&&& 2.53\\
  &&&& 0.894 &&&& -0.27\\[-0.3cm]
  & Ref.~\citenum{Tsirelson03} & RT & $(3, -1)$ & 0.106 &
  0.19 & 1.47 & 26 & -0.01\\[-0.3cm]
  &&&& 0.253 &&&& 1.75\\
  &&&& 0.868 &&&&-0.39\\[-0.3cm]
  & DFT\textsuperscript{b} & -- & $(3, +1)$ & 0.132 & 0.17 & 1.97 & -- & 0.00\\[-0.3cm]
  &&&& 0.294 &&&& 2.36\\
  &&&& 0.867 &&&& -0.41\\[-0.3cm]
  & EHC\textsuperscript{c} & -- & $(3, +1)$ & 0.133 & 0.17 & 1.90 & -- & 0.00\\[-0.3cm]
  &&&& 0.295 &&&& 2.31\\
  \hline
\end{longtable*}

\newpage

\begin{longtable*}[c]{>{\centering}p{0.04\textwidth}>{\centering}p{0.08\textwidth}
    >{\centering}p{0.1\textwidth}>{\centering}p{0.08\textwidth}
    >{\centering}p{0.08\textwidth}>{\centering}p{0.08\textwidth}
    >{\centering}p{0.08\textwidth}>{\centering}p{0.08\textwidth}
    >{\centering\arraybackslash}p{0.08\textwidth}}
  \hline &&&& \textbf{pos.} &&&& $H$ \\[-0.2cm]
  \textbf{CP} & \multirow{2}{*}{\textbf{model}} & $T$ & \multirow{2}{*}{\textbf{rank}}
  & $x$ & \rhoc & \Lc & \multirow{2}{*}{$\epsilon$} & $\lambda_1$\\[-0.2cm]
  \textbf{\#} && [K] && $y$ & [\eA] & [\eAA] && $\lambda_2$\\[-0.2cm]
  &&&& $z$ &&&& $\lambda_3$\\
  \hline\hline\endhead
  \multirow{18}{*}{3} &&&& 0 &&&& -0.02\\[-0.3cm]
  & EHC\textsuperscript{a} & 100(2) &
  $(3, +1)$ & 0 & 0.06 & 1.80 & -- & 0.91\\[-0.3cm]
  &&&& 0 &&&& 0.91\\
  &&&& 0 &&&& -0.07\\[-0.3cm]
  & SHC & 298.3(1) & $(3, +1)$ & 0 & 0.09 & 1.31 & -- & 0.69\\[-0.3cm]
  &&&& 0 &&&& 0.69\\
  &&&& 0 &&&& -0.06\\[-0.3cm]
  & SHC & 45.009(7) & $(3, +1)$ & 0 & 0.08 & 1.39 & -- & 0.72\\[-0.3cm]
  &&&& 0 &&&& 0.72\\
  &&&& 0 &&&& -0.05\\[-0.3cm]
  & SHC & 25.086(5) & $(3, +1)$ & 0 & 0.08 & 1.36 & -- & 0.70\\[-0.3cm]
  &&&& 0 &&&& 0.70\\
  &&&& 0 &&&& -0.05\\[-0.3cm]
  & SHC & 13.5(5) & $(3, +1)$ & 0 & 0.09 & 1.22 & -- & 0.64\\[-0.3cm]
  &&&& 0 &&&& 0.64\\
  &&&& 0 &&&& -0.05\\[-0.3cm]
  & Ref.~\citenum{Tsirelson03} & RT & $(3, +1)$ & 0 & 0.17 & 0.77 & -- & 0.41\\[-0.3cm]
  &&&& 0 &&&& 0.41\\
  &&&& 0 &&&&-0.01\\[-0.3cm]
  & DFT\textsuperscript{b} & -- & $(3, +1)$ & 0 & 0.10 & 1.13 & -- &0.57\\[-0.3cm]
  &&&& 0 &&&&0.57\\
  &&&& 0 &&&& -0.04\\[-0.3cm]
  & EHC\textsuperscript{c} & -- & $(3, +1)$ & 0 & 0.10 & 1.16 & -- & 0.60\\[-0.3cm]
  &&&& 0 &&&& 0.60\\
  \hline
\end{longtable*}

\newpage

\begin{longtable*}[c]{>{\centering}p{0.04\textwidth}>{\centering}p{0.08\textwidth}
    >{\centering}p{0.1\textwidth}>{\centering}p{0.08\textwidth}
    >{\centering}p{0.08\textwidth}>{\centering}p{0.08\textwidth}
    >{\centering}p{0.08\textwidth}>{\centering}p{0.08\textwidth}
    >{\centering\arraybackslash}p{0.08\textwidth}}
  \hline &&&& \textbf{pos.} &&&& $H$ \\[-0.2cm]
  \textbf{CP} & \multirow{2}{*}{\textbf{model}} & $T$ & \multirow{2}{*}{\textbf{rank}}
  & $x$ & \rhoc & \Lc & \multirow{2}{*}{$\epsilon$} & $\lambda_1$\\[-0.2cm]
  \textbf{\#} && [K] && $y$ & [\eA] & [\eAA] && $\lambda_2$\\[-0.2cm]
  &&&& $z$ &&&& $\lambda_3$\\
  \hline\hline\endhead
  \multirow{18}{*}{4} &&&& 0.5 &&&& -0.02\\[-0.3cm]
  & EHC\textsuperscript{a} & 100(2) &
  $(3, +1)$ & 0.5 & 0.07 & 0.90 & -- & 0.14\\[-0.3cm]
  &&&& 0.5 &&&& 0.78\\
  &&&& 0.5 &&&& -0.09\\[-0.3cm]
  & SHC & 298.3(1) & $(3, +1)$ & 0.5 & 0.10 & 0.40 & -- & 0.01\\[-0.3cm]
  &&&& 0.5 &&&& 0.47\\
  &&&& 0.5 &&&& -0.10\\[-0.3cm]
  & SHC & 45.009(7) & $(3, +1)$ & 0.5 & 0.10 & 0.40 & -- & 0.02\\[-0.3cm]
  &&&& 0.5 &&&& 0.49\\
  &&&& 0.5 &&&& -0.09\\[-0.3cm]
  & SHC & 25.086(5) & $(3, +1)$ & 0.5 & 0.10 & 0.40 & -- & 0.01\\[-0.3cm]
  &&&& 0.5 &&&& 0.48\\
  &&&& 0.5 &&&& -0.09\\[-0.3cm]
  & SHC & 13.5(5) & $(3, +1)$ & 0.5 & 0.10 & 0.40 & -- & 0.01\\[-0.3cm]
  &&&& 0.5 &&&& 0.47\\
  &&&& 0.5 &&&& -0.04\\[-0.3cm]
  & Ref.~\citenum{Tsirelson03} & RT & $(3, +1)$ & 0.5 & 0.10 & 0.39 & -- & 0.03\\[-0.3cm]
  &&&& 0.5 &&&& 0.40\\
  &&&& 0.5 &&&&-0.05\\[-0.3cm]
  & DFT\textsuperscript{b} & -- & $(3, +1)$ & 0.5 & 0.11 & 0.38 & -- &0.01\\[-0.3cm]
  &&&& 0.5 &&&&0.42\\
  &&&& 0.5 &&&& -0.03\\[-0.3cm]
  & EHC\textsuperscript{c} & -- & $(3, +1)$ & 0.5 & 0.10 & 0.41 & -- & 0.00\\[-0.3cm]
  &&&& 0.5 &&&& 0.44\\
  \hline
\end{longtable*}

\newpage

\begin{longtable*}[c]{>{\centering}p{0.04\textwidth}>{\centering}p{0.08\textwidth}
    >{\centering}p{0.1\textwidth}>{\centering}p{0.08\textwidth}
    >{\centering}p{0.08\textwidth}>{\centering}p{0.08\textwidth}
    >{\centering}p{0.08\textwidth}>{\centering}p{0.08\textwidth}
    >{\centering\arraybackslash}p{0.08\textwidth}}
  \hline &&&& \textbf{pos.} &&&& $H$ \\[-0.2cm]
  \textbf{CP} & \multirow{2}{*}{\textbf{model}} & $T$ & \multirow{2}{*}{\textbf{rank}}
  & $x$ & \rhoc & \Lc & \multirow{2}{*}{$\epsilon$} & $\lambda_1$\\[-0.2cm]
  \textbf{\#} && [K] && $y$ & [\eA] & [\eAA] && $\lambda_2$\\[-0.2cm]
  &&&& $z$ &&&& $\lambda_3$\\
  \hline\hline\endhead
  \multirow{18}{*}{5} &&&& 0.667 &&&& 0.04\\[-0.3cm]
  & EHC\textsuperscript{a} & 100(2) &
  $(3, +3)$ & 0.333 & 0.07 & 0.87 & -- & 0.04\\[-0.3cm]
  &&&& 0.5 &&&& 0.79\\
  &&&& 0.667 &&&& -0.04\\[-0.3cm]
  & SHC & 298.3(1) & $(3, -1)$ & 0.333 & 0.10 & 0.41 & 0.00 & -0.04\\[-0.3cm]
  &&&& 0.5 &&&& 0.49\\
  &&&& 0.667 &&&& -0.05\\[-0.3cm]
  & SHC & 45.009(7) & $(3, -1)$ & 0.333 & 0.10 & 0.42 & 0.00 & -0.05\\[-0.3cm]
  &&&& 0.5 &&&& 0.52\\
  &&&& 0.667 &&&& -0.04\\[-0.3cm]
  & SHC & 25.086(5) & $(3, -1)$ & 0.333 & 0.10 & 0.42 & 0.00 & -0.04\\[-0.3cm]
  &&&& 0.5 &&&& 0.50\\
  &&&& 0.667 &&&& -0.04\\[-0.3cm]
  & SHC & 13.5(5) & $(3, -1)$ & 0.333 & 0.10 & 0.41 & 0.00 & -0.04\\[-0.3cm]
  &&&& 0.5 &&&& 0.49\\
  &&&& 0.667 &&&& -0.04\\[-0.3cm]
  & Ref.~\citenum{Tsirelson03} & RT & $(3, -1)$ & 0.333 & 0.11 & 0.41 & 0.00 & -0.04\\[-0.3cm]
  &&&& 0.5 &&&& 0.49\\
  &&&& 0.667 &&&&-0.01\\[-0.3cm]
  & DFT\textsuperscript{b} & -- & $(3, -1)$ & 0.333 & 0.11 & 0.44 & 0.00 &-0.01\\[-0.3cm]
  &&&& 0.5 &&&&0.45\\
  &&&& 0.667 &&&& -0.01\\[-0.3cm]
  & EHC\textsuperscript{c} & -- & $(3, -1)$ & 0.333 & 0.11 & 0.45 & 0.00 & -0.01\\[-0.3cm]
  &&&& 0.5 &&&& 0.47\\
  \hline
\end{longtable*}

\newpage

\begin{longtable}[c]{>{\centering}p{0.04\textwidth}>{\centering}p{0.08\textwidth}
    >{\centering}p{0.1\textwidth}>{\centering}p{0.08\textwidth}
    >{\centering}p{0.08\textwidth}>{\centering}p{0.08\textwidth}
    >{\centering}p{0.08\textwidth}>{\centering}p{0.08\textwidth}
    >{\centering\arraybackslash}p{0.08\textwidth}}
  \hline &&&& \textbf{pos.} &&&& $H$ \\[-0.2cm]
  \textbf{CP} & \multirow{2}{*}{\textbf{model}} & $T$ & \multirow{2}{*}{\textbf{rank}}
  & $x$ & \rhoc & \Lc & \multirow{2}{*}{$\epsilon$} & $\lambda_1$\\[-0.2cm]
  \textbf{\#} && [K] && $y$ & [\eA] & [\eAA] && $\lambda_2$\\[-0.2cm]
  &&&& $z$ &&&& $\lambda_3$\\
  \hline\hline\endhead
  \multirow{18}{*}{6} &&&& 0.766 &&&& -0.74\\[-0.3cm]
  & EHC\textsuperscript{a} & 100(2) &
  $(3, +1)$ & 0 & 0.19 & 1.93 & -- & 0.00\\[-0.3cm]
  &&&& 0.305 &&&& 2.67\\
  &&&& 0.763 &&&& -0.40\\[-0.3cm]
  & SHC & 298.3(1) & $(3, -1)$ & 0 & 0.15 & 2.23 & 13 & -0.03\\[-0.3cm]
  &&&& 0.298 &&&& 2.66\\
  &&&& 0.760 &&&& -0.41\\[-0.3cm]
  & SHC & 45.009(7) & $(3, -1)$ & 0 & 0.15 & 2.28 & 16 & -0.02\\[-0.3cm]
  &&&& 0.302 &&&& 2.71\\
  &&&& 0.760 &&&& -0.41\\[-0.3cm]
  & SHC & 25.086(5) & $(3, -1)$ & 0 & 0.15 & 2.25 & 13 & -0.03\\[-0.3cm]
  &&&& 0.301 &&&& 2.69\\
  &&&& 0.763 &&&& -0.39\\[-0.3cm]
  & SHC & 13.5(5) & $(3, -1)$ & 0 & 0.15 & 2.19 & 11 & -0.03\\[-0.3cm]
  &&&& 0.298 &&&& 2.61\\
  &&&& 0.816 &&&& -0.28\\[-0.3cm]
  & Ref.~\citenum{Tsirelson03} & RT & $(3, +1)$ & 0 &
  0.19 & 1.47 & 26 & 0.01\\[-0.3cm]
  &&&& 0.246 &&&& 1.74\\
  &&&& 0.771 &&&&-0.39\\[-0.3cm]
  & DFT\textsuperscript{b} & -- & $(3, -1)$ & 0 & 0.17 & 1.97 & 1697 &-0.00\\[-0.3cm]
  &&&& 0.294 &&&&2.36\\
  &&&& 0.771 &&&& -0.41\\[-0.3cm]
  & EHC\textsuperscript{c} & -- & $(3, -1)$ & 0 & 0.17 & 1.89 & 282 & -0.00\\[-0.3cm]
  &&&& 0.294 &&&& 2.31\\
  \hline
  \caption{Selected critical points (CP) as obtained from a QTAIM
    analysis of the electron density distribution $\rho(\mathbf{r})$
    in \Mg. Values were obtained from multipolar refinements based on
    the experimental XRD data and the static theoretical structure
    factors in this paper, from Tsirelson \textit{et
      al.}\cite{Tsirelson03}, and from a periodic DFT calculation.
    \textsuperscript{a}model E-EHCM1 in Tab.~\ref{tab:exp-ehc};
    \textsuperscript{b}unit cell parameters taken from E-EHCM1;
    \textsuperscript{c}model T-EHCM1 in Tab.~\ref{tab:theo-ehc}.}
  \label{tab:qtaim-analysis}
\end{longtable}

\newpage

\subsection{Laplacian maps}
\label{sec:lapl}

\subsubsection{SHC models using temperature-dependent experimental XRD
  data}

\begin{figure}
  \centering
  \includegraphics[width=0.8\textwidth]{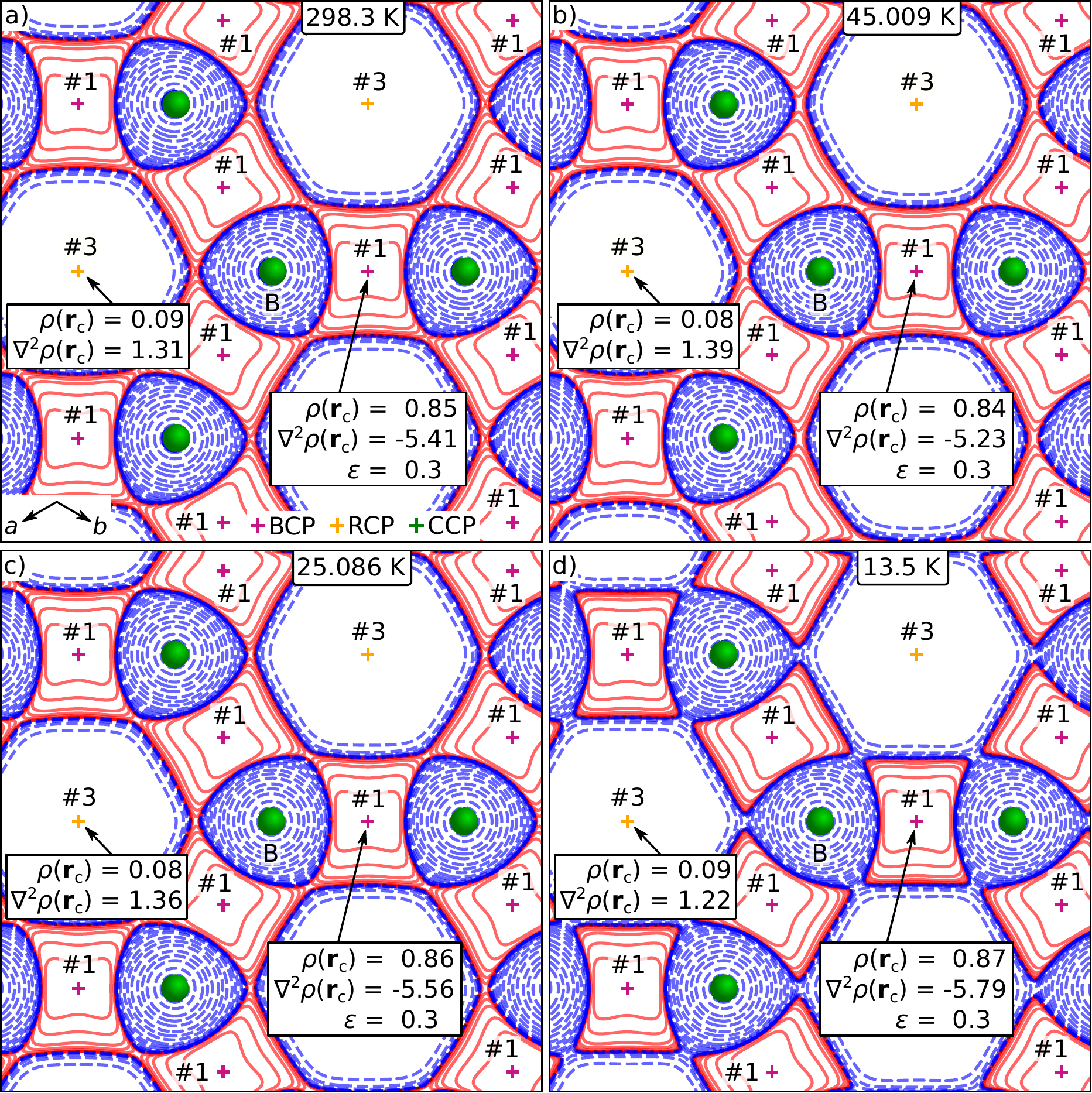}
  \caption{Maps of the Laplacian of the electron density,
    $\nabla^2 \rho(\mathbf{r})$, in the boron network parallel to the
    $a$-$b$ plane.  The underlying $\rho(\mathbf{r})$ were obtained
    from SHC multipolar models (Tab.~\ref{tab:lt-shc-1} and
    Tab.~\ref{tab:lt-shc-2}) refined using experimental XRD data
    collected at various temperatures between a)~298.3(1)~K and
    d)~13.5(5)~K. Contour lines are drawn for negative (solid red
    lines) and positive values of $\nabla^2 \rho(\mathbf{r})$ (dashed
    blue lines) at levels of $\pm 2 \cdot 10^n$, $\pm 4 \cdot 10^n$,
    and $\pm 8 \cdot 10^n$ with $n \in \{-2, \dots, 3\}$. Values of
    the electron density \rhoc\ (in \eA), the Laplacian \Lc\ (in
    \eAA), and the bond ellipticity (only for BCPs) are specified at
    the locations of critical points (indicated by crosses).  Their
    numbering refers to Tab.~\ref{tab:qtaim-analysis}.}
  \label{fig:lapl-crystal1-b-plane}
\end{figure}

\begin{figure}
  \centering
  \includegraphics[width=0.8\textwidth]{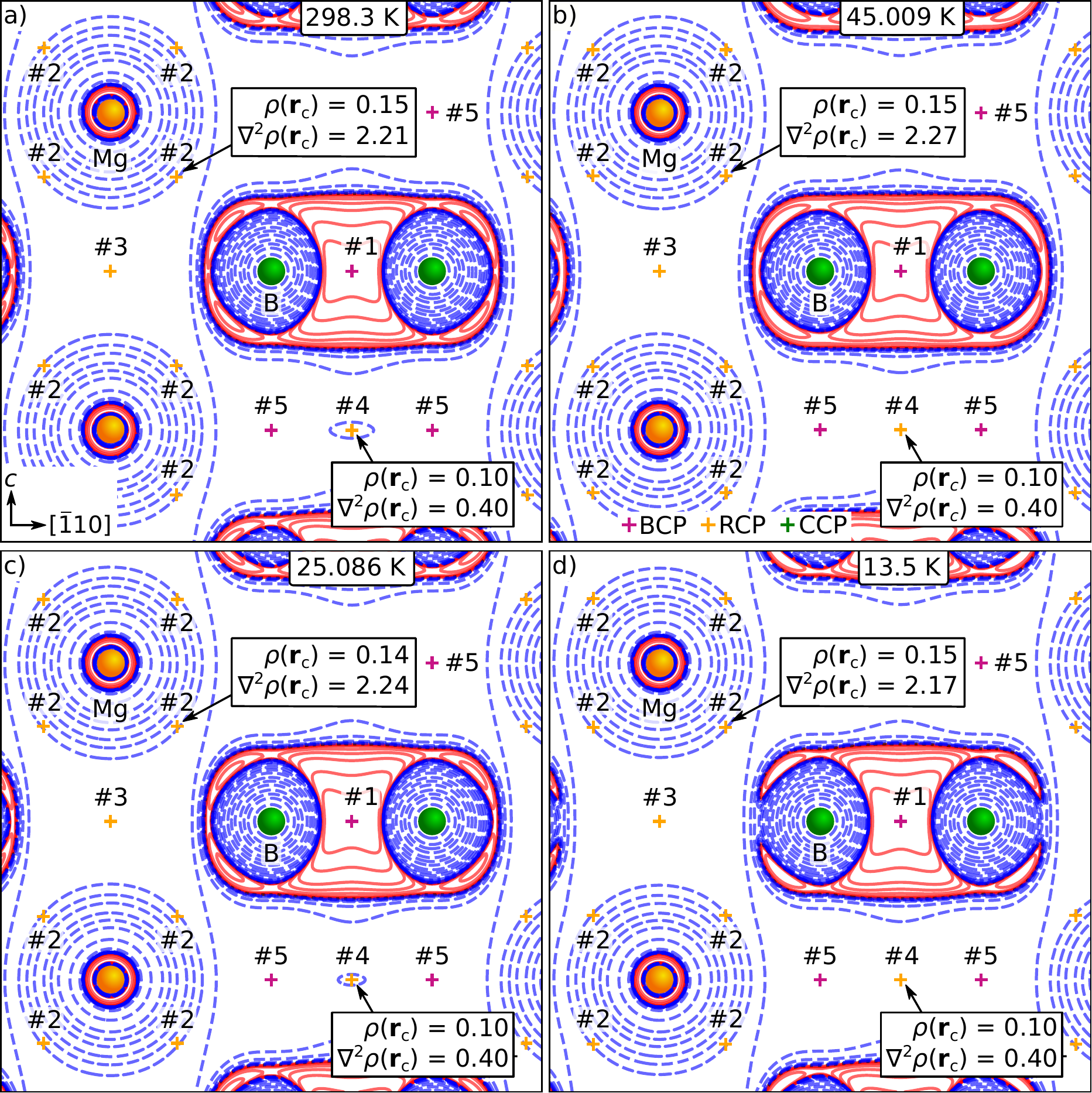}
  \caption{Maps of the Laplacian of the electron density,
    $\nabla^2 \rho(\mathbf{r})$, in a plane containing magnesium and
    boron atoms.  The underlying $\rho(\mathbf{r})$ were obtained from
    SHC multipolar models (Tab.~\ref{tab:lt-shc-1} and
    Tab.~\ref{tab:lt-shc-2}) refined using experimental XRD data
    collected at various temperatures between a)~298.3(1)~K and
    d)~13.5(5)~K. Contour lines are drawn for negative (solid red
    lines) and positive values of $\nabla^2 \rho(\mathbf{r})$ (dashed
    blue lines) at levels of $\pm 2 \cdot 10^n$, $\pm 4 \cdot 10^n$,
    and $\pm 8 \cdot 10^n$ with $n \in \{-2, \dots, 3\}$. Values of
    the electron density \rhoc\ (in \eA), the Laplacian \Lc\ (in
    \eAA), and the bond ellipticity (only for BCPs) are specified at
    the locations of critical points (indicated by crosses).  Their
    numbering refers to Tab.~\ref{tab:qtaim-analysis}.}
  \label{fig:lapl-crystal1-mg-b-plane}
\end{figure}

\FloatBarrier
\newpage

\section{Ab-initio ADP values}
\label{sec:DFT-ADPs}

\begin{longtable}[c]{>{\centering}p{0.15\textwidth}
    >{\centering}p{0.15\textwidth}>{\centering}p{0.15\textwidth}
    >{\centering}p{0.15\textwidth}
    >{\centering\arraybackslash}p{0.15\textwidth}}
    \hline
    $T$ [K] & $U_{11}$(Mg) [\AA$^2$] & $U_{33}$(Mg) [\AA$^2$] & $U_{11}$(B) [\AA$^2$] & $U_{33}$(B) [\AA$^2$]\\
    \hline\hline\endhead
    298.3 & 0.00488 & 0.00511 & 0.00365 & 0.00520\\
    45.009 & 0.00260 & 0.00263 & 0.00275 & 0.00351\\
    25.086 & 0.00258 & 0.00261 & 0.00273 & 0.00349\\
    13.5 & 0.00258 & 0.00260 & 0.00273 & 0.00348\\
    \hline
    \caption{Ab-initio ADP values as derived from lattice 
    dynamic DFT calculations using the finite-difference approach at temperatures corresponding to 
    the conducted low-temperature XRD studies.}
  \label{tab:DFT-ADPs}
\end{longtable}

\FloatBarrier
\newpage

\section{Electron density decomposition}
\label{sec:density-decomposition}

\begin{figure}[htb]
  \centering
  \includegraphics[width=0.9\textwidth]{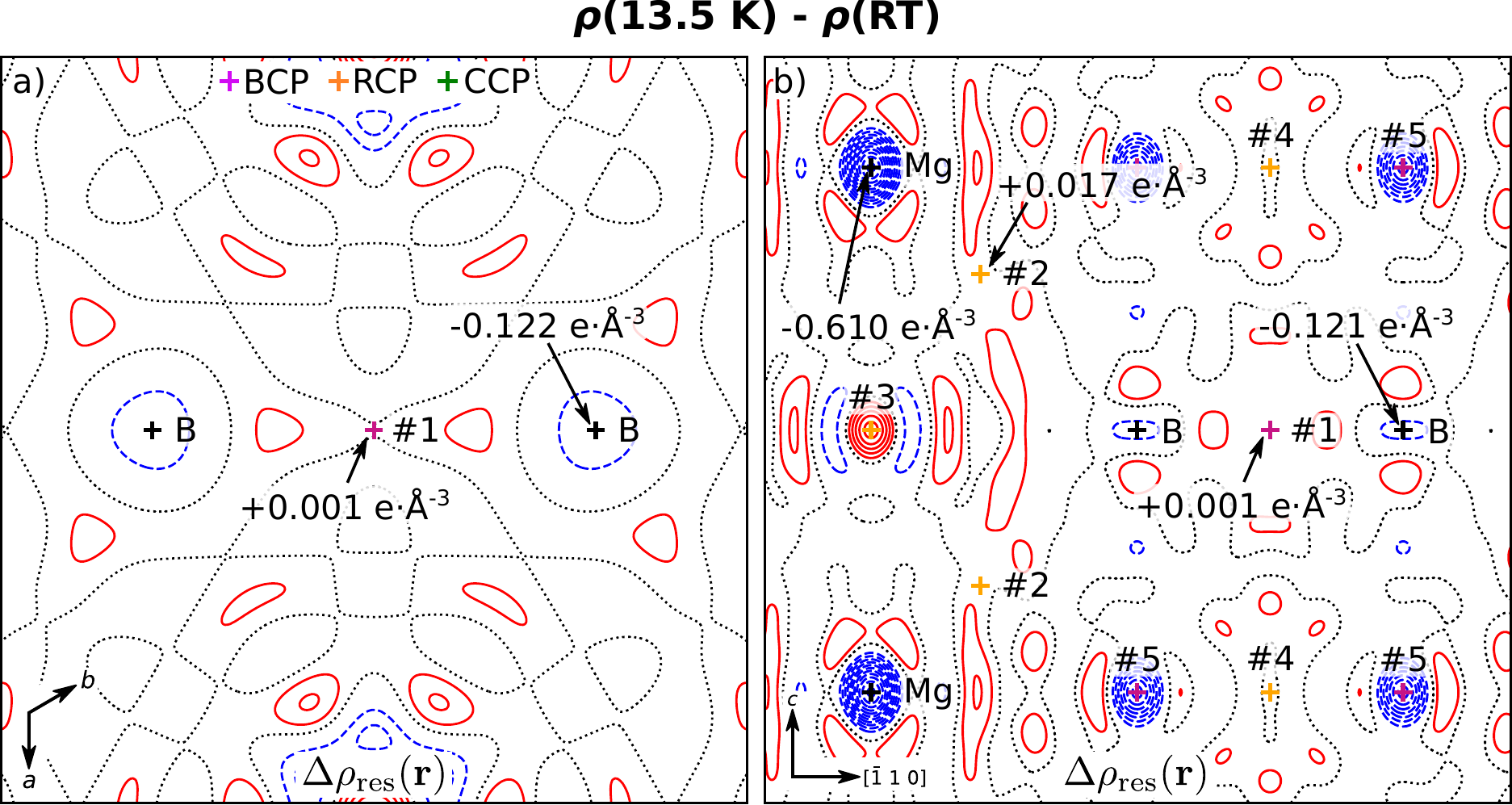}
  \caption{Iso-contour maps of the residual differences between
    electron density distributions at 13.5~K and RT in planes
    a)~parallel and b)~perpendicular to the hexagonal boron layers in
    \Mg. Contour lines at positive (red; solid), zero (black; dotted)
    and negative values (blue; dashed lines) are equally spaced by
    increments of $\pm$0.05~\eA. Crosses indicate the location of the
    QTAIM critical points specified in Tab.~\ref{tab:qtaim-analysis}.}
  \label{fig:rho-diff-decomposition-13p5K-RT-supp}
\end{figure}

\begin{figure}[htb]
  \centering
  \includegraphics[width=0.9\textwidth]{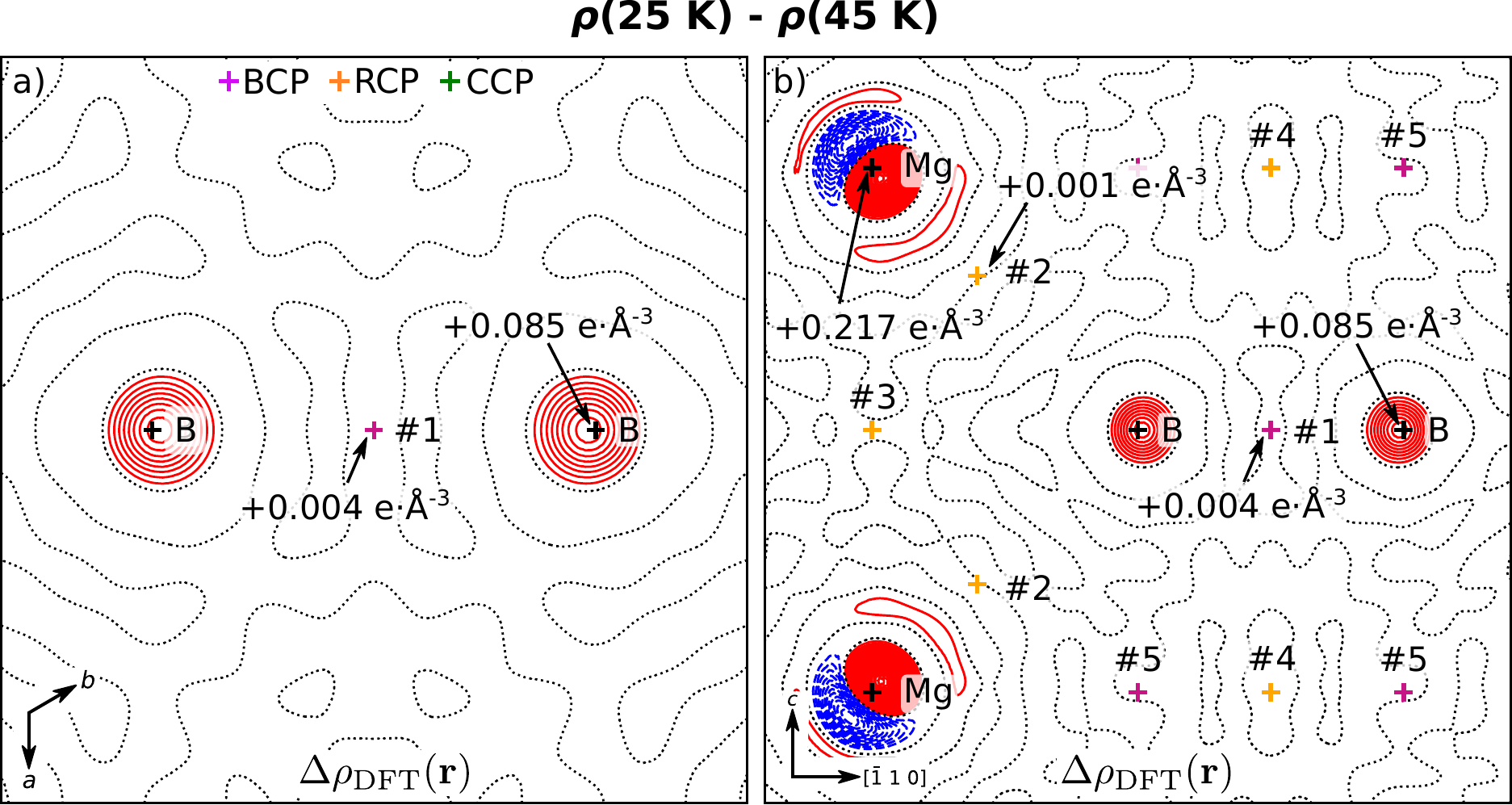}
  \caption{Iso-contour maps of the total differences between electron
    density distributions at 25~K and 45~K as derived by standard DFT
    calculations. Maps are oriented a)~parallel and b)~perpendicular
    to the hexagonal boron layers in \Mg.  Contour lines at positive
    (red; solid), zero (black; dotted) and negative values (blue;
    dashed lines) are equally spaced by increments of
    $\pm$\textbf{0.01}~\eA. Crosses indicate the location of the QTAIM critical
    points specified in Tab.~\ref{tab:qtaim-analysis}.}
  \label{fig:rho-diff-decomposition-25K-45K-supp}
\end{figure}

\FloatBarrier
\hspace{1em}
\newpage

\bibliography{manuscript}